\PassOptionsToPackage{dvipsnames}{xcolor}
\documentclass[11pt,a4paper]{article}
 
 \ifx\pdfoutput\not\undefined
 \pdfoutput=1
\fi
\IfFileExists{jheppub.sty}{%
\usepackage{jheppub}%
}{%
\usepackage{article_jhep_fallback}%
}
\usepackage{ifpdf}
\usepackage{mdframed}

\usepackage{graphicx,subcaption,comment}
\usepackage{float}
\usepackage{mathrsfs}
\usepackage{physics}
\graphicspath{{images/}{figures/}{Essay template/images/}}
\usepackage{mathtools}
\usepackage{epsfig}
\usepackage{pdfpages}
\usepackage{bbm}
\usepackage{graphicx,epstopdf}
\usepackage[numbers,sort&compress]{natbib}
\usepackage[makeroom]{cancel}
\usepackage{hyperref}
\hypersetup{pdftex,colorlinks=true,allcolors=blue}
\usepackage{array}
\usepackage[export]{adjustbox}

\usepackage[normalem]{ulem}

\usepackage[utf8]{inputenc}

\usepackage{tocloft}
\usepackage{bbold}
\usepackage{tikz}
\usetikzlibrary{backgrounds}
\usetikzlibrary{patterns}
\usetikzlibrary{arrows.meta}
\usetikzlibrary{3d,calc}
\usetikzlibrary{shapes,snakes}
\tikzstyle{bag} = [align=center]
\usetikzlibrary{decorations.pathmorphing}
\def\bea{\begin{eqnarray}}
\def\eea{\end{eqnarray}}

\usepackage{hyperref}
\usepackage{subcaption}

\newcommand{\eg}{{\it e.g.,}\ }
\newcommand{\ie}{{\it i.e.,}\ }
\newcommand{\mt}[1]{\textrm{\tiny #1}}

\newcommand{\ruv}[1]{r_{\mt{UV}}}

 \newcommand{\badat}{\begin{alignedat}}
 \newcommand{\eadat}{\end{alignedat}}
 
 \def\be{\begin{equation}}
\def\ee{\end{equation}}

\usepackage{xcolor}

\newcommand{\pink}[1]{\textcolor{\pink}{#1}}

\definecolor{dblue}{rgb}{0.2,0.50,0.80}

\def\H{\mathcal{H}}

\def\W{\mathcal{W}}

\def\T{\mathcal{T}}

\def\ruv{r_\mt{UV}}

\DeclareFontFamily{OT1}{pzc}{}
\DeclareFontShape{OT1}{pzc}{m}{it}{<-> s * [1.10] pzcmi7t}{}
\DeclareMathAlphabet{\mathpzc}{OT1}{pzc}{m}{it}

\definecolor{vert}{rgb}{0.1367 0.543 0.1367}
\providecolor{Green}{rgb}{0,0.5,0}
\providecolor{LimeGreen}{RGB}{50,205,50}

\title{Flat Space Entanglement from Wedge Holography}

\author[a]{Robert C. Myers}
\author[a,b,c]{and Bruno Sivilotti} 

\affiliation[a]{Perimeter Institute for Theoretical Physics, Waterloo, ON N2L 2Y5, Canada}
\affiliation[b]{Dept.~of Physics $\&$ Astronomy, University of Waterloo, Waterloo, ON N2L 3G1, Canada}
\affiliation[c]{Universidad de Buenos Aires, Departamento de F\'isica and IFIBA - CONICET\\
1428 Buenos Aires, Argentina}

\emailAdd{rmyers.perimeter@gmail.com}
\emailAdd{b.sivilotti@df.uba.ar}

\date{\today}

\abstract{We study holographic entanglement entropy in three-dimensional flat space by combining wedge holography with a controlled flat limit of AdS end-of-the-world branes. The limit takes a wedge of AdS$_4$ bounded by two AdS$_3$ branes to a Randall-Sundrum-like configuration of two flat branes, and the entanglement entropy of an interval in the two-dimensional corner CFT becomes that of an interval at null infinity of the branes. The result has the functional form of the entanglement entropy of a two-dimensional Carrollian CFT, the proposed holographic dual of gravity in asymptotically flat three-dimensional spacetimes. However, the inferred central charges yield $c_M=0$ and $c_L\to\infty$ in the flat limit. Adding a gravitational Chern-Simons term produces a $c_M$-like contribution, but we still find $c_L\to\infty$ and $c_M\to0$ in this limit. Moreover, the corresponding extremal surfaces do not approach the swing surfaces proposed by Song and collaborators. Instead, they collapse onto null rays along null infinity. We argue that both features are generic for flat limits of AdS/CFT. They follow from entanglement entropy being a property of the domain of dependence and from the equality of causal and entanglement wedges for intervals in the vacuum.}

\begin{document}

\maketitle

\section{Introduction}\label{sec:intro}

One of the most striking lessons of quantum gravity is the deep connection between spacetime geometry and quantum information. This connection finds its sharpest expression in holography, where a gravitational theory admits an equivalent description in terms of a lower-dimensional quantum theory without gravity. The best understood example is, of course, the AdS/CFT correspondence which relates quantum gravity in an asymptotically anti-de Sitter (AdS) spacetime to a conformal field theory (CFT) on its asymptotic boundary \cite{Maldacena:1997re, Witten:1998qj}. This duality translates questions about quantum gravity into questions about a non-gravitational quantum field theory, and vice versa.

A particularly useful entry in the holographic dictionary is the Ryu-Takayanagi (RT) prescription for entanglement entropy \cite{Ryu:2006bv,Ryu:2006ef}. Given a spacelike subregion $A$ in the boundary CFT, its entanglement entropy is computed by the area of an extremal or minimal-area surface $\Gamma_A$ in the bulk which is homologous to $A$,
\begin{equation}\label{RT}
    S_A =\min_{\Gamma_A} \frac{\mathrm{Area}(\Gamma_A)}{4G_N}.
\end{equation}
This formula gives a direct relation between a quantum information quantity in the boundary theory and the geometry of the bulk spacetime. In the case of $\text{AdS}_3/\text{CFT}_2$, this expression precisely reproduces numerous universal results for entanglement entropy in two-dimensional CFTs \cite{Holzhey:1994we,Calabrese:2004eu}. Further, if the boundary $\text{CFT}_2$ has a gravitational anomaly, \ie distinct left- and right-moving central charges, eq.~\eqref{RT} can be extended  to reproduce the anomalous contributions to the entanglement entropy, using topologically massive gravity in the bulk \cite{Castro:2014tta} -- see section \ref{chern-simons} for further discussion.

It is then natural to ask whether similar ideas can be applied beyond asymptotically AdS spacetimes. In particular, one would like to understand holography in asymptotically flat spacetimes. This problem is more subtle than in AdS, because flat spacetime does not have a timelike conformal boundary where an ordinary relativistic CFT naturally lives. Instead, the natural asymptotic regions are null infinity, spacelike infinity, and timelike infinity,  and the relevant asymptotic symmetry group is the $\text{BMS}$ group \cite{Sachs:1962wk, Bondi:1962px}. The putative dual theory at null infinity is often described as a Carrollian conformal field theory (CCFT), or equivalently a BMS-invariant field theory, of one dimension less \cite{Bagchi:2010zz,Bagchi:2012cy}. 

Using twist operators to compute the R\'enyi entropies, the authors of \cite{Bagchi:2014iea} computed the vacuum entanglement entropy of an interval in a two-dimensional Carrollian CFT.\footnote{We note that implicitly \cite{Bagchi:2014iea} employs the highest-weight representation of the CCFT$_2$.} The result consists of two terms governed by the central charges $c_L$ and $c_M$ of the corresponding BMS$_3$ algebra. The first, proportional to $c_L$, is a logarithmic term similar to that found in a relativistic $\text{CFT}_2$, while the second, proportional to $c_M$, is a term linear in the Carrollian time separation of the endpoints -- see eq.~\eqref{CCFT_EE_intro}. A goal of flat holography is then to reproduce this result geometrically in asymptotically flat three-dimensional spacetimes, and the swing-surface prescription of \cite{Jiang:2017ecm,Apolo:2020bld} accomplishes precisely this. Instead of the RT prescription \eqref{RT}, this construction uses null rays emanating from the interval endpoints at null infinity, connected by a spacelike geodesic through the bulk. 

Given our extensive understanding of holographic entanglement entropy (HEE) in the context of the AdS/CFT, a natural way to approach understanding flat space entanglement is through a limit of AdS/CFT in which the bulk curvature scale is sent to infinity, \eg see \cite{Lipstein:2025jfj,Eden:2026ulx,Alday:2024yyj,Compere:2019bua,Compere:2020lrt,Fitzpatrick:2010zm,Hijano:2019qmi,Polchinski:1999ry,Susskind:1998vk,Berenstein:2025tts,deGioia:2022fcn,deGioia:2024yne,Bagchi:2025vri,Hijano:2020szl, Nguyen:2026rwt, Nguyen:2026dpa}. A recent study \cite{Caminiti:2026pma} carefully examined this limit for HEE in the context of $\text{AdS}_3/\text{CFT}_2$. An alternative  route was proposed in \cite{Hao:2025ocu},\footnote{Another approach recently used Coulomb branch solutions in the context of D$p$-brane holography \cite{Jorstad:2026jlg}.} where they examined the Randall-Sundrum framework \cite{Randall:1999ee,Randall:1999vf} with two flat end-of-the-world (EOW) branes from the perspective of wedge holography \cite{Akal:2020wfl}. From the effective gravitational theory on the three-dimensional branes, it was suggested that this approach recovered the swing-surface prescription as the description of the HEE. Our goal in this paper is to examine this suggestion in detail. In particular, we show that the configuration with two flat branes can be reached as a controlled limit of the well understood wedge holography framework with AdS branes, and we examine the behaviour of HEE in this limit. 

Our starting point is wedge holography \cite{Akal:2020wfl}, in which gravity on a wedge of $\text{AdS}_{d+1}$ bounded by two EOW branes is dual to a $\text{CFT}_{d-1}$ on the corner where the branes meet asymptotically. With a Kaluza-Klein reduction on the interval between the branes, one finds that the  spectrum of the effective $d$-dimensional theory contains an exactly massless graviton for any choice of tensions, and at leading order it yields Einstein gravity on the branes \cite{Miao:2020oey,Hu:2022lxl}.
Next we combine wedge holography with the controlled flat limit of EOW branes introduced in \cite{Neuenfeld:2025wnl}. The brane tensions are taken to their critical values while a broken isometry of $\text{AdS}_{d+1}$ keeps the branes at a finite position in the bulk. The wedge then becomes a Randall-Sundrum~I configuration \cite{Randall:1999ee} with two flat branes,\footnote{We should add that in this limit, the spectrum also contains a massless scalar, the radion  \cite{Charmousis:1999rg,Chacko:2001radion} -- see appendix \ref{brane description} for further details.
\label{footy1}} and the corner where the $\text{CFT}_{d-1}$ lives becomes null infinity of the branes. Since every stage of the limit is isometric to an ordinary wedge, the HEE can be computed with standard holographic tools throughout. The flat-space entanglement entropy is then the limit of a well-defined $\text{CFT}_{d-1}$ entropy. Our focus is on $d=3$, where we can examine the connection to the swing-surface prescription \cite{Jiang:2017ecm,Apolo:2020bld} for HEE in asymptotically flat three-dimensional spacetimes.

Our main results are as follows: First, the HEE of an interval at null infinity of the branes has the expected form of entanglement entropy in a $\text{CCFT}_2$. However, the central charges appearing in this expression are peculiar, \ie the coefficient of the log term diverges as the effective curvature scale is taken to infinity (\ie as the branes become flat). Meanwhile the term linear in the time separation vanishes at all stages of the limit. Second, adding a gravitational Chern-Simons term on one of the branes produces the linear-in-time-separation term, but it still vanishes in the $\ell_{\text{eff}}\to\infty$ limit. 
Third, the extremal surfaces do not approach swing surfaces. Instead, their intersection with the branes collapses onto null generators of null infinity, running from the endpoints of the boundary interval to timelike infinity. Two different versions of the controlled flat limit yield the same HEE results and the same limiting surfaces. 
The divergence of the HEE and its failure to become a swing surface agree with the direct AdS$_3$/CFT$_2$ analysis of Caminiti and Pasterski \cite{Caminiti:2026pma}. Their work also identifies distinct observables associated with swing surfaces, which complements our present wedge construction.
Furthermore, the picture of bulk reconstruction emerging from our analysis is consistent with the algebraic approach of \cite{Laddha:2020kvp, Chen:2026zxc}.

The remainder of the paper is organized as follows: Section \ref{setup_section} reviews EOW branes in AdS, wedge holography and the HEE of an interval in the corner CFT, and then constructs a controlled flat limit of the wedge, the Poincar\'e-slab limit. Section \ref{results_section} takes the flat limit of the HEE, describes the fate of the extremal surface, and adds a gravitational Chern-Simons term on one brane. Section \ref{null-anchored-appendix} rederives these results with a second controlled flat limit, the null-anchored limit, in which the approach of the extremal surface to null infinity can be followed more easily. We close in section \ref{sec:disc} with a discussion of the implications of our results and the future outlook. 
Appendix \ref{brane description} explains various technical points for the effective three-dimensional gravity description of the AdS$_4$ wedge. Appendices \ref{Poin-slab-appendix} and \ref{NA_appendix} collect various technical details of the flat limits, including general non-symmetric intervals, the regulator at null infinity, and the frame dependence of the anomalous entropy functional. Appendix~\ref{wedge_symmetry_appendix} examines the realization of asymptotic symmetries as bulk diffeomorphisms and their behaviour in the flat limit.


\section{Setup: Wedge Holography and the Flat Limit}\label{setup_section}

In this section, we introduce the ingredients needed to take the controlled flat limit of the holographic entanglement entropy (HEE) in section \ref{results_section}. We begin by reviewing end-of-the-world (EOW) branes in AdS and their role in wedge holography. For $d=3$, we then compute the HEE of an interval in the corner CFT$_2$, and conclude by presenting a prescription for taking a controlled flat limit of the EOW-brane configuration.

\subsection{End-of-the-world branes in $\text{AdS}$}\label{branes in AdS}

We start by discussing the addition of codimension-one EOW branes in $\text{AdS}_{d+1}$. Although wedge holography requires two such branes, here we focus on a single one to highlight the essential features. We begin with the gravitational action
\begin{align}\label{action}
    S= \frac{1}{16\pi G_N}\int_{M} d^{d+1}x \sqrt{-g}\left(R+
    \frac{d(d-1)}{L^2}\right)+\frac{1}{8\pi G_N}\int_{Q} d^dx \sqrt{-h}\,(K-T)\,,
\end{align}
where $M$ is the $(d+1)$-dimensional bulk, $R$ is its Ricci scalar, $L$ is the bulk $\text{AdS}_{d+1}$ curvature scale, and $Q$ denotes the EOW brane.\footnote{With a single brane, the action would also include a GHY term \cite{York:1972sj,Gibbons:1976ue}, as well as various boundary counterterms \cite{Emparan:1999pm}, on the asymptotic AdS$_{d+1}$ boundary region. However, this asymptotic region is reduced to a codimension-two corner and hence these boundary terms  are removed in wedge holography.} 
Throughout this work, we use Greek letters for bulk indices and Latin letters for indices intrinsic to the brane. The induced metric on the brane is $h_{ab}$, and its extrinsic curvature is
\begin{align}\label{extrinsic_curv}
    K_{ab}={h_{a}}^{\lambda}\,{h_{b}}^{\sigma}\,\nabla_\lambda n_{\sigma}\,,
\end{align}
with $n_{\sigma}$ the outward-pointing unit normal to $Q$. In eq.~\eqref{action}, $K$ is the trace of $K_{ab}$ and $T$ is the brane tension. Hence the brane action is the Gibbons-Hawking-York (GHY) term \cite{York:1972sj,Gibbons:1976ue} supplemented by a tension term, and when several branes are present we include one such term for each brane. These conventions, including the sign of the tension, coincide with those of the AdS/BCFT literature \cite{Takayanagi:2011zk,Fujita:2011fp}. We work with EOW branes which act as a boundary where spacetime ends, rather than as two-sided branes with spacetime on both sides, for which the relations below between tension and geometry would be modified accordingly.

Varying the action with respect to the metric gives the bulk Einstein equations. Imposing Neumann boundary conditions on $Q$, \ie allowing the induced metric to fluctuate, yields the additional equation of motion
\begin{align}\label{brane eom}
    K_{ab}= h_{ab}(K-T)= \frac{T}{d-1}\, h_{ab}\,.
\end{align}
Combining eq.~\eqref{brane eom} with the Gauss--Codazzi equation \cite{Speranza:2019hkr} and the Riemann tensor of $\text{AdS}_{d+1}$, one finds that the intrinsic Riemann tensor of the brane is
\begin{align}\label{intrinsic riemann}
R^\mt{$(Q)$}_{abcd}=-\left(\frac{1}{L^2}-\frac{T^2}{(d-1)^2}\right)\left(h_{ac}\,h_{bd}-h_{ad}\,h_{bc}\right)\,.
\end{align}
That is, the brane is locally maximally symmetric. Depending on its tension, its geometry is $\text{AdS}_d$ for $|T|<T_c$, $\text{Mink}_d$ for $|T|=T_c$, and $\text{dS}_d$ for $|T|>T_c$, where the critical tension is given by $T_c=(d-1)/L$. In the following, we restrict to AdS or flat branes, \ie $|T|\leq T_c$. Comparing eq.~\eqref{intrinsic riemann} with the Riemann tensor of $\text{AdS}_d$, the curvature scale of the brane is
\begin{align}\label{length_scale_T}
    \ell_{\text{brane}}=L\,\frac{T_c}{\sqrt{T_c^2-T^2}}.
\end{align}

We can also relate the tension of the brane, and therefore its curvature scale, to the position of the brane in $\text{AdS}_{d+1}$. In Poincar\'e coordinates,
\begin{align}\label{Poincare metric}
    ds^2=\frac{L^2}{z^2}\left(dz^2-dt^2+dx_i dx^i\right)\,, \hspace{40pt} {\rm with}\ \ i=1, \dots, d-1,
\end{align}
where we denote $x\equiv x_1$, it is straightforward to see that a brane at $x=\alpha z$ solves eq.~\eqref{brane eom}. Its induced metric is that of $\text{AdS}_d$ with curvature scale $\ell_{\text{brane}}=L \sqrt{1+\alpha^2}$. The sign of the tension depends on which side of the brane is removed:
\begin{align}\label{alpha(T)}
    &\text{Removing $x<\alpha z$: }\,\,\,T=-T_c \frac{\alpha}{\sqrt{1+\alpha^2}}\,,\hspace{25pt}\text{Removing $x>\alpha z$: }\,\,\,T=T_c \frac{\alpha}{\sqrt{1+\alpha^2}}\,.
\end{align}
In particular, a positive-tension brane removes less than half of the Poincar\'e patch, while a negative-tension brane removes more than half.

As $|T|\to T_c$, the curvature scale \eqref{length_scale_T} diverges and the brane becomes flat. At fixed Poincar\'e coordinates, eq.~\eqref{alpha(T)} shows that this limit drives the brane towards the asymptotic boundary, with $|\alpha|\to\infty$. For $T\to+T_c$, the brane approaches the part of the boundary ``behind'' it, removing less and less of the $\text{AdS}_{d+1}$ spacetime. For $T\to-T_c$, instead, the region that is kept shrinks onto the boundary. Both cases will be relevant below, since in the wedge one of the two branes ends up with negative tension. We return to this point in section \ref{controlled_limit_section}.

\subsection{Wedge holography}\label{wedge_holo_section}

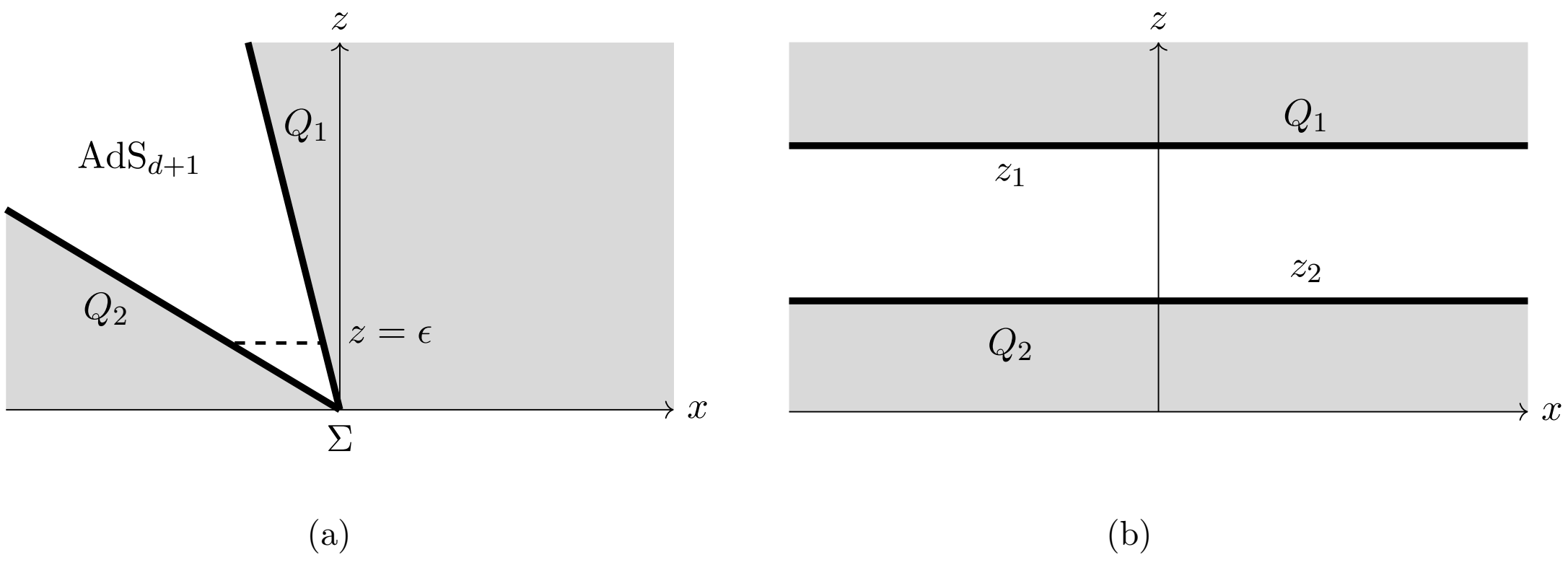
\begin{figure}
\centering
\resizebox{0.9\textwidth}{!}{%
\begin{minipage}{\textwidth}
\centering
\hspace{-1cm}
\begin{subfigure}[c]{0.44\textwidth}
\centering
\begin{tikzpicture}[scale=1.7]
\def\alpha{0.8}
\def\xzero{0}

\fill[gray!30]
(-2,0) --
plot[domain=-2:\xzero] (\x,{-0.6*(\x-\xzero)}) --
(-2,0) -- cycle;

\fill[gray!30]
(0,0) --
plot[domain=\xzero:2] (0,2.2) --(2,2.2)--
(2,0) -- cycle;
\fill[gray!30]
(0,0) --
plot[domain=\xzero:2] (0,0) --(-0.55,2.2)--
(0,2.2) -- cycle;

\draw[line width=2pt,black,domain=-2:\xzero]
plot (\x,{-0.6*(\x-\xzero)});

\draw[line width=2pt,black,domain=-0.55:\xzero]
plot (\x,{-\alpha*5*(\x-\xzero)});

\draw[dashed, line width=1pt,black,domain=-0.63:-0.1]
plot (\x,0.4);

\node at (0,-0.17) {$\Sigma$};
\node at (0.3,0.45) {$z=\epsilon$};

\draw[->] (-2,0) -- (2,0) node[right] {$x$};
\draw[->] (0,0) -- (0,2.2) node[above] {$z$};

\node at (-1.2, 1.5) {$\mathrm{AdS}_{d+1}$};
\node at (-1.4, 0.6) {$Q_2$};
\node at (-0.2, 1.7) {$Q_1$};
\end{tikzpicture}%

\vspace{3mm}
{\small (a)}

\phantomcaption
\label{fig:wedge_setup_a}
\end{subfigure}
\hspace{0.062\textwidth}
\begin{subfigure}[c]{0.48\textwidth}
\centering
\raisebox{0.32cm}{
\begin{tikzpicture}[scale=1.505]
\def\zA{0.75}
\def\zB{1.8}

\fill[gray!30]
(-2.5,0) --
plot[domain=-2.5:2.5] (\x,\zA) --
(2.5,0) -- cycle;

\fill[gray!30]
(-2.5,2.5) --
plot[domain=-2.5:2.5] (\x,\zB) --
(2.5,2.5) -- cycle;

\draw[line width=2pt,black,domain=-2.5:2.5]
plot (\x,\zA);

\draw[line width=2pt,black,domain=-2.5:2.5]
plot (\x,\zB);

\draw[->] (-2.5,0) -- (2.5,0) node[right] {$x$};
\draw[->] (0,0) -- (0,2.5) node[above] {$z$};

\node at (1,\zA+0.2) {$z_2$};
\node at (-1,\zA-0.3) {$Q_2$};

\node at (-1,\zB-0.2) {$z_1$};
\node at (1,\zB+0.2) {$Q_1$};
\end{tikzpicture}
}

\vspace{3mm}
{\small (b)}

\phantomcaption
\label{fig:wedge_setup_b}
\end{subfigure}
\end{minipage}%
}
\caption{(a) Wedge holography setup. Cut of $\mathrm{AdS}_{d+1}$ in Poincar\'e coordinates bounded by two AdS EOW branes $Q_1$ and $Q_2$ that intersect at $\Sigma$, with cutoff surface $z=\epsilon$. The shaded regions are removed. (b) Controlled flat limit of the wedge configuration. The two branes become flat and form a Randall-Sundrum~I configuration, with the UV brane $Q_2$ at $z=z_2$ and the IR brane $Q_1$ at $z=z_1>z_2$. (In section \ref{controlled_limit_section}, these final positions are denoted $z_{*,2}$ and $z_{*,1}$.)}
\label{fig:wedge_setup_combined}
\end{figure}

Wedge holography is a codimension-two extension of the $\text{AdS/CFT}$ correspondence in which a gravitational theory in a wedge-shaped region of AdS is dual to a lower-dimensional CFT localized at the corner of the wedge \cite{Akal:2020wfl}.\footnote{The reader can find discussions of these wedge configurations from alternative perspectives in \cite{Geng:2020fxl, Kogan:2000vb}.} The wedge is bounded by two EOW branes that intersect at the asymptotic boundary on a codimension-two surface $\Sigma$, where the dual CFT resides, as shown in Figure \ref{fig:wedge_setup_a}. Unlike standard $\text{AdS/CFT}$, where a $(d+1)$-dimensional bulk is dual to a $d$-dimensional non-gravitating theory, wedge holography relates the bulk gravitational theory to a non-gravitating theory two dimensions lower. In this correspondence, there are three useful descriptions of the theory \cite{Akal:2020wfl}
\begin{equation}\label{wedge_triality}
    \text{Gravity on a wedge of AdS$_{d+1}$}\;\simeq\;\text{KK-reduced gravity on AdS$_d$}\;\simeq\;\text{CFT on $\Sigma_{d-1}$}\,.
\end{equation}
The intermediate description above follows from a Kaluza-Klein (KK) reduction on the interval between the EOW branes, yielding a $d$-dimensional gravitational theory on the common AdS$_d$ slices. This description can be seen as closely related to double holography, \eg see \cite{Chen:2020uac,Chen:2020hmv}, where an EOW brane supports an effective gravitational theory in the spirit of Karch and Randall \cite{Karch:2000ct,Karch:2001cw}. The second EOW brane bounding the wedge removes the remaining asymptotic boundary, and hence it removes the non-gravitating bath in the dual theory. 

There is, however, an important difference between this KK reduction and the description associated with a single brane. For a single $\text{AdS}_d$ brane, the $d$-dimensional graviton is only quasi-localized and acquires a small mass, which vanishes only as $T\to T_c$ \cite{Karch:2000ct}. In the wedge, by contrast, the transverse interval is compact, and with Neumann boundary conditions on both branes, the spectrum contains an exactly massless, normalizable graviton for any choice of tensions, together with a tower of massive KK modes \cite{Akal:2020wfl,Miao:2020oey,Hu:2022lxl}. The massless graviton is therefore naturally understood as the zero mode of the KK reduction on the interval, rather than as a mode localized near either brane. At leading order, this reduction yields $d$-dimensional Einstein gravity, as described in appendix \ref{brane description} and shown in eq.~\eqref{3d params} below. The massive modes and corrections beyond leading order are analyzed in \cite{Miao:2020oey,Hu:2022lxl}. The leading-order description is reliable on distance scales large compared to the inverse mass of the first massive mode. In the Randall-Sundrum~I configuration reached by our flat limit (see section \ref{controlled_limit_section}), this mass scale depends on the warp factor between the two branes \cite{Randall:1999ee}. As noted in footnote \ref{footy1}, the lowest-lying KK scalar in the wedge bounded by AdS branes becomes massless in the flat limit and hence also contributes to the low energy theory in this limit -- see appendix \ref{brane description}.

\subsubsection*{Entanglement entropy in the corner theory} 

\begin{figure}
\centering
\begin{tikzpicture}[
    scale=0.90,
    line join=round,
    line cap=round,
    >=stealth
]

\colorlet{brane}{Cerulean!25}
\colorlet{region}{gray!35}

\filldraw[fill=brane, opacity=0.6]
(-2,-2.75) -- (-2,1.8) -- (0,2.6) -- (0,-2) -- cycle;

\filldraw[fill=brane, opacity=0.6]
(0,-2) -- (0,2.6) -- (2,1.8) -- (2,-2.75) -- cycle;

\draw[line width=1pt] (0,-2) -- (0,2.6);

\draw[line width=1.5pt,color=red] (0,-0.8) -- (0,1.4);

\draw[line width=1pt, color=Green]
(0,1.45)
arc[start angle=110,end angle=180,
x radius=1.91,y radius=1.25];
\draw[line width=1pt, color=Green]
(0,-0.8)
arc[start angle=285,end angle=180,
x radius=1,y radius=1.15];

\draw[line width=1pt, color=Green]
(0,1.45)
arc[start angle=70,end angle=0,
x radius=1.91,y radius=1.25];
\draw[line width=1pt, color=Green]
(0,-0.8)
arc[start angle=-105,end angle=0,
x radius=1,y radius=1.15];

\draw[dashed, line width = 1pt, color=LimeGreen]
(-1.25,0.2)
arc[start angle=180,end angle=360,
x radius=1.25,y radius=0.38];

\draw[dashed,line width = 1pt, color=LimeGreen]
(-1.1,-0.3)
arc[start angle=190,end angle=350,
x radius=1.12,y radius=0.38];
\draw[dashed, line width = 1pt,color=LimeGreen]
(-1.1,0.73)
arc[start angle=190,end angle=350,
x radius=1.12,y radius=0.38];

\draw[dashed, line width = 1pt,color=LimeGreen]
(-0.75,-0.7)
arc[start angle=210,end angle=330,
x radius=0.85,y radius=0.38];

\node at (-0.3,0.7) {\Large \textcolor{red}{$A$}};
\node[below] at (-1.4,-1.6) {\Large $Q_1$};
\node[below] at (1.4,-1.6) {\Large $Q_2$};
\node[below] at (0,-2.4) {\Large $W$};
\node[right] at (0,2.05) {\Large $\Sigma$};

\node[left] at (-0.9,1.1)
{\large \textcolor{Green}{$\gamma_A^{(1)}$}};
\node[right] at (0.9,1.1)
{\large \textcolor{Green}{$\gamma_A^{(2)}$}};

\draw[->, thick]
(-2.2,-0.15) -- (-0.5,-0.28);
\node[right] at (-3.2,-0.15)
{\Large $\Gamma_A$};

\end{tikzpicture}
\caption{Constant-time slice used to compute the HEE of $A\subset\Sigma$ in wedge holography. The two EOW branes are $Q_1$ and $Q_2$, and the minimal surface $\Gamma_A$ intersects them on $\gamma_A^{(1,2)}$.}
\label{fig:wedge_HEE}
\end{figure}
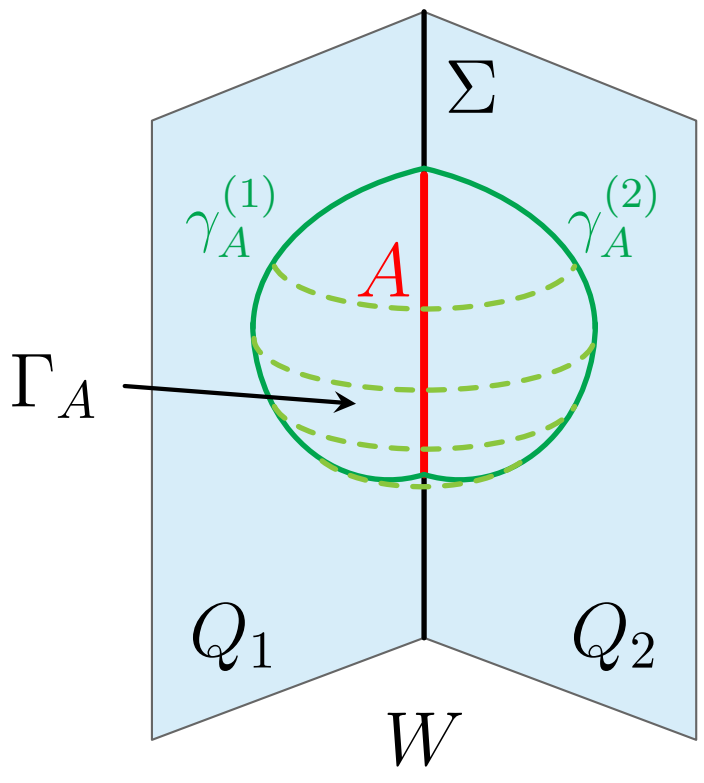

In a holographic duality between a gravitational theory and a non-gravitating CFT, a useful probe of the bulk geometry is the holographic entanglement entropy (HEE) of a subregion of the CFT \cite{Ryu:2006bv, Ryu:2006ef, Hubeny:2007xt}. For a $(d-2)$-dimensional region $A$ on a constant-time slice of the corner $\Sigma$, it was proposed in \cite{Akal:2020wfl} that the HEE in wedge holography is computed by the two-step minimization
\begin{align}\label{Minimization_Wedge}
    S_{A} = \min_{\text{ $\gamma_A^{(i)}$}}\left[\min_{\Gamma_A}\left[\frac{A(\Gamma_A)}{4G_N}\right]\right], \hspace{30pt} \text{with }\Bigg\{\begin{aligned}\partial\Gamma_A=&\ \gamma_A^{(1)}\cup\gamma_A^{(2)},\\ \partial\gamma_A^{(1)}=&\ \partial\gamma_A^{(2)}=\partial A.
    \end{aligned}
\end{align}
As shown in Figure \ref{fig:wedge_HEE}, we first choose two $(d-2)$-dimensional spacelike surfaces $\gamma_A^{(i)}\subset Q_i$ on the branes with $\partial\gamma_A^{(i)}=\partial A$, and a $(d-1)$-dimensional spacelike bulk surface $\Gamma_A$ with $\partial\Gamma_A=\gamma_A^{(1)}\cup\gamma_A^{(2)}$. We first minimize the area over variations of $\Gamma_A$ with fixed $\gamma_A^{(i)}$, and then we vary over the choices of $\gamma_A^{(i)}$. One finds that the second minimization is equivalent to requiring that $\Gamma_A$ meets both branes orthogonally, \eg \cite{Chen:2020uac}. For subregions that are not on a constant-time slice, minimal surfaces are replaced by extremal ones, as in the covariant HRT prescription \cite{Hubeny:2007xt}.

From this point forward, we work with $d=3$. That is, the bulk wedge is a portion of $\text{AdS}_4$ bounded by two $\text{AdS}_3$ EOW branes of the form \eqref{alpha(T)}, and the boundary theory is a $\text{CFT}_2$ on $\Sigma$. In this case, we apply the above prescription to evaluate the HEE for an interval in the two-dimensional corner theory.  However we note that much of our discussion carries over to higher dimensions.

It is convenient to slice $\text{AdS}_4$ into $\text{AdS}_3$ slices in Poincar\'e coordinates. Starting from eq.~\eqref{Poincare metric}, we change coordinates as
\begin{align}\label{change of variables}
    z= \frac{w}{\cosh\zeta},\hspace{30pt}x=-w \tanh{\zeta},\hspace{30pt}\zeta\in \mathbb{R},\,\,\,w\in(0,\infty),
\end{align}
in terms of which the metric becomes
\begin{align}\label{Poincare slicing}
    ds^2=L^2\left[d\zeta^2+\cosh^2\!\zeta\,\frac{dw^2-dt^2+dy^2}{w^2}\right]\,,
\end{align}
where we have set $y\equiv x_2$.
Each slice of constant $\zeta$ is an $\text{AdS}_3$ with curvature scale $L\cosh\zeta$, and the branes \eqref{alpha(T)} sit at constant $\zeta=\zeta_b$ with $\alpha=-\sinh\zeta_b$. We place the two EOW branes at $0<\zeta_1<\zeta_2$ and keep the region $\zeta\in[\zeta_1,\zeta_2]$. Since $x$ decreases with $\zeta$ at fixed $w$, the brane $Q_1$ at $\zeta_1$ lies to the right of the brane $Q_2$ at $\zeta_2$ in the $x$-$z$ plane, as shown in Figure \ref{fig:wedge_setup_a}. Using eq.~\eqref{alpha(T)}, their tensions and curvature scales are
\begin{align}\label{tensions_mu}
    T_1=-T_c\tanh\zeta_1,\hspace{30pt}T_2=T_c\tanh\zeta_2,\hspace{30pt}\ell_{\text{brane},i}=L\cosh\zeta_i.
\end{align}
The asymptotic boundary of the wedge is the corner $w=0$, \ie $z=x=0$. 
Hence, $t$ and $y$ parameterize the directions along the corner $\Sigma$, on which the boundary CFT$_2$ lives.

Now on the corner, we consider an interval of length $\sigma$,
\begin{align}\label{static_interval}
    I : \left\{\hspace{2pt}y\in\left[-\frac{\sigma}{2},\frac{\sigma}{2}\right],\hspace{2pt} w=0,\hspace{2pt} t=0\right\}.
\end{align}
The minimal surface for this interval is the part of the standard $\text{AdS}_4$ hemisphere $x^2+y^2+z^2=\sigma^2/4$ that lies inside the wedge \cite{Akal:2020wfl}, as illustrated in Figure \ref{fig:wedge_HEE}. In slicing coordinates, its position and induced metric are
\begin{align}\label{induced minimal surf}
\begin{aligned}
    \Gamma :\Bigg\{w^2 +y^2 =&\frac{\sigma^2}{4}, \hspace{10pt} y\in\left[-\frac{\sigma}{2},\frac{\sigma}{2}\right], \hspace{10pt} \zeta \in [\zeta_1, \zeta_2]\Bigg\}\,,\\
    ds^2_{\Gamma}=L^2&\left[d\zeta^2 +\cosh^2\!\zeta\,\frac{(\sigma/2)^2\,dw^2}{w^2[(\sigma/2)^2-w^2]}\right]\,.
    \end{aligned}
\end{align}
The hemisphere is the well-known extremal surface in AdS$_4$ anchored on the circular entangling surface $x^2+y^2=\sigma^2/4$ at the $z=0$ boundary \cite{Ryu:2006ef}. Hence, it also satisfies the bulk extremality condition within the AdS$_4$ wedge considered here. Since $\Gamma$ is independent of $\zeta$, the vector $\partial_\zeta$ is tangent to the surface. Therefore, $\Gamma$ meets both branes orthogonally, as required by the minimization procedure outlined in eq.~\eqref{Minimization_Wedge}. Moreover, its intersection with each brane is a semicircle in the $w$--$y$ plane, \ie a geodesic on the $\text{AdS}_3$ branes. We will use  this property in section \ref{chern-simons}.

To compute the area of $\Gamma$, we introduce the usual cutoff $z=\epsilon$, which translates into the $\zeta$-dependent cutoff $w\geq\epsilon\cosh\zeta$. We then find \cite{Akal:2020wfl}
\begin{align}\label{int_entropy}
\begin{aligned}
    S =& \frac{A(\Gamma)}{4G_N} = \frac{L^2}{2G_N}\int_{\zeta_1}^{\zeta_2}d\zeta\int_{\epsilon \cosh\zeta}^{\sigma/2}\!\!\!dw \hspace{5pt}\cosh\zeta\,\frac{\sigma/2}{w\sqrt{(\sigma/2)^2-w^2}} \\
    =&\frac{L^2}{2G_N}\left[\sinh\zeta_2-\sinh\zeta_1\right] \log\!\left(\frac{\sigma}{\epsilon}\right)-\frac{L^2}{2G_N}\int_{\zeta_1}^{\zeta_2}d\zeta \cosh\zeta\log(\cosh\zeta)+O(\epsilon)\,.
    \end{aligned}
\end{align}
The second term depends on the choice of regulator; for instance, a cutoff at fixed $w$ would remove it. It is therefore a non-universal constant, a point we return to in section \ref{results_section}. The first term matches the universal vacuum entanglement entropy of an interval of length $\sigma$ in a $\text{CFT}_2$ \cite{Calabrese:2004eu,Holzhey:1994we}, 
\begin{equation}\label{universal_entropy}
    S = \frac{c}{3}\,\log\left(\frac{\sigma}{\epsilon}\right) 
\end{equation}
with central charge
\begin{align}\label{holo_central_charge}
    c = \frac{3L^2}{2G_N}\,\big(\sinh\zeta_2-\sinh\zeta_1\big)= \frac{3L^2}{2G_N}\left[\frac{T_2}{\sqrt{T_c^2-T_2^2}}+\frac{T_1}{\sqrt{T_c^2-T_1^2}}\right].
\end{align}
Note that $c>0$ for any $\zeta_1<\zeta_2$, independently of the signs of the tensions.

In the above computation, we assumed that the intersection of $\Gamma$ with each brane $Q_i$ does not lie entirely in the region $z<\epsilon$, \ie that the lower limit of the $w$ integral is below $\sigma/2$ for all $\zeta\in[\zeta_1,\zeta_2]$. Since the maximum of $\cosh\zeta$ on this range is attained on one of the branes, this condition reads
\begin{align}\label{effective_cutoff_0}
    \frac{\sigma}{2}>\,\epsilon\,\cosh\zeta_i=\epsilon\,\frac{T_c}{\sqrt{T_c^2-T_i^2}}=\epsilon\,\frac{\ell_{\text{brane},i}}{L},\hspace{30pt}i=1,2,
\end{align}
where $\ell_{\text{brane},i}$ denotes the curvature scale on each of the branes, as given in eq.~\eqref{length_scale_T}. This reflects the fact that the bulk cutoff $\epsilon$ induces a cutoff on each brane, which in the Poincar\'e coordinates $(w,t,y)$ of the brane reads $\epsilon_{i}=\epsilon\,\ell_{\text{brane},i}/{L}$.
When we take the flat limit $\ell_{\text{brane},i}\to\infty$, it is these brane cutoffs that we keep finite. Since $\epsilon_i$ is a coordinate cutoff, this statement refers to the Poincar\'e coordinates of the wedge before the isometry that accompanies the limit is applied. For instance, in the Poincar\'e-slab limit of section \ref{controlled_limit_section}, holding $\epsilon$ fixed in the dilated coordinates is equivalent to keeping the $\epsilon_i$ finite in the original ones.

We can also consider a spacelike interval tilted in the time direction, with endpoints $\pm(\Delta y/2, \Delta t/2)$. By Lorentz invariance of the vacuum, its entropy is again given by eq.~\eqref{int_entropy}, now with $\sigma=\sqrt{\Delta y^2-\Delta t^2}$. The corresponding extremal HRT surface \cite{Hubeny:2007xt} is obtained by boosting eq.~\eqref{induced minimal surf}:
\begin{align}\label{min_surf_tilted}
    \Gamma :\bigg\{w^2 +y^2\left(1-\frac{\Delta t^2}{\Delta y^2}\right) =\frac{\sigma^2}{4},\hspace{10pt}t=\frac{\Delta t}{\Delta y}\,y,\hspace{10pt}y\in\left[-\frac{\Delta y}{2},\frac{\Delta y}{2}\right], \hspace{10pt} \zeta \in [\zeta_1, \zeta_2]\bigg\}.
\end{align}
Note that since the boost is an isometry of each $\text{AdS}_3$ slice, the intersection of this surface with each brane is again a geodesic on the brane.

Above, we wrote our results in terms of four-dimensional bulk quantities. To connect with the three-dimensional brane description, we express them in terms of brane parameters. Integrating out the bulk of the wedge induces, at leading order, an Einstein-Hilbert action on the branes (appendix \ref{brane description}). Choosing the induced metric on $Q_2$ as representative, this yields
\begin{align}\label{3d params}
\begin{aligned}
    G_N^{\text{eff}}=&\,\frac{G_N}{L}\frac{\cosh\zeta_2}{\sinh\zeta_2-\sinh\zeta_1}=\frac{G_N}{L}\frac{T_c}{\sqrt{T_c^2-T_2^2}}\frac{1}{\frac{T_2}{\sqrt{T_c^2-T_2^2}}+\frac{T_1}{\sqrt{T_c^2-T_1^2}}}\,,\\
    \ell_{\text{eff}}=&\,\ell_{\text{brane},2}=L\frac{T_c}{\sqrt{T_c^2-T_2^2}}\,,
    \end{aligned}
\end{align}
where $G_N^{\text{eff}}$ is the induced Newton's constant of the three-dimensional description and $\ell_{\text{eff}}$ is the curvature scale of $Q_2$. The central charge \eqref{holo_central_charge} then takes the standard Brown-Henneaux form \cite{Brown:1986nw}
\begin{align}\label{eff_c}
    c=\frac{3\,\ell_{\text{eff}}}{2\,G_N^{\text{eff}}}\,.
\end{align}
The choice of $Q_2$ as the representative above is a convention. The induced metrics on $Q_1$ and $Q_2$ differ only by the constant Weyl factor $\cosh^2\!\zeta_1/\cosh^2\!\zeta_2$. Choosing $Q_1$ instead would change $G_N^{\text{eff}}$ and $\ell_{\text{eff}}$ individually, but not their ratio 
and hence not the central charge $c$ of the corner theory.

\subsection{Flat limit of the EOW branes}\label{controlled_limit_section}

In this section, we construct the flat limit of the branes, which is used in the next section. Starting from the wedge of section \ref{wedge_holo_section}, we take the tensions of the branes to their critical values while simultaneously acting with a broken isometry of $\text{AdS}_4$, so that the branes become flat while remaining at finite positions in the bulk. The final setup, which we call the Poincar\'e-slab limit, is a Randall-Sundrum~I configuration \cite{Randall:1999ee}: a positive-tension (UV) brane and a negative-tension (IR) brane at constant values of the Poincar\'e coordinate $z$, with the corner of the wedge, where the dual CFT lives, pushed to null infinity of the flat branes. Our construction is based on the controlled flat limit of a single EOW brane introduced by Neuenfeld \cite{Neuenfeld:2025wnl}, which we extend to the wedge here. In section \ref{null-anchored-appendix}, we introduce an alternative controlled limit, the null-anchored limit, which keeps the asymptotic boundary of the branes at a finite position in Poincar\'e coordinates and is better suited to visualizing the extremal HRT surfaces near null infinity.

Eq.~\eqref{length_scale_T} shows that tuning $|T|\to T_c$ yields $\ell_{\text{brane}}\to\infty$ as desired but, as further discussed in section \ref{branes in AdS}, this tuning alone also pushes the brane to the asymptotic boundary. For $T\to+T_c$ the brane approaches the part of the boundary behind it, while for $T\to-T_c$ the region it bounds collapses onto the boundary. In either case the naive limit is degenerate, \ie the branes end up on the conformal boundary, where their induced metric diverges. Further, nothing in this simple limit makes the corner $\Sigma$, where the dual CFT lives, become null infinity of a flat region of the brane. For a single brane, the naive limit also renders the brane-localized graviton non-normalizable \cite{Neuenfeld:2025wnl}. Therefore, we must approach the flat limit differently.

One can think that the flat limit constructed by \cite{Neuenfeld:2025wnl} proceeds through a sequence of small increments, each involving two steps. First, we vary the brane tensions towards their critical values, which changes the positions of the branes and their intrinsic curvature. Second, we apply an AdS$_4$ isometry to move the resulting brane configuration to a convenient position in Poincar\'e coordinates, without changing the tensions. To understand this second step, recall that a single brane, or two branes meeting at a common corner, breaks the $\mathrm{SO}(3,2)$ isometry group of AdS$_4$ down to the $\mathrm{SO}(2,2)$ isometries of the AdS$_3$ slices in eq.~\eqref{Poincare slicing}. Acting with a broken generator maps a brane configuration to another that appears different in Poincar\'e coordinates but has the same tensions. Since an isometry acting on the entire configuration (branes, boundary subregion, and regulator) does not change any physical quantities, the physical content of the limit lies in the change of tensions. The accompanying isometry only selects which region of the branes remains at finite coordinate position as $|T|\to T_c$. We choose it so that the corner of the wedge becomes null infinity of the limiting flat branes. A practical consequence is that, at every stage of the limit, the configuration is isometric to a wedge of the form discussed in the previous section with the corresponding tensions. Hence, throughout the limit, all quantities can be computed using the results of that section, provided the interval and regulator are transformed accordingly.

The Poincar\'e-slab limit and the null-anchored limit of section \ref{null-anchored-appendix} are generated by two different broken isometries. Both lead to the same flat brane configuration and they only differ in their description in terms of the original Poincar\'e patch. That is, they differ in which coordinates are held fixed and hence in how null infinity is approached. The Poincar\'e-slab limit below provides the most direct route to the flat-space HEE, which is the subject of section \ref{results_section}.

\subsubsection*{Poincar\'e-slab limit}

Following \cite{Neuenfeld:2025wnl}, we work in the Poincar\'e coordinates \eqref{Poincare metric}, in which a single brane anchored at $x=x_0$ on the boundary is
\begin{align}\label{brane_eq_setup2}
        z=-\frac{\sqrt{T_c^2-T^2}}{T}(x-x_0)\,.
\end{align}
Taking $T\to T_c$ alone pushes the brane to the asymptotic boundary at $z=0$ and $x<x_0$ (assuming $T>0$ above). However, making use of a broken AdS$_4$ isometry, we can also vary $x_0$ while changing the tension, in a way that prevents the brane from being pushed to the boundary.

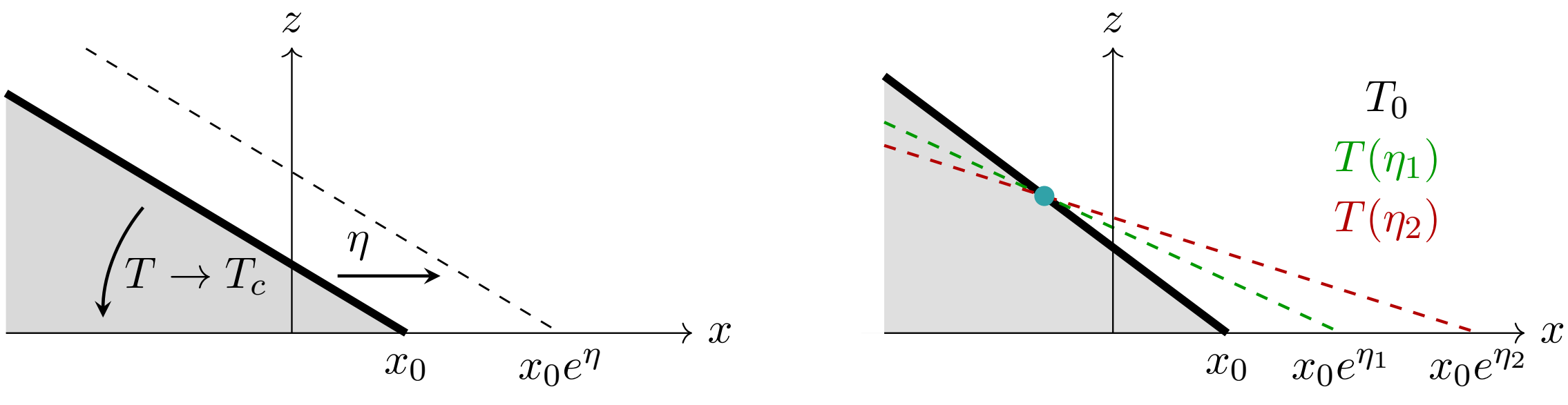
\begin{figure}

\begin{subfigure}[t]{0.48\textwidth}

\resizebox{1\linewidth}{!}{%
\centering
\begin{tikzpicture}[scale=1]
\useasboundingbox (-2.7,-0.7) rectangle (3.8,2.8);
\def\alpha{0.6}
\def\xzero{1}

\fill[gray!30]
(-2.5,0) --
plot[domain=-2.5:\xzero] (\x,{-\alpha*(\x-\xzero)}) --
(-2.5,0) -- cycle;

\draw[line width=2pt,black,domain=-2.5:\xzero]
plot (\x,{-\alpha*(\x-\xzero)});

\draw[->] (-2.5,0) -- (3.5,0) node[right] {$x$};
\draw[->] (0,0) -- (0,2.5) node[above] {$z$};

\node at (\xzero,-0.3) {$x_0$};

\draw[->,line width=0.8pt,>=stealth]
    (-1.3,1.1) arc[start angle=140,end angle=180,radius=1.5]
    node[pos=0.65,right,yshift=0pt] {$T \to T_c$} ;
\draw[->,line width=0.8pt,>=stealth]
    (0.4,0.5) --(1.3,0.5)
    node[pos=0.2,above,yshift=0pt] {$\eta$} ;
\draw[dashed, line width=0.5pt,black,domain=-1.8:\xzero+1.35]
plot (\x,{-\alpha*(\x-(\xzero+1.35))});

\node at (\xzero+1.35,-0.3) {$x_0 e^{\eta}$};
\end{tikzpicture}
}
\caption{}
\label{fig:brane transform_setup2_a}
\end{subfigure}
\hfill
\begin{subfigure}[t]{0.48\textwidth}

\resizebox{1\linewidth}{!}{%
\centering
\begin{tikzpicture}[scale=1]
\useasboundingbox (-2.7,-0.7) rectangle (3.8,2.8);
\def\xstar{-0.6}
\def\zstar{1.2}

\def\xA{1}
\def\xB{2}
\def\xC{3.2}

\pgfmathsetmacro{\alphaA}{\zstar/(\xA-\xstar)}
\pgfmathsetmacro{\alphaB}{\zstar/(\xB-\xstar)}
\pgfmathsetmacro{\alphaC}{\zstar/(\xC-\xstar)}

\fill[gray!25]
(-2,0) --
plot[domain=-2:\xA] (\x,{\alphaA*(\xA-\x)}) --
(-2.2,0) -- cycle;

\draw[line width=2pt,black,domain=-2:\xA]
plot (\x,{\alphaA*(\xA-\x)});

\node at (\xA,-0.3) {$x_0$};

\draw[dashed, line width=0.75pt,black!40!green,domain=-2:\xB]
plot (\x,{\alphaB*(\xB-\x)});
\node at (\xB,-0.3) {$x_0 e^{\eta_1}$};

\draw[dashed, line width=0.75pt,black!30!red,domain=-2:\xC]
plot (\x,{\alphaC*(\xC-\x)});
\node at (\xC,-0.3) {$x_0 e^{\eta_2}$};

\fill[black!20!Aquamarine] (\xstar,\zstar) circle (2.5pt);
\node[align=center] at (\xC-0.8,1.5) {$T_0$\\
\textcolor{black!40!green}{$T(\eta_1)$}\\
\textcolor{black!30!red}{$T(\eta_2)$}};

\draw[->] (-2,0) -- (3.6,0) node[right] {$x$};
\draw[->] (0,0) -- (0,2.5) node[above] {$z$};

\end{tikzpicture}
}
\caption{}
\label{fig:brane transform_setup2_b}
\end{subfigure}
\caption{(a) Cut of $\text{AdS}_4$ in Poincar\'e coordinates \eqref{Poincare metric} with an EOW brane of the form eq.~\eqref{brane_eq_setup2}. For a positive tension brane, we remove the grey region to the left of the brane. The figure also shows the effect of  applying the dilatation \eqref{dilatation}, and of increasing the tension of the brane. (b) We show how this configuration varies under the joint transformation \eqref{brane_eq_joint}. The solid line represents the initial brane
while the dashed lines represent the brane after the transformation for two different values of $\eta$ ($\eta_1<\eta_2$). 
The green dot indicates the fixed bulk point $(x_*,z_*)$.}
\end{figure}

The broken isometry that we implement is a dilatation of the Poincar\'e coordinates. For $x_0\neq0$, this isometry is broken by the brane \eqref{brane_eq_setup2} and therefore generates new brane configurations. After the transformation
\begin{align}\label{dilatation}
    t=e^{-\eta}t', \hspace{15pt}x=e^{-\eta}x',\hspace{15pt}y=e^{-\eta}y',\hspace{15pt}z=e^{-\eta}z',
\end{align}
the brane equation
\eqref{brane_eq_setup2} becomes
\begin{align}\label{brane_eq_setup3}
    z'=-\frac{\sqrt{T_c^2-T^2}}{T}(x'-x_0e^{\eta}).
\end{align}
Therefore, at fixed $x'$, the dilatation pushes the brane away from the asymptotic boundary. We show this effect, together with the effect of changing the tension, in Figure \ref{fig:brane transform_setup2_a}. In the following, we omit the primes on the coordinates after the dilatation for simplicity.

\paragraph{Joint flat limit:} The dilatation \eqref{dilatation} pushes the brane to larger $x$, while changing its tension changes its slope in the $x$--$z$ plane. We can combine these two to keep a bulk point $(x_*,z_*)$ on the brane fixed, as shown in Figure \ref{fig:brane transform_setup2_b}. To this end, we choose the tension as a function of the dilatation parameter as\footnote{At $\eta=0$, the brane interpolates between $(x,z)=(x_0,0)$ and $(x_*,z_*)$.}
\begin{align}\label{T(eta)_setting2}
    T(z_*,x_*,x_0,\eta)=T_c\frac{x_0e^{\eta}-x_*}{\sqrt{z_*^2+\left[x_0e^{\eta}-x_*\right]^2}},
\end{align}
with $x_*<x_0$, so that the tension remains positive for all $\eta\geq0$.
By construction, $\eta\to\infty$ implements $T\to T_c$. With this choice, the brane equation \eqref{brane_eq_setup3} becomes
\begin{align}\label{brane_eq_joint}
    z=\frac{z_*}{x_*-x_0e^{\eta}}(x-x_0e^{\eta}),\hspace{30pt}\lim_{\eta\to\infty}z = z_*.
\end{align}
Therefore, in this joint limit, the brane ends up sitting at constant $z=z_*$.
From eq.~\eqref{Poincare metric}, the induced metric on the limiting brane is $\text{Mink}_3$ rescaled by the constant factor $L^2/z_*^2$. This final configuration is analogous to the Randall-Sundrum~II (RS2) model \cite{Randall:1999vf}
(if we remove the region $z<z_*$ for a positive tension brane).\footnote{The interested reader can see how the brane transforms in global AdS$_4$ coordinates in appendix \ref{Poin-slab-appendix}.}

\paragraph{Flat limit of the wedge:}
Following eq.~\eqref{brane_eq_setup2} for the wedge geometry, we begin with the two branes $Q_1$ and $Q_2$ at
\begin{align}\label{brane_eq_setup4}
        z=-\frac{\sqrt{T_c^2-T_i^2}}{|T_i|}(x-x_0)\,,
\end{align}
so that they both meet the boundary at $x=x_0>0$. We will assume $T_2>0$ and $T_1<0$ and so we remove the regions $z>z_{Q_1}(x)$ and $z<z_{Q_2}(x)$, as in Figure \ref{fig:wedge_setup_a}. 
For simplicity, we anchor the branes to the bulk points $(x_{*,i}=0,z_{*,i})$, with $z_{*,1}>z_{*,2}$.\footnote{With this choice, $Q_1$ has negative tension throughout the limit. This is harmless, since the central charge \eqref{holo_central_charge} is positive for any $\zeta_1<\zeta_2$.}
Applying the joint limit for each of the branes, eqs.~\eqref{brane_eq_joint} and \eqref{T(eta)_setting2}
become
\begin{align}\label{wedge_branes_eta}
    Q_1:&\ \ z=z_{*,1}\left(1-\frac{x}{x_0e^{\eta}}\right),\hspace{50pt}Q_2:\ \,z=z_{*,2}\left(1-\frac{x}{x_0e^{\eta}}\right)\,,\\
    &\ \ T_1=-T_c\,\frac{x_0e^{\eta}}{\sqrt{z_{*,1}^2+x_0^2e^{2\eta}}}\,,\hspace{50pt}\ \, T_2=T_c\,\frac{x_0e^{\eta}}{\sqrt{z_{*,2}^2+x_0^2e^{2\eta}}}\,.
    \nonumber
\end{align}
Since $z_{*,1}>z_{*,2}$, $Q_1$ lies to the right of $Q_2$ for all $z>0$ and the branes never cross.\footnote{In terms of the slicing coordinates \eqref{Poincare slicing} centered at the corner (\ie with $x=x_0e^{\eta}-w\tanh\zeta$), the branes sit at $\sinh\zeta_i=x_0e^{\eta}/z_{*,i}$.} The central charge \eqref{holo_central_charge} and the brane parameters \eqref{3d params} are therefore
\begin{align}\label{c_and_params_eta}
    c=\frac{3L^2}{2G_N}\,x_0e^{\eta}\left(\frac{1}{z_{*,2}}-\frac{1}{z_{*,1}}\right)\,,\hspace{16pt}
    \ell_{\text{eff}}=L\sqrt{1+\frac{x_0^2e^{2\eta}}{z_{*,2}^2}}\,,\hspace{16pt}
    G_N^{\text{eff}}=\frac{G_N}{L}\,\frac{\sqrt{1+x_0^2e^{2\eta}/z_{*,2}^2}}{x_0e^{\eta}\left(1/z_{*,2}-1/z_{*,1}\right)}\,.
\end{align}

In the flat limit $\eta\to\infty$, the brane $Q_1$ ends up at $z=z_{*,1}$ and $Q_2$ at $z=z_{*,2}<z_{*,1}$, and the wedge becomes the region $z\in [z_{*,2},z_{*,1}]$ in between the two flat branes, as shown in Figure \ref{fig:wedge_setup_b}. From eq.~\eqref{wedge_branes_eta}, $Q_2$ finishes with tension $T_2=T_c$, while $Q_1$ ends up with $T_1=-T_c$. 
The final configuration is identical to a Randall-Sundrum~I (RS1) model \cite{Randall:1999ee}, with $Q_2$ being the UV brane and $Q_1$, the IR brane. Meanwhile, the corner of the wedge moves off to $x\to\infty$, \ie towards null infinity of the boundary Minkowski space at the edge of the Poincar\'e patch. From eq.~\eqref{c_and_params_eta}, we see that in the limit, $c\to\infty$ and $\ell_{\text{eff}}\to\infty$, while $G_N^{\text{eff}}\to G_N/[L(1-z_{*,2}/z_{*,1})]$ remains finite.

An advantage of this way of taking the controlled flat limit is that the limiting geometry of the branes is very simple. The induced metric on the branes is $\text{Mink}_3$ multiplied by the constant $L^2/z_{*,i}^2$. A drawback is that the asymptotic boundary of the branes is pushed to $x\to\infty$, which obscures the near-boundary behaviour and complicates the visualization of the extremal surfaces anchored there. For this reason, in section \ref{null-anchored-appendix}, we describe a second flat limit generated by a different broken isometry of $\text{AdS}_4$. In this ``null-anchored limit'', the boundary of the branes remains anchored in the Poincar\'e coordinates, so the region that becomes null infinity is easy to study, at the cost of a less transparent description of the limiting Minkowski geometry.

%
%
%

\section{Flat Limit of Holographic Entanglement Entropy}\label{results_section}

Having assembled the ingredients in section \ref{setup_section}, we now compute the entanglement entropy of an interval at null infinity of the flat branes. The computation is an ordinary AdS/CFT one at every stage. As emphasized at the start of section \ref{controlled_limit_section}, each step of the controlled limit is isometric to a wedge of the form studied in section \ref{wedge_holo_section}, so the entropy is always given by eq.~\eqref{int_entropy}. What changes along the limit are the tensions of the branes, together with the position of the interval and of the regulator. Our task is therefore to identify the family of boundary intervals which flows to a given interval at null infinity, and to follow eq.~\eqref{int_entropy} along that family.

We will compare the result with the vacuum entanglement entropy of an interval in a $\text{CCFT}_2$ on a circle of circumference $a$, which was computed with the replica trick in a highest-weight representation in \cite{Bagchi:2014iea},
\begin{equation}
    S =
    \frac{c_L}{6}\log\left(\frac{a}{\pi\epsilon}\sin\frac{\Delta \phi}{2}\right)
    + \frac{c_M}{12}\,
    \Delta u \cot\frac{\Delta \phi}{2}\,,
    \label{CCFT_EE_intro}
\end{equation}
where $u$ is the Carrollian time, $\phi\sim\phi+2\pi$ is the angular coordinate, and $c_L$ and $c_M$ are the central charges of the $\text{BMS}_3$ algebra \eqref{BMS3}.  For Einstein gravity, the asymptotic symmetry analysis gives $c_L=0$ and $c_M=3/G_N^{(3)}$ \cite{Barnich:2006av}, and the swing surfaces of \cite{Jiang:2017ecm,Apolo:2020bld} reproduce eq.~\eqref{CCFT_EE_intro} with these values. We examine the distinction between the symmetries acting on local operators and the globally defined gravitational asymptotic symmetries in appendix~\ref{wedge_symmetry_appendix}, including their different contractions and associated central charges.

\subsection{HEE in the Poincar\'e-slab limit}\label{results_Setting2}

To reduce the clutter in our expressions, we simplify our notation as follows: We abbreviate
\begin{align}\label{Xdef}
    X\equiv x_0\,e^{\eta}\,,
\end{align}
for the position of the corner of the wedge. Further, we write the anchor points as $(x_{*,i},z_{*,i})=(0,z_i)$, \ie we replace $z_{*,i}\to z_i$ in eq.~\eqref{wedge_branes_eta}. Then, the flat limit is $X\to\infty$, at the end of which the UV brane $Q_2$ sits at $z=z_2$ and the IR brane $Q_1$, at $z=z_1>z_2$.\footnote{Let us add that as in the discussion of the Poincar\'e slab limit above, we work here in the Poincar\'e coordinates after the dilatation \eqref{dilatation} with the primes dropped.} 

\paragraph{The interval at null infinity:}
Since the boundary of the branes moves during the limit, we must first determine which boundary interval flows to the desired interval at null infinity. It is useful to introduce the Bondi-like coordinates for the boundary directions
\begin{align}\label{null_coordinates}
    \tilde r=\sqrt{x^2+y^2},\hspace{20pt}\tan\phi=\frac{y}{x},\hspace{20pt}\tilde u=t-\sqrt{x^2+y^2}\,,
\end{align}
in terms of which the bulk AdS$_4$ metric \eqref{Poincare metric} becomes
\begin{align}\label{bondi_bulk}
    ds^2=\frac{L^2}{z^2}\left[dz^2-d\tilde u^2-2\,d\tilde u\,d\tilde r+\tilde r^2d\phi^2\right]\,.
\end{align}
Hence with the Poincar\'e-slab limit, the induced metric on $Q_2$ becomes $(L/z_2)^2$ times the flat $\text{Mink}_3$ metric in Bondi coordinates. We can absorb this overall scale factor by writing the induced metric in terms of  proper Bondi coordinates \begin{align}\label{bondi_physical}
    u=\frac{L}{z_2}\,\tilde u\,,\hspace{40pt} r=\frac{L}{z_2}\,\tilde r\,.
\end{align}
We reserve $u$ and $r$ for the physical Bondi (or equivalently, Carrollian) time and radius on the brane and use the tilded coordinates only in the intermediate steps.\footnote{As in eq.~\eqref{3d params}, we use the metric on $Q_2$ as the representative of the three-dimensional description. Using $Q_1$ instead would rescale $u$ and $r$ by the constant $z_2/z_1$ and would not change the entropy.}

An interval at null infinity $\tilde r\to\infty$ of the flat branes with endpoints $(\tilde u,\phi)=(\tilde u_\pm,\phi_\pm)$ therefore descends, at each stage of the limit, from the interval on the boundary of the wedge $z=0$, $x=X$ with endpoints
\begin{align}\label{endpoints_poinc_slab}
    (t,y)_{\pm}=\left(\tilde{u}_{\pm}+\frac{X}{\cos\phi_{\pm}},\,\,X\tan\phi_{\pm}\right).
\end{align}
Below, we simplify our analysis by placing the interval symmetrically about $\phi=0$, \ie
\begin{align}\label{symmetric_interval}
    \phi_\pm=\pm\frac{\Delta\phi}{2}\,.
\end{align}

\paragraph{Regulators:}
 Before proceeding, it is worth being explicit about the regulator. At each stage of the limit we cut off the bulk at $z=\epsilon$, in the dilated coordinates \eqref{dilatation}. By eq.~\eqref{effective_cutoff_0}, this induces a cutoff $\epsilon_i=\epsilon\,\ell_{\text{brane},i}/L$ on the brane $Q_i$ in the Poincar\'e coordinates of the brane. Keeping $\epsilon$ fixed along the limit means that the original cutoff in the non-dilated coordinates is taken to zero and $\epsilon\ell_{\text{brane},i}/L$ is finite in those coordinates. Equivalently, in the dilated coordinates $\epsilon_i$ is taken to infinity, but so is the proper length of the interval $\sigma$ so that eq.~\eqref{effective_cutoff_0} is still satisfied. 
 
 In contrast, in flat space one would normaly regulate an interval at null infinity on a cut of fixed radius $\tilde r=R_\epsilon$, with a short-distance cutoff along that cut. These two prescriptions are not related by a uniform rescaling. The surface $z=\epsilon$ meets the brane on an angle-dependent cut of flat space (see eq.~\eqref{angle_dependent_cutoff}), and the corner metric differs from the metric on a cut of fixed $\tilde r$ by the $\phi$-dependent Weyl factor discussed below eq.~\eqref{standard_null_metric}. The latter contributes an interval-dependent term to the entropy, which we compute in eq.~\eqref{shift_weyl_factor}. For a non-symmetric interval, the two endpoints are moreover regulated on different $\tilde{r}$-cuts, which produces the additional $\bar\phi$-dependent terms discussed in appendix \ref{Poin-slab-appendix}. Once these effects are accounted for, the remaining freedom in the regulator is the usual interval-independent rescaling $\epsilon\to\lambda\epsilon$, which shifts $S$ by $-\frac{c}{3}\log\lambda$. In our limit, we choose the (dimensionless) regulator at null infinity to be 
\begin{align}\label{regulator_def}
    \hat\epsilon\equiv\frac{\epsilon}{X}\,e^{\Lambda-1}\,,\hspace{40pt}
    \Lambda\equiv\frac{z_1\log(X/z_2)-z_2\log(X/z_1)}{z_1-z_2}\,,
\end{align}
which remains finite as $X\to\infty$ with $\epsilon$ fixed, since $\Lambda=\log X+\mathcal{O}(1)$. As shown in appendix \ref{Poin-slab-appendix}, the factor $e^{\Lambda-1}$ is precisely what is needed to absorb the non-universal constant of eq.~\eqref{int_entropy}, which diverges in the flat limit as well. With this choice, we hold $\hat\epsilon$  fixed as we take the limit\footnote{The dimensionless $\hat\epsilon$ is the UV cutoff measured in units of the proper radius of the cutoff surface near null infinity on which it is imposed. From this perspective, the universal CFT$_2$ result for the entanglement entropy of an interval on a circle of circumference $a$ becomes \cite{Holzhey:1994we,Calabrese:2004eu}
$$
    S = \frac{c}{3}\,\log\!\left(\frac{a}{\pi \epsilon_{\text{prop}}}\sin\frac{\Delta\phi}{2}\right) = \frac{c}{3}\,\log\!\left(\frac{2}{\hat\epsilon}\sin\frac{\Delta\phi}{2}\right)\,.
$$
That is, we set $\hat\epsilon=2\pi\epsilon_{\text{prop}}/a$.} and there are no remaining divergent constants in the resulting entropy. 


\paragraph{Holographic entanglement entropy:}  Next we examine the flat limit of our result \eqref{int_entropy} for the holographic entanglement entropy calculated using wedge holography. Recall that this expression applies for an interval that is tilted in time, as described by eq.~\eqref{min_surf_tilted}.
For the symmetric interval \eqref{symmetric_interval}, the invariant separation of the endpoints \eqref{endpoints_poinc_slab} is
\begin{align}\label{sigma_sym}
    \sigma^2=4X^2\tan^2\frac{\Delta\phi}{2}-\Delta\tilde u^2\,,
\end{align}
with $\Delta\phi=\phi_+-\phi_-$ and $\Delta\tilde u=\tilde u_+-\tilde u_-$. Hence, expanding in the flat limit $X\to\infty$ with fixed $\Delta\tilde u$,
\begin{align}\label{intermediate_log_PS}
    \log\left(\frac{\sigma}{\epsilon}\right)\simeq\log\left(\frac{2X}{\epsilon}\tan\frac{\Delta\phi}{2}\right)-\frac{\Delta \tilde{u}^2}{8X^2}\cot^2\frac{\Delta\phi}{2}+\mathcal{O}(X^{-4})\,.
\end{align}
The argument of the logarithm here involves $\tan(\Delta\phi/2)$, whereas the (expected) Carrollian result \eqref{CCFT_EE_intro} involves the logarithm of $\sin(\Delta\phi/2)$. 
We now explain this apparent mismatch and show how the two expressions are in fact compatible.

The key point is that,
in our construction the holographic CFT sits at fixed $x$ rather than at fixed $\tilde r$. That is, the corner of the wedge lies on the boundary of the branes at $z=0$, $x=X$, so that at the end of the limit the asymptotic boundary of the branes is formally at $x\to\infty$, while null infinity is reached with $\tilde r\to\infty$ in eq.~\eqref{bondi_bulk}. The induced metrics on these two surfaces are different, as we discuss in more detail in appendix \ref{Poin-slab-appendix}. The two limits yield
\begin{align}\label{standard_null_metric}
    \lim_{\tilde r\to\infty}\frac{ds^2|_{\tilde r}}{\tilde r^2}=d\phi^2\,,\hspace{40pt}\lim_{x\to\infty}\frac{ds^2|_{x}}{x^2}=\sec^2\!\phi\,\,d\phi^2\,,
\end{align}
which differ by the Weyl factor $e^{2w}=\sec^2\!\phi$. Under a Weyl rescaling $g\to e^{2w}g$, the entanglement entropy of an interval in a two-dimensional CFT changes by \cite{Calabrese:2004eu}
\begin{align}\label{weyl_entropy_contribution}
    \Delta S=\frac{c}{6}\left[w(\text{1st endpoint})+w(\text{2nd endpoint})\right]\,.
\end{align}
These new contributions arise from the change $\epsilon\to\epsilon\,e^{w}$ of the {proper} cutoff. For the symmetric interval \eqref{symmetric_interval}, the endpoints are at $\phi=\pm\Delta\phi/2$ and so
\begin{align}\label{shift_weyl_factor}
    S_{\text{Fixed }\tilde r}=S_{\text{Fixed }x}+\frac{c}{3}\log\left(\cos\frac{\Delta\phi}{2}\right),
\end{align}
which replaces the $\tan(\Delta\phi/2)$ in eq.~\eqref{intermediate_log_PS} by $\sin(\Delta\phi/2)$. Note that the overall factor $L^2/z_2^2$, which relates the induced metric on $Q_2$ to $\text{Mink}_3$, multiplies both metrics and therefore does not enter their relative Weyl factor.

Two comments are in order before assembling the result. First, our calculations above were simplified by considering the symmetric interval \eqref{symmetric_interval}. Appendix \ref{Poin-slab-appendix} considers the general case where the resulting entropy appears to contain terms which depend on the angular position $\bar\phi=(\phi_++\phi_-)/2$ of the interval.
As explained there, this $\bar\phi$-dependence arises because the cutoff surfaces $z=\epsilon$ and $\tilde{r}=R$ are inequivalent and is removed by properly accounting for the appropriate conformal transformation, similar to the discussion leading to eq.~\eqref{shift_weyl_factor}.\footnote{The corresponding analysis of the Chern-Simons contribution in section \ref{chern-simons} appears in appendix \ref{Poin-slab-appendix}.}
Second, since each stage of the limit is isometric to a standard wedge, and since the transformation \eqref{dilatation} is a dilatation, the extremal HRT surface is not deformed along the limit, but merely rescaled. We return to its geometry in section \ref{geometrical_picture_section}.

We can now assemble the entropy. Combining eqs.~\eqref{int_entropy}, \eqref{intermediate_log_PS} and \eqref{shift_weyl_factor}, we find
\begin{align}\label{S_PS_sym}
\begin{aligned}
    S=&\,\frac{L^2}{2G_N}X\left(\frac{1}{z_2}-\frac{1}{z_1}\right)\log\left(\frac{2X}{\epsilon}\sin\frac{\Delta\phi}{2}\right)\\
    &-\frac{L^2}{16G_N X}\left(\frac{1}{z_2}-\frac{1}{z_1}\right)\Delta \tilde{u}^2\cot^2\frac{\Delta\phi}{2}+\mathcal{O}(X^{-3})+\text{constants}\,.
    \end{aligned}
\end{align}
It remains to express this in terms of the parameters of the three-dimensional brane description. From eq.~\eqref{c_and_params_eta}, the effective Newton constant remains finite in the flat limit,
\begin{align}\label{RS1_eff_G}
    G_N^{\text{eff}}\big|_{\text{flat}}=\frac{G_N}{L}\frac{1}{1-\frac{z_2}{z_1}}\,,
\end{align}
where again $z_2$ and $z_1$ are the positions of the UV and IR branes, respectively. Of course, the curvature scale of $Q_2$ diverges in the flat limit, with
\begin{align}\label{leff_X}
    \ell_{\text{eff}}=L\sqrt{1+\frac{X^2}{z_2^2}}=\frac{L}{z_2}X+\mathcal{O}(1/X)\,.
\end{align}
Using these together with the physical Carrollian time shift $\Delta u=(L/z_2)\Delta\tilde u$ from eq.~\eqref{bondi_physical}, and absorbing the interval-independent terms into $\hat\epsilon$ as described above, eq.~\eqref{S_PS_sym} becomes\footnote{The leftover `constants' in eq.~\eqref{final_result_set2} after defining $\hat{\epsilon}$ are of order $\mathcal{O}(L/\ell_{\text{eff}})$ and thus vanish in the limit. They could also have been absorbed in the definition of $\hat{\epsilon}$.}
\begin{align}\label{final_result_set2}
    S= \frac{\ell_{\text{eff}}}{2G_N^{\text{eff}}}\log\left(\frac{2}{\hat\epsilon}\sin\frac{\Delta\phi}{2}\right)-\frac{1}{16 G_N^{\text{eff}}\ell_{\text{eff}}}\,\Delta u^2\cot^2\frac{\Delta\phi}{2}+\text{constants}+\mathcal{O}\big(\epsilon,(L/\ell_{\text{eff}})^{2}\big)\,.
\end{align}
The leading term has precisely the $\Delta\phi$ dependence of the $c_L$ term in the Carrollian result \eqref{CCFT_EE_intro}, with $c_L=3\ell_{\text{eff}}/G_N^{\text{eff}}\to\infty$.
However, there is no term proportional to a single power of $\Delta u$ as in eq.~\eqref{CCFT_EE_intro} and hence we conclude $c_M=0$.  

Note that the second term in eq.~\eqref{final_result_set2} vanishes in the strict flat limit, but we keep it for two reasons. First, it is of the same order in $1/\ell_{\text{eff}}$ as the Chern-Simons contribution computed in section \ref{chern-simons}, so the two must be discussed together. Second, it is not a failure to reproduce the Carrollian answer, but rather a correction away from the Carrollian limit. As discussed in appendix \ref{Poin-slab-appendix}, its precise angular dependence depends on how the boundary frame at null infinity is fixed at finite $\ell_{\text{eff}}$. In the null-anchored limit of section \ref{null-anchored-appendix}, it becomes precisely the first correction in the ultrarelativistic expansion of the relativistic $\text{CFT}_2$ answer. Finally, let us note that the `constants' in eq.~\eqref{final_result_set2} and in our subsequent HEE expressions are terms that are independent of the geometry of the interval. Further, they are also dependent on our regulator scheme but with our choice of $\hat\epsilon$, these constants vanish in the flat limit as $1/\ell_{\text{eff}}$.

\subsection{Geometrical picture}\label{geometrical_picture_section}

We now examine what happens to the extremal surface in the limit, from the perspective of the three-dimensional branes.

At each stage of the limit, the extremal surface for the interval \eqref{endpoints_poinc_slab} is the boosted hemisphere \eqref{min_surf_tilted}. In the Poincar\'e coordinates where the corner sits at $(z,x)=(0,X)$ and where the endpoints \eqref{endpoints_poinc_slab} of the symmetric interval \eqref{symmetric_interval} sit at $(t,y)=(t_0\pm\Delta t/2,\pm\Delta y/2)$, we have
\begin{align}\label{min_surf_slab}
\begin{aligned}
    &(x-X)^2+y^2\left(1-\frac{\Delta t^2}{\Delta y^2}\right)+z^2=\frac{\sigma^2}{4}\,,\hspace{20pt}
    t=t_0+\frac{\Delta t}{\Delta y}\,y\,,
    \end{aligned}
\end{align}
where we keep only the points lying between the two branes.   The UV brane sits at $z=z_2(1-x/X)$  by eq.~\eqref{wedge_branes_eta} and so the intersection of the extremal surface with $Q_2$ can be written as
\begin{align}
   \left\{ \begin{aligned}
        x(\lambda)=&\,X-\frac{\sigma}{2}\frac{\sin\lambda}{\sqrt{1+z_2^2/X^2}}\,,\hspace{40pt}
        &y(\lambda)=&\,-\frac{\Delta y}{2}\cos\lambda\,,\\
        z(\lambda)=&\frac{z_2\,\sigma}{2\,X}\frac{\sin\lambda}{\sqrt{1+z_2^2/X^2}}\,,\hspace{40pt}
        &t(\lambda)=&\,t_0-\frac{\Delta t}{2}\cos\lambda\,,  \end{aligned}\right.\hspace{20pt}\lambda\in[0,\pi]\,,
        \label{eq:lavender}
\end{align}
with the endpoints at $\lambda=0,\pi$. 
As noted below eq.~\eqref{min_surf_tilted}, this curve is a geodesic of the $\text{AdS}_3$ geometry of the brane.

In the Bondi coordinates \eqref{null_coordinates}, and recalling that the midpoint and the displacements scale with $X$ as in eq.~\eqref{endpoints_poinc_slab}, every point of this curve runs off to $\tilde r(\lambda)\to\infty$ in the flat limit. The endpoints remain at $\tilde u=\tilde u_\pm$, while all the interior points are pushed to $\tilde u\to\infty$. The intersection of the extremal surface with the UV brane therefore collapses entirely onto null infinity. That is, it becomes a pair of null rays which emanate from the endpoints of the interval, run along null infinity at $\phi=\phi_\pm$ and end at timelike infinity $i^+$. In particular, the limiting surface does not enter the bulk of the brane, and so is not the swing surface of \cite{Jiang:2017ecm,Apolo:2020bld} -- see Figure \ref{fig:Penrose_min_surf:a}. This near-boundary behaviour of the extremal surface is easier to visualize in the null-anchored limit, as illustrated in Figure \ref{fig:joint limit for the min surf} in section \ref{null-anchored-appendix}.

This behaviour is not an accident of the limit. As we discuss in section \ref{sec:disc}, it follows from the equality of causal and entanglement wedges for intervals in the vacuum, which holds at every stage of the flat limit.

\subsection{Gravitational Chern-Simons term and the $c_M$ structure}\label{chern-simons}

The symmetry algebra \eqref{BMS3} of a $\text{CCFT}_2$ carries two central charges, both of which appear in the entanglement entropy \eqref{CCFT_EE_intro}. Above, from the three-dimensional perspective, 
we implicitly had Einstein gravity at leading order (see appendix \ref{brane description}), and the holographic entanglement entropy \eqref{final_result_set2} only showed the $c_L$ term. To reveal the $c_M$ contribution, we introduce a gravitational anomaly in the two-dimensional boundary theory \cite{Castro:2014tta} by adding a gravitational Chern-Simons term on the positive tension brane,\footnote{We could add such a term on both branes but for simplicity, we only add it on $Q_2$ here. One might also consider adding a four-dimensional Pontryagin term to the bulk action. Since this term is a total derivative, it induces gravitational Chern-Simons terms on both EOW branes but with equal and opposite couplings. Their contributions cancel in the effective three-dimensional theory, yielding $1/\mu_{\mathrm{eff}}=0$.} so that the induced three-dimensional theory is topologically massive gravity \cite{Deser:1981wh,Deser:1982vy}. With this addition to the gravitational action and the corresponding anomaly in the boundary theory, the left- and right-moving central charges of the $\text{CFT}_2$ differ before the flat limit is taken. 

The full bulk action becomes
\begin{align}\label{new_action}
\begin{aligned}
    I[g_{\mu\nu}]=&\frac{1}{16\pi G_N}\int_{M} d^{4}x \sqrt{-g}\left(R+\frac{6}{L^2}\right)+\sum_{i=1}^2\frac{1}{8\pi G_N}\int_{Q_i} d^3x \sqrt{-h}\,(K-T_i)\\
    &+\frac{1}{32 \pi G_N\mu}\int_{Q_2}d^3x \hspace{2pt}\sqrt{-h}\,\epsilon^{abc}{\Gamma^{e}}_{ad}\left(\partial_b {\Gamma^{d}}_{ec}+\frac{2}{3}{\Gamma^{d}}_{bf}{\Gamma^{f}}_{ce}\right),
    \end{aligned}
\end{align}
where the first two terms are those in eq.~\eqref{action}, and the Levi-Civita connection ${\Gamma^a}_{bc}$ in the Chern-Simons term is constructed with the induced metric on $Q_2$. The new term modifies the equations of motion \eqref{brane eom} of the $Q_2$ brane to be \cite{Castro:2014tta}
\begin{align}\label{new eom}
K_{ab}-h_{ab}(K-T_2)=-\frac{ C_{ab}}{\mu}\hspace{30pt}{\rm with}\ \ C_{ab}= {\epsilon_{a}}^{cd}\nabla_c\left( R^{(Q_2)}_{db}-\frac{1}{4}h_{db}\,R^{(Q_2)}\right)\,,
\end{align}
where $C_{ab}$ is the Cotton tensor and it is again constructed with the induced metric on $Q_2$. Since the covariant derivative is compatible with the induced metric, the Cotton tensor vanishes when the intrinsic geometry on $Q_2$ has constant curvature, which by eq.~\eqref{intrinsic riemann} holds for all of the cases considered above. Hence every brane configuration used so far remains a solution after the Chern-Simons term is included.\footnote{Topologically massive gravity admits other solutions with a nonvanishing Cotton tensor, with warped $\text{AdS}_3$ being the standard example, \eg see \cite{Anninos:2008fx}. These new solutions will play no role in our analysis.}

As shown in appendix \ref{brane description}, with the bulk action \eqref{action} the leading-order three-dimensional description is Einstein gravity, where the central charge of the dual CFT$_2$ is $c=\frac{3\ell_{\text{eff}}}{2G_N^{\text{eff}}}$. As noted above, with the gravitational Chern-Simons term in eq.~\eqref{new_action}, the leading-order description instead becomes topologically massive gravity. In this case, the three-dimensional effective Chern-Simons coupling is
\begin{align}\label{lam_eff_def}
    \frac{1}{\mu_{\text{eff}}}=\frac{1}{\mu L}\,\frac{\cosh\zeta_2}{\sinh\zeta_2-\sinh\zeta_1}\,,
\end{align}
in the slicing variables of eqs.~\eqref{change of variables}--\eqref{Poincare slicing}, so that $\mu G_N=\mu_{\text{eff}}G_N^{\text{eff}}$. The left- and right-moving central charges are then \cite{Castro:2014tta}
\begin{align}\label{grav_anomaly}
    c_-=\frac{3\ell_{\text{eff}}}{2G_N^{\text{eff}}}\left(1-\frac{1}{\mu_{\text{eff}}\ell_{\text{eff}}}\right),\hspace{40pt}c_+=\frac{3\ell_{\text{eff}}}{2G_N^{\text{eff}}}\left(1+\frac{1}{\mu_{\text{eff}}\ell_{\text{eff}}}\right).
\end{align}
With the branes anchored as in eq.~\eqref{wedge_branes_eta}, \ie $\sinh\zeta_i=X/z_i$, eq.~\eqref{lam_eff_def} gives in the flat limit
\begin{align}\label{mu_eff_flat}
\frac{1}{\mu_{\text{eff}}}\bigg|_{\text{flat}}=\frac{1}{\mu L}\,\frac{1}{1-\frac{z_2}{z_1}}\,,
\end{align}
which is finite. Hence the anomaly does not disappear, but it becomes relatively negligible as $\ell_{\text{eff}}\to\infty$, \ie by eq.~\eqref{grav_anomaly}, we have
\begin{align}\label{eq:cdiffer}
    \frac{c_+-c_-}{c_++c_-}=\frac{1}{\mu_{\text{eff}}\,\ell_{\text{eff}}}\ \longrightarrow\ 0\,.
\end{align}

\subsubsection*{Anomalous entropy functional}

For a $\text{CFT}_2$ with a gravitational anomaly, dual to topologically massive gravity in $\text{AdS}_3$, the authors of \cite{Castro:2014tta} showed that the holographic entanglement entropy is computed by
\begin{align}\label{CS entropy}
    S_{\text{HEE}}=S_{\text{RT}} + S_{\text{CS}}=\frac{\mathrm{Area}(\mathcal{C})}{4G_N^{\text{eff}}}+\frac{1}{4G_N^{\text{eff}} \mu_{\text{eff}}}\int_{\mathcal{C}}ds \hspace{3pt} \tilde{ n}\cdot\nabla n\,,
\end{align}
where $\mathcal{C}$ is the bulk curve extremizing the (total) functional and the first term is, of course, the standard Ryu-Takayanagi contribution. In the second Chern-Simons term,
$n$ and $\tilde n$ are, respectively, a timelike and a spacelike unit normal to the curve and orthogonal to one another. Further, the derivative is taken along the curve. Geometrically, this anomalous term measures the twist of the normal frame along the curve. It is invariant under local $\text{SO}(1,1)$ boosts of the frame in the interior of the curve, but not under boosts at the endpoints, \ie it measures the relative boost between the two endpoint frames. The specification of those endpoint frames is therefore part of the definition of the functional, and not an auxiliary choice. That is, it is the entropy functional's version of the statement that the anomalous theory is not invariant under local Lorentz transformations. The endpoint frames are fixed by the boundary regulator and time slicing \cite{Castro:2014tta}. How the anomalous contribution changes under a change of endpoint frames is worked out in appendix \ref{Poin-slab-appendix}.

Given the endpoint frames $n_i$ and $n_f$, the integral in eq.~\eqref{CS entropy} is evaluated by introducing any parallel transported normal frame $(q,\tilde q)$ along $\mathcal{C}$, $\nabla q=\nabla\tilde q=0$, and writing
\begin{align}\label{q_expansion}
n(s)=\cosh[\vartheta(s)]\,q(s)+\sinh[\vartheta(s)]\,\tilde{q}(s)\,,
\end{align}
so that
\begin{align}\label{CS_in_rapidity}
    S_{\text{CS}}=\frac{1}{4G_N^{\text{eff}}\mu_{\text{eff}}}\int_{\mathcal{C}}ds \,\,\dot{\vartheta}=\frac{1}{4G_N^{\text{eff}}\mu_{\text{eff}}}\left[\vartheta(s_f)-\vartheta(s_i)\right].
\end{align}
The frame $(q,\tilde q)$ is a computational device. The answer depends only on $n_i$ and $n_f$. In locally $\text{AdS}_3$ spacetimes, the bulk geodesic connecting the two boundary endpoints extremizes eq.~\eqref{CS entropy}, and is believed to be the dominant saddle point \cite{Castro:2014tta}. For a tilted in time interval on the boundary of $\text{AdS}_3$, with endpoints at $(t,y)=(t_0\pm\Delta t/2,\pm\Delta y/2)$ as in the previous subsection, and with the endpoint frames chosen to be $n_i=n_f=(\partial_t)_{\text{CFT}}$, the anomalous contribution to the vacuum entanglement entropy is \cite{Castro:2014tta}
\begin{align}\label{CS tilted}
    S^{(t)}_{\text{CS}}= \frac{1}{2G_N^{\text{eff}}\mu_{\text{eff}}} \,\text{arctanh}\left(\frac{\Delta t}{\Delta y}\right),
\end{align}
consistent with the central charges \eqref{grav_anomaly}. Combined with the RT term, it reproduces the vacuum entanglement entropy of an interval in a $\text{CFT}_2$ with left- and right-moving central charges $c_-$ and $c_+$ \cite{Castro:2014tta},
\begin{equation}\label{TMG HEE}
    S=\frac{c_++c_-}{6}\log\left(\frac{\sigma}{\epsilon}\right)+\frac{c_+-c_-}{6}\,\text{arctanh}\left(\frac{\Delta t}{\Delta y}\right)\,,
\end{equation}
where $\sigma^2=\Delta y^2-\Delta t^2$.\footnote{The corresponding result on the circle is given in eq.~\eqref{eq:complicated}.} Using $\mu_{\text{eff}}G_N^{\text{eff}}=\mu G_N$, this can equally be written in terms of the four-dimensional parameters.

In our setup the Chern-Simons term lives only on $Q_2$, and we take the entanglement entropy to be the wedge prescription \eqref{Minimization_Wedge} supplemented by a term of the form \eqref{CS entropy} evaluated on the curve $\gamma=\Gamma\cap Q_2$ on which the bulk extremal surface ends. This is what the Lewkowycz-Maldacena-Dong replica argument \cite{Lewkowycz:2013nqa,Dong:2013qoa} gives for a term in the action localized on the brane, \ie the brane contribution to the replica partition function localizes on the intersection of the replica fixed-point set with the brane, in exactly the same way that a brane-localized Einstein-Hilbert term contributes an area term on the brane in the double-holography constructions of \cite{Chen:2020uac}. The configuration also remains extremal, since $\gamma$ is a geodesic of the $\text{AdS}_3$ geometry of $Q_2$, as noted below eq.~\eqref{min_surf_tilted}, and such geodesics extremize the entropy functional \eqref{CS entropy} in locally $\text{AdS}_3$ backgrounds \cite{Castro:2014tta}. 

Finally, eq.~\eqref{CS tilted} applies at every stage of the flat limit. It requires only that the geometry on the brane remains $\text{AdS}_3$ and that the intersection of the bulk extremal surface with the brane is the brane geodesic connecting the two endpoints, and both hold throughout the limiting process. That is, every stage of the limit is isometric to the wedge of section \ref{wedge_holo_section}, where these two requirements were established.

\subsubsection*{Flat-space entropy}

We now apply this to the Poincar\'e-slab limit, using the symmetric interval \eqref{symmetric_interval}. In this case, the displacements between the endpoints \eqref{endpoints_poinc_slab} are
\begin{align}\label{displacement_ratio_sym}
    \Delta t=\Delta\tilde u\,,\hspace{30pt}\Delta y=2X\tan\frac{\Delta\phi}{2}\,,\hspace{30pt}{\rm so\ that}\ \ \frac{\Delta t}{\Delta y}=\frac{\Delta\tilde u}{2X}\,\cot\frac{\Delta\phi}{2}\,,
\end{align}
at every stage of the limit. Eq.~\eqref{CS tilted} then gives
\begin{align}\label{CS_t_sym}
    S^{(t)}_{\text{CS}}=\frac{1}{2G_N^{\text{eff}}\mu_{\text{eff}}}\,\text{arctanh}\left(\frac{\Delta\tilde u}{2X}\cot\frac{\Delta\phi}{2}\right)=\frac{1}{4G_N^{\text{eff}}\mu_{\text{eff}}X}\,\Delta\tilde u\cot\frac{\Delta\phi}{2}+\mathcal{O}(X^{-3})\,,
\end{align}
where the superscript $(t)$ records that the endpoint frames are $n_i=n_f=\partial_t$.

This already has the $c_M$ form, but the choice of endpoint frames requires some care. The frame $n=\partial_t$ used in eq.~\eqref{CS tilted} is natural for the $\text{CFT}_2$, but not for the limiting $\text{CCFT}_2$, where the natural timelike normal at each endpoint is the unit normal to the constant-$u$ cuts of null infinity. This change of frames is discussed in detail in appendix \ref{Poin-slab-appendix}. However, it is trivial in the present case with the symmetric interval and  the anomalous contribution in the Carrollian frame is simply given by
\begin{align}\label{CS_slab_result}
    S^{(u)}_{\text{CS}}=S^{(t)}_{\text{CS}}=\frac{1}{4G_N^{\text{eff}}\mu_{\text{eff}}\ell_{\text{eff}}}\,\Delta u\,\cot\frac{\Delta\phi}{2}+\mathcal{O}\left((L/\ell_{\text{eff}})^3\right),
\end{align}
where we used eqs.~\eqref{bondi_physical} and \eqref{leff_X}, \ie $\Delta u=(L/z_2)\Delta\tilde u$ and $\ell_{\text{eff}}= XL/z_2 + {\cal O}(1/X)$.

Collecting eqs.~\eqref{final_result_set2} and \eqref{CS_slab_result}, the flat limit of the wedge-holographic entanglement entropy of an interval at null infinity of the three-dimensional branes is
\begin{align}\label{combined_result}
\begin{aligned}
    S=&\,\frac{\ell_{\text{eff}}}{2G_N^{\text{eff}}}\log\left(\frac{2}{\hat\epsilon}\sin\frac{\Delta\phi}{2}\right)+\frac{1}{4G_N^{\text{eff}}\ell_{\text{eff}}}\left[\frac{1}{\mu_{\text{eff}}}\Delta u\cot\frac{\Delta\phi}{2}-\frac{\Delta u^2}{4}\cot^2\frac{\Delta\phi}{2}\right]\\
    &\qquad+\text{constants}+\mathcal{O}\big(\epsilon,(L/\ell_{\text{eff}})^{2}\big)\,,
    \end{aligned}
\end{align}
which has the functional form of the Carrollian result \eqref{CCFT_EE_intro}, together with the additional $\Delta u^2$ term, which appears at the same order in an expansion in $1/\ell_{\text{eff}}$. Comparing with eq.~\eqref{CCFT_EE_intro},
the central charges inferred from our entanglement entropy expression above are
\begin{align}\label{central_charges_result}
    c_L=\frac{3\ell_{\text{eff}}}{G_N^{\text{eff}}}\ \longrightarrow\ \infty\,,\hspace{40pt}c_M=\frac{3}{G_N^{\text{eff}}\mu_{\text{eff}}\ell_{\text{eff}}}\ \longrightarrow\ 0\,,
\end{align}
Hence, both central charges are driven to extreme values in the strict flat limit, albeit in opposite directions. 

It is worth noting how eq.~\eqref{combined_result} organizes itself in terms of the parent $\text{CFT}_2$ data. With $c_\pm$ given by eq.~\eqref{grav_anomaly} and the speed of light\footnote{Recall that $v_{\gamma}$ has units of inverse length and is related to the usual dimensionless speed of light as $v_{\gamma}=2\pi\,c_{\gamma}/a$ with $a$ the circumference of the cylinder on which the CFT is placed -- see footnote \ref{footy33}.} $v_\gamma=1/\ell_{\text{eff}}$ as in eq.~\eqref{cgamma_identification}, the two subleading terms in eq.~\eqref{combined_result} are
\begin{align}\label{cgamma_organization}
    \frac{(c_+-c_-)\,v_\gamma}{12}\,\Delta u\cot\frac{\Delta\phi}{2}\,,\hspace{40pt}-\frac{(c_++c_-)}{48}\,v_\gamma^2\,\Delta u^2\cot^2\frac{\Delta\phi}{2}\,,
\end{align}
The first is precisely the first-order term in the expansion of the relativistic answer \eqref{eq:complicated} in the ultrarelativistic parameter $v_\gamma$, under the identifications $c_L=c_++c_-$ and $c_M=(c_+-c_-)v_\gamma$. The second has the coefficient of the second-order term, but the relativistic expansion \eqref{cgamma_expansion} comes with $\csc^2(\Delta\phi/2)$ rather than $\cot^2(\Delta\phi/2)$. As discussed in appendix \ref{Poin-slab-appendix}, this angular dependence is fixed by how the boundary frame at null infinity is chosen at finite $\ell_{\text{eff}}$, and the null-anchored limit of section \ref{null-anchored-appendix} reproduces the $\csc^2$ form. The identification $c_M=(c_+-c_-)v_\gamma$ is precisely the central charge matching that one finds by contracting the symmetry algebra to go from the $\text{CFT}_2$ to a $\text{CCFT}_2$ (both in the highest-weight representation), and it is what makes $c_M\to0$ unavoidable here: the anomaly $c_+-c_-$ remains finite while $v_\gamma\to0$. We return to this in section \ref{sec:disc}.

\section{Null-anchored limit}\label{null-anchored-appendix}

In this section, we rederive the results of section \ref{results_section} using a second controlled flat limit mentioned in section \ref{controlled_limit_section}, which is generated by a different broken isometry of $\text{AdS}_4$. Since the two limits reach the same flat brane configuration, this provides an independent check, and the reader interested only in the results may skip ahead to section \ref{sec:disc}. The null-anchored limit is nonetheless instructive for three reasons. First, the corner of the wedge remains at a finite position in the Poincar\'e coordinates: its $t=0$ slice is fixed, and it tends to the null plane $t=x$ of the asymptotic boundary, which becomes past null infinity of the branes. This allows the fate of the extremal surface near null infinity to be followed explicitly. Second, this limit fixes the boundary frame at null infinity differently from the Poincar\'e-slab limit at subleading order in $1/\ell_{\text{eff}}$. With this frame, the $\mathcal{O}(1/\ell_{\text{eff}})$ term in the entropy takes precisely the form expected from the ultrarelativistic expansion of the relativistic $\text{CFT}_2$ result. Third, the change of endpoint frame needed for the anomalous contribution, which was trivial for the symmetric interval in section \ref{chern-simons}, is nontrivial here. The detailed computations are collected in appendix \ref{NA_appendix}.

\subsection{The null-anchored limit of the wedge}\label{NA_limit_subsection}

\begin{figure}
    \centering
    \begin{subfigure}[t]{0.3\textwidth}
        \centering
        \includegraphics[width=1.15\linewidth]{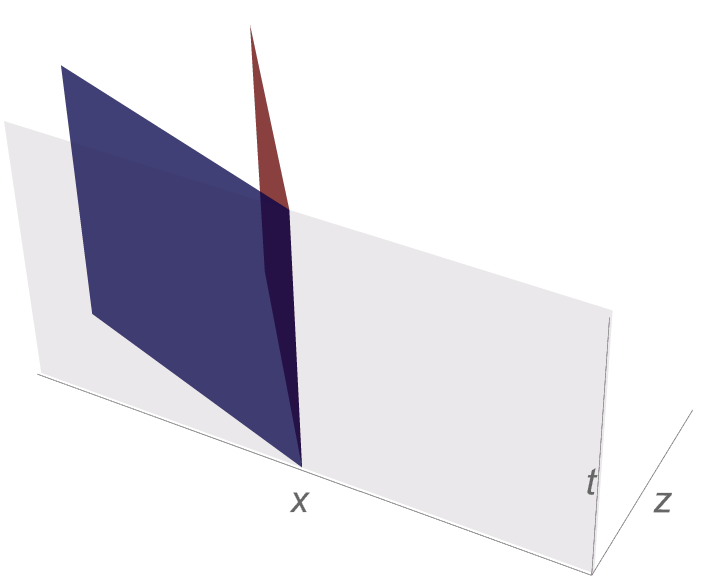}
        \caption{\small $\eta=0$}
        \label{fig:poincare-branes-comparison-a}
    \end{subfigure}
    \hfill
    \begin{subfigure}[t]{0.3\textwidth}
        \centering
        \includegraphics[width=1.15\linewidth]{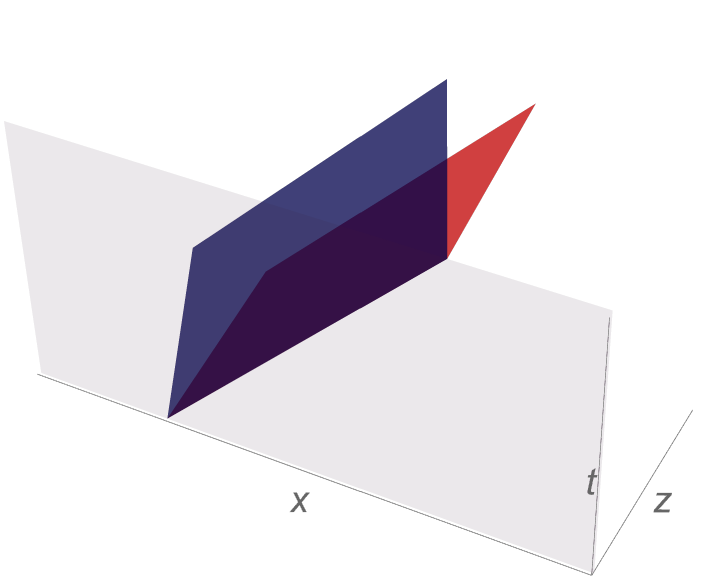}
        \caption{\small boost, $\eta=2.5$}
        \label{fig:poincare-branes-comparison-b}
    \end{subfigure}
    \hfill
    \begin{subfigure}[t]{0.3\textwidth}
        \centering
        \includegraphics[width=1.15\linewidth]{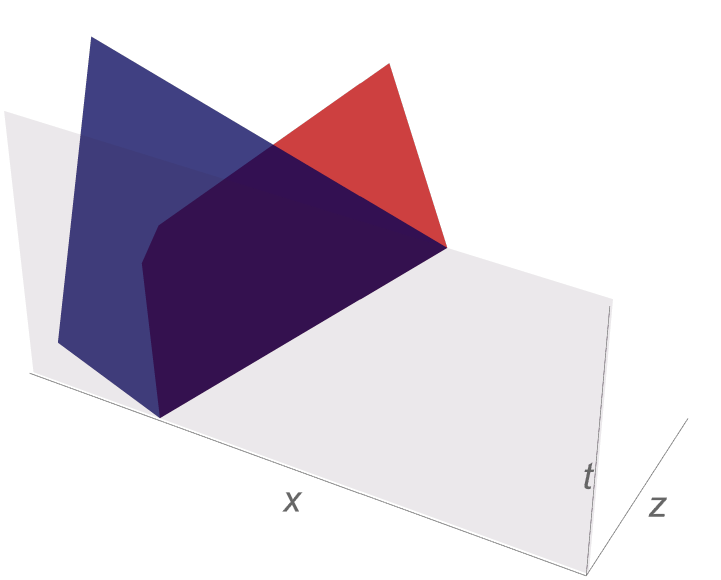}
        \caption{\small joint limit, $\eta=2.5$}
        \label{fig:poincare-branes-comparison-c}
    \end{subfigure}
    \caption{Fixed $y$ slice of $\text{AdS}_4$ in Poincar\'e coordinates, showing two single branes of the form \eqref{brane equation poincare v2} with tensions $T_{\text{blue}}=0.95\,T_c$ and $T_{\text{red}}=0.6\,T_{\text{c}}$. (a) The branes before the limit. (b) The effect of the boost \eqref{boost_NA} alone, with $\eta=2.5$. (c) The joint limit \eqref{brane_eq_joint_null_anchored} with $\eta=2.5$.}
    \label{fig:poincare-branes-comparison}
\end{figure}

We start from the wedge of section \ref{wedge_holo_section}, \ie with the branes $Q_i$ at $x=-\sinh\zeta_i\,z$ in the Poincar\'e coordinates \eqref{Poincare metric} and the corner at $z=x=0$. The broken isometry that we now use is a boost in the $x$--$t$ plane,
\begin{align}\label{boost_NA}
    t'=\cosh\eta\;t+\sinh\eta\;x\,,\hspace{40pt}x'=\cosh\eta\;x+\sinh\eta\;t\,,
\end{align}
with $y$ and $z$ unchanged. Under this boost, a brane at constant $\zeta$ in coordinates eq.~\eqref{change of variables} is mapped to
\begin{align}\label{brane equation poincare v2}
    x'=-\frac{\sinh\zeta}{\cosh\eta}\,z'+\tanh\eta\;t'\,.
\end{align}
On its own, the boost leaves the tension of the brane unchanged, and as $\eta\to\infty$, the brane collapses onto the null plane $x'=t'$, independent of its tension. This is illustrated in Figure \ref{fig:poincare-branes-comparison-b}. Meanwhile, the asymptotic boundary of the brane, $z'=0$, is the line $x'=\tanh\eta\,t'$, which becomes null in the limit.

\paragraph{Joint flat limit:} As in the Poincar\'e-slab limit, we combine the broken isometry with a change of the tensions. Here we choose the brane tensions so that the coefficient of $z'$ in eq.~\eqref{brane equation poincare v2} is held fixed,
\begin{align}\label{jointlimit}
    \sinh\zeta_i(\eta)=\cosh\eta\,\sinh\zeta_i^{(0)}\equiv\beta_i\cosh\eta\,,
\end{align}
where $\zeta_i^{(0)}$ denote the positions of the branes before the limit.\footnote{In terms of the tensions \eqref{tensions_mu}, this reads $T_2(\eta)=T_c\,T_2^{(0)}\cosh\eta\big/\sqrt{T_c^2+(T_2^{(0)})^2\sinh^2\eta}$, and similarly for $T_1$ with the replacement $T_1\to|T_1|$.} The branes are then
\begin{align}\label{brane_eq_joint_null_anchored}
    Q_i:\ \ x'=-\beta_i\,z'+\tanh\eta\;t'\,,
\end{align}
and so the slice $t'=0$ of each brane is left unchanged by the joint transformation. The branes therefore remain anchored in the bulk while their tensions approach the critical values, $T_2\to T_c$ and $T_1\to-T_c$, where we have taken $0<\zeta_1^{(0)}<\zeta_2^{(0)}$ so that the final tensions agree with those of the Poincar\'e-slab limit.\footnote{One can also start from $\zeta_1^{(0)}<0$, \ie from a wedge bounded by two positive-tension branes. Then $\sinh\zeta_1(\eta)\to-\infty$ and $T_1\to+T_c$, the limiting brane $Q_1$ lies on the other side of the null plane $t=x$, and the corner becomes past null infinity of $Q_2$ but future null infinity of $Q_1$. Since we use $Q_2$ as the representative of the three-dimensional description, the results below are unchanged, apart from replacing $\beta_1\log\beta_1\to\beta_1\log|\beta_1|$ in eq.~\eqref{regulator_NA} and an additional interval-independent constant. However, if one does this, in the limiting effective description all the KK modes become massless, and thus the three-dimensional effective description is no longer Einstein gravity. See Appendix \ref{brane description} for more details.} 
Dropping the primes, with the limit $\eta\to\infty$, we reach the configuration
\begin{align}\label{brane_eq_Setting1_flat}
    Q_i:\ \ z=\frac{t-x}{\beta_i}\,,
\end{align}
with the wedge being the region $(t-x)/\beta_2\leq z\leq(t-x)/\beta_1$. Both branes are anchored on the null plane $t=x$ of the asymptotic boundary, which is the origin of our name for this limit. In contrast with the boost alone, the branes are not pushed onto the null plane $x=t$ in the bulk, as shown in Figure \ref{fig:poincare-branes-comparison-c}. The corresponding configurations in global coordinates are shown in appendix \ref{NA_appendix}.

Using eqs.~\eqref{holo_central_charge}, \eqref{3d params} and \eqref{jointlimit}, the central charge and the brane parameters are
\begin{align}\label{c_and_params_NA}
\begin{gathered}
    c=\frac{3L^2}{2G_N}\,(\beta_2-\beta_1)\cosh\eta\,,\hspace{30pt}
    \ell_{\text{eff}}=L\sqrt{1+\beta_2^2\cosh^2\eta}\,,\\
    G_N^{\text{eff}}=\frac{G_N}{L}\,\frac{\sqrt{1+\beta_2^2\cosh^2\eta}}{(\beta_2-\beta_1)\cosh\eta}\,,
\end{gathered}
\end{align}
as the limit proceeds.
Hence $c$ and $\ell_{\text{eff}}$ diverge in the flat limit, while $G_N^{\text{eff}}\to(G_N/L)\,\beta_2/(\beta_2-\beta_1)$ and, from eq.~\eqref{lam_eff_def}, $1/\mu_{\text{eff}}\to(1/\mu L)\,\beta_2/(\beta_2-\beta_1)$ remain finite if we introduce the gravitational Chern-Simons term. These agree with eqs.~\eqref{RS1_eff_G} and \eqref{mu_eff_flat} under the identification $z_2/z_1\to\beta_1/\beta_2$, as expected since the two limits reach the same flat configuration.

\paragraph{Null infinity of the branes:} Since $|T_i|\to T_c$, the limiting branes are flat by eq.~\eqref{intrinsic riemann}, but this is not manifest in the Poincar\'e description. The induced metric on $Q_i$ is
\begin{align}\label{induced_NA}
    ds^2=\frac{L^2}{(t-x)^2}\Big[(dt-dx)^2+\beta_i^2\left(-dt^2+dx^2+dy^2\right)\Big]\,,
\end{align}
and in appendix \ref{NA_appendix} we construct the coordinate transformation which brings it to the $\text{Mink}_3$ form, $ds^2=-dv^2+2\,dv\,dr+r^2d\phi^2$, in advanced Bondi coordinates. The asymptotic boundary $z\to0$ of the brane, \ie $t-x\to0$, is past null infinity $\mathscr{I}^-$ of $Q_2$, and the corner is reached at $r\to\infty$ with
\begin{align}\label{change_at_asymptotic}
    t=x=\frac{v}{1+\cos\phi}\,,\hspace{40pt}y=L\beta_2\tan\frac{\phi}{2}\,.
\end{align}
The coordinates $v$ and $r$ are normalized so that the metric on $Q_i$ is exactly $\text{Mink}_3$, \ie $v$ is the physical (advanced) Carrollian time, playing the role of $u$ in eq.~\eqref{bondi_physical}. The orientation of $\phi$ is chosen so that it increases with $y$, as in section \ref{results_Setting2}. Let us note that the generator $\phi=\pi$ of $\mathscr{I}^-$  is mapped to $y\to\infty$.

\subsection{Flat limit of HEE}\label{results_setting1}

Since the boost is an isometry, at every stage of the limit we can work in the coordinates before the boost, where the wedge is the standard one of section \ref{wedge_holo_section} with branes at $\zeta_i(\eta)$, and the entropy is given by eq.~\eqref{int_entropy} (in the absence of a Chern-Simons term). An interval at $\mathscr{I}^-$ of $Q_2$ with endpoints $(v_\pm,\phi_\pm)$, where $\phi_+>\phi_-$, then descends from the interval on the corner $z=x=0$ with endpoints
\begin{align}\label{initial endpoints null anchored}
    (t,y)_\pm=\left(\frac{1}{\cosh\eta}\,\frac{v_\pm}{1+\cos\phi_\pm}\,,\ L\beta_2\tan\frac{\phi_\pm}{2}\right)\,.
\end{align}
The boost \eqref{boost_NA} then maps to $t'=v_\pm/(1+\cos\phi_\pm)$ and $x'=\tanh\eta\,t'$, reproducing eq.~\eqref{change_at_asymptotic} as $\eta\to\infty$. Note that in these coordinates, the time separation of the endpoints decreases as $1/\cosh\eta$.

The cutoff surface $z=\epsilon$ meets the limiting brane $Q_2$ at
\begin{align}\label{angle_dep_cutoff_NA}
    r=R_\epsilon(v,\phi)=\frac{1}{1+\cos\phi}\left(v+\frac{L^2\beta_2}{\epsilon}\right),
\end{align}
which is not a surface of fixed $r$. As in the Poincar\'e-slab limit, comparing with the flat-space result regulated at a fixed radius requires both endpoints to lie on the same surface, \ie $R_\epsilon(v_+,\phi_+)=R_\epsilon(v_-,\phi_-)$. As $\epsilon\to0$ with $v_\pm$ fixed, this is simply achieved with the symmetric interval \eqref{symmetric_interval}, \ie
\begin{align}\label{symmetric_NA}
    \phi_\pm=\pm\frac{\Delta\phi}{2}\,.
\end{align}
We restrict to this case in the following, and treat the general interval in appendix \ref{NA_appendix}.

\paragraph{Holographic entanglement entropy:} For the symmetric interval, the endpoints \eqref{initial endpoints null anchored} are separated by
\begin{align}\label{displacements_NA_sym}
    \Delta y=2L\beta_2\tan\frac{\Delta\phi}{4}\,,\hspace{40pt}\frac{\Delta t}{\Delta y}=\frac{\Delta v}{2L\beta_2\cosh\eta}\,\csc\frac{\Delta\phi}{2}\,,
\end{align}
with $\Delta v=v_+-v_-$. The invariant separation $\sigma^2=\Delta y^2-\Delta t^2$ therefore gives
\begin{align}\label{preliminary log}
    \log\left(\frac{\sigma}{\epsilon}\right)=\log\left(\frac{2L\beta_2}{\epsilon}\tan\frac{\Delta\phi}{4}\right)-\frac{\Delta v^2}{8L^2\beta_2^2\cosh^2\eta}\,\csc^2\frac{\Delta\phi}{2}+\mathcal{O}(\cosh^{-4}\eta)\,.
\end{align}
As in section \ref{results_Setting2}, the argument of the leading logarithm differs from the Carrollian $\sin(\Delta\phi/2)$ because the boundary metric of the corner differs from the standard metric at null infinity by a Weyl factor. Since $\Delta t\to0$, the corner metric $-dt^2+dy^2$ reduces on $\mathscr{I}^-$ to $dy^2=(L\beta_2/2)^2\sec^4(\phi/2)\,d\phi^2$. Up to the constant factor $(L\beta_2/2)^2$, which is absorbed into the regulator below, this differs from $d\phi^2$ by $e^{2w}=\sec^4(\phi/2)$. Eq.~\eqref{weyl_entropy_contribution} then removes $\frac{c}{3}\log\sec^2(\Delta\phi/4)$ from the entropy, which turns $2\tan(\Delta\phi/4)$ into $\sin(\Delta\phi/2)$ in the leading term. Finally, we define the regulator at null infinity by
\begin{align}\label{regulator_NA}
\begin{aligned}
    \hat\epsilon&\equiv\frac{2\epsilon}{L\beta_2}\,e^{\Lambda-1}\,,\\
    \Lambda&\equiv\frac{\sinh\zeta_2\log\sinh\zeta_2-\sinh\zeta_1\log\sinh\zeta_1}{\sinh\zeta_2-\sinh\zeta_1}=\log\cosh\eta+\frac{\beta_2\log\beta_2-\beta_1\log\beta_1}{\beta_2-\beta_1}\,,
\end{aligned}
\end{align}
where $\Lambda$ is the same function of the brane positions as in eq.~\eqref{regulator_def}, as is seen using $\sinh\zeta_i=X/z_i$. As shown in appendix \ref{NA_appendix}, the factor $e^{\Lambda-1}$ absorbs the leading non-universal constant in eq.~\eqref{int_entropy}, which again diverges in the flat limit. Since $\Lambda=\log\cosh\eta+\mathcal{O}(1)$, holding $\hat\epsilon$ fixed here requires $\epsilon\propto1/\cosh\eta$. This is equivalent to holding fixed the induced cutoff $\epsilon_2=\epsilon\,\ell_{\text{eff}}/L$ on $Q_2$, cf.~eq.~\eqref{effective_cutoff_0}.

Combining eqs.~\eqref{int_entropy}, \eqref{c_and_params_NA} and \eqref{preliminary log} with the Weyl shift and the regulator \eqref{regulator_NA}, we find
\begin{align}\label{rot inv NA cl}
\begin{aligned}
    S=&\,\frac{L^2}{2G_N}(\beta_2-\beta_1)\cosh\eta\,\log\left(\frac{2}{\hat\epsilon}\sin\frac{\Delta\phi}{2}\right)\\
    &-\frac{\beta_2-\beta_1}{16G_N\beta_2^2\cosh\eta}\,\Delta v^2\csc^2\frac{\Delta\phi}{2}+\text{constants}+\mathcal{O}(\cosh^{-3}\eta)\,,
\end{aligned}
\end{align}
or, in terms of the three-dimensional parameters \eqref{c_and_params_NA}\footnote{As in the Poincar\'e-slab case, the leftover constants after defining $\hat{\epsilon}$ are of order $\mathcal{O}(L/\ell_{\text{eff}})$ and thus vanish in the limit. They could also have been absorbed in the definition of $\hat{\epsilon}$.}
\begin{align}\label{result_setting_1}
    S=\frac{\ell_{\text{eff}}}{2G_N^{\text{eff}}}\log\left(\frac{2}{\hat\epsilon}\sin\frac{\Delta\phi}{2}\right)-\frac{\Delta v^2}{16G_N^{\text{eff}}\ell_{\text{eff}}}\csc^2\frac{\Delta\phi}{2}+\text{constants}+\mathcal{O}\big(\epsilon,(L/\ell_{\text{eff}})^{2}\big)\,.
\end{align}
The leading term coincides with that of eq.~\eqref{final_result_set2}, and so again reproduces the $c_L$ term of the Carrollian result \eqref{CCFT_EE_intro} with $c_L=3\ell_{\text{eff}}/G_N^{\text{eff}}\to\infty$, and $c_M=0$. The subleading term, however, comes with $\csc^2(\Delta\phi/2)$ rather than the $\cot^2(\Delta\phi/2)$ of eq.~\eqref{final_result_set2}. The difference, $\Delta v^2/(16G_N^{\text{eff}}\ell_{\text{eff}})$, depends on the interval, so the two results are not related by a shift of the constant terms. Rather, as discussed in appendix \ref{Poin-slab-appendix}, the two limits agree on the leading term but fix the boundary frame at null infinity differently at $\mathcal{O}(1/\ell_{\text{eff}})$. In the null-anchored limit, the subleading term in eq.~\eqref{result_setting_1} is precisely the first correction in the ultrarelativistic expansion \eqref{cgamma_expansion} of the relativistic $\text{CFT}_2$ answer, using $v_\gamma=1/\ell_{\text{eff}}$.

\paragraph{Anomalous contribution:} We now add the gravitational Chern-Simons term \eqref{new_action} on $Q_2$. It turns out that the analysis is more transparent if we consider a general interval here. As explained in section \ref{chern-simons}, the brane configurations are unchanged, and eq.~\eqref{CS tilted} applies at every stage of the limit with the endpoint frames $n=\partial_t$. In contrast with the area term, the anomalous contribution is insensitive to the shape of the cutoff surface. Hence proceeding with a general interval with endpoints as in eq.~\eqref{initial endpoints null anchored}, the displacements satisfy the exact relation
\begin{align}\label{fraction_NA_appendix}
    \frac{\Delta t}{\Delta y}=\frac{1}{2L\beta_2\cosh\eta}\left[\Delta v\cot\frac{\Delta\phi}{2}+v_+\tan\frac{\phi_+}{2}+v_-\tan\frac{\phi_-}{2}\right].
\end{align}
The last two terms depend on the positions of the individual endpoints, and they do not cancel even for the symmetric interval \eqref{symmetric_NA}, where they sum to $\Delta v\tan(\Delta\phi/4)$. As discussed in appendix \ref{Poin-slab-appendix}, they are an artifact of the endpoint frames. The natural frame for the $\text{CCFT}_2$ at each endpoint is the unit normal to the constant-$v$ cut of $\mathscr{I}^-$. In appendix \ref{NA_appendix}, we show that this frame is related to $n=\partial_t$ by a local boost \eqref{related_normal_frames} of rapidity
\begin{align}\label{chi_NA}
    \tanh\chi(v,\phi)=\frac{v}{L\beta_2\cosh\eta}\,\tan\frac{\phi}{2}\,.
\end{align}
By eq.~\eqref{entropy_rel_dif_frames}, the change of frame subtracts precisely the last two terms in eq.~\eqref{fraction_NA_appendix} at leading order, and we find
\begin{align}\label{CS_NA_result}
    S^{(v)}_{\text{CS}}=\frac{1}{4G_N^{\text{eff}}\mu_{\text{eff}}\ell_{\text{eff}}}\,\Delta v\,\cot\frac{\Delta\phi}{2}+\mathcal{O}\left((L/\ell_{\text{eff}})^3\right),
\end{align}
for an arbitrary interval, in agreement with eq.~\eqref{CS_slab_result}. Hence, unlike in the Poincar\'e-slab limit, the change of frame is essential here, even for the symmetric interval.

Collecting eqs.~\eqref{result_setting_1} and \eqref{CS_NA_result}, the null-anchored limit gives
\begin{align}\label{combined_result_NA}
\begin{aligned}
    S=&\,\frac{\ell_{\text{eff}}}{2G_N^{\text{eff}}}\log\left(\frac{2}{\hat\epsilon}\sin\frac{\Delta\phi}{2}\right)+\frac{1}{4G_N^{\text{eff}}\ell_{\text{eff}}}\left[\frac{1}{\mu_{\text{eff}}}\,\Delta v\cot\frac{\Delta\phi}{2}-\frac{\Delta v^2}{4}\csc^2\frac{\Delta\phi}{2}\right]\\
    &\qquad+\text{constants}+\mathcal{O}\big(\epsilon,(L/\ell_{\text{eff}})^{2}\big)\,.
\end{aligned}
\end{align}
Again comparing with eq.~\eqref{CCFT_EE_intro},
we infer the same central charges as in eq.~\eqref{central_charges_result} coming from the Poincar\'e-slab analysis, \ie $c_L=3\ell_{\text{eff}}/G_N^{\text{eff}}\to\infty$ and $c_M=3/(G_N^{\text{eff}}\mu_{\text{eff}}\ell_{\text{eff}})\to0$. We note again that the above expression differs from the Poincar\'e-slab result \eqref{combined_result} in the form of the $\Delta v^2$ term, as discussed above.

\subsection{Extremal surface near null infinity}\label{min_surf_NA_sec}

\begin{figure}[t]
    \centering
    \begin{subfigure}[t]{0.32\textwidth}
        \centering
        \includegraphics[width=1.05\linewidth]{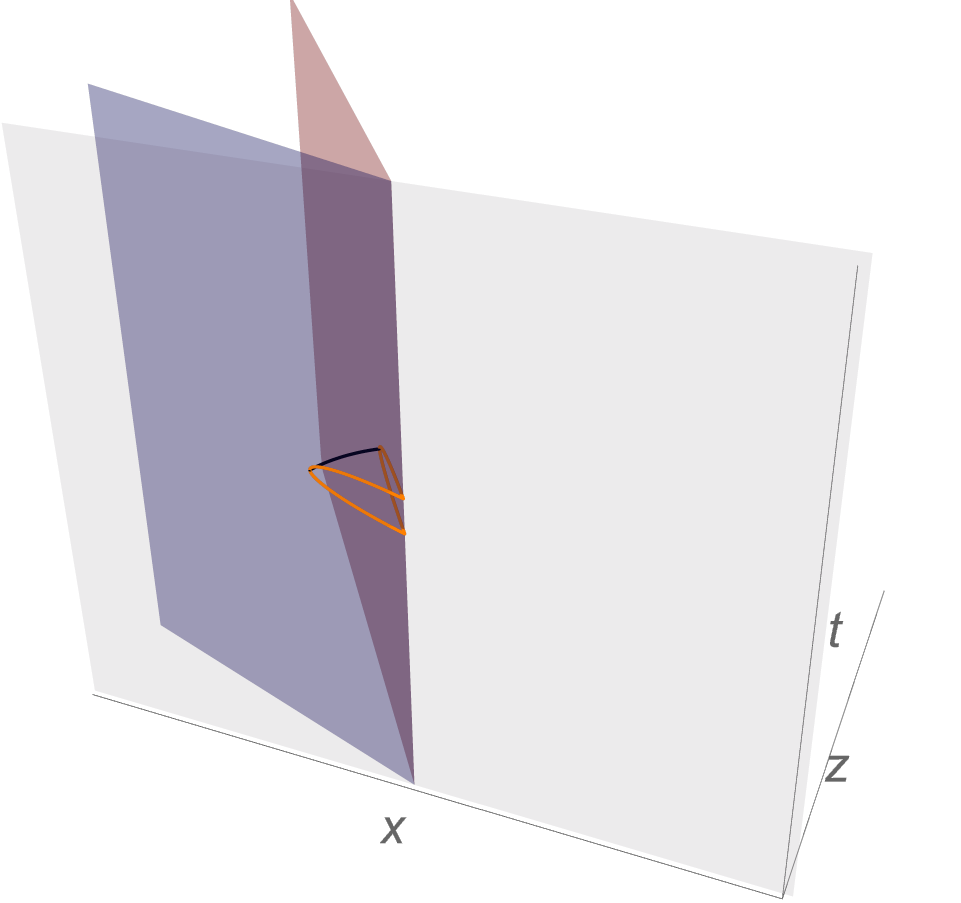}
    \end{subfigure}
    \hfill
    \begin{subfigure}[t]{0.32\textwidth}
        \centering
        \includegraphics[width=1.05\linewidth]{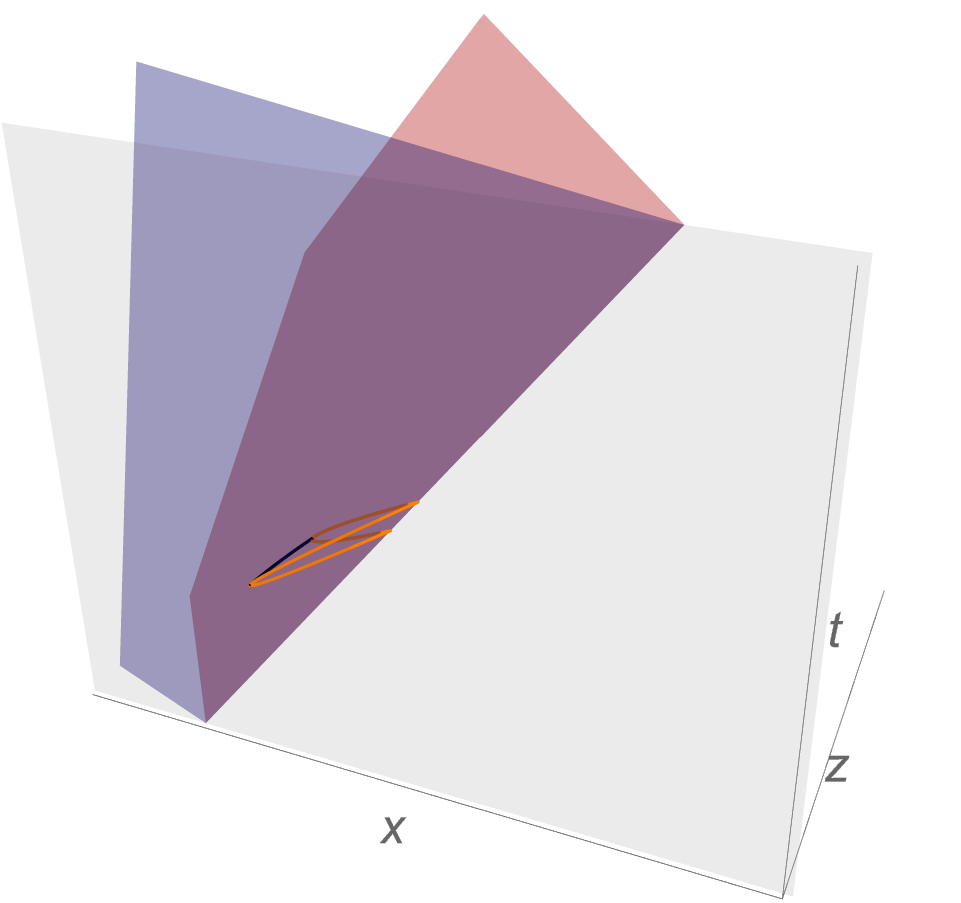}
    \end{subfigure}
    \hfill
    \begin{subfigure}[t]{0.32\textwidth}
        \centering
        \includegraphics[width=1.05\linewidth]{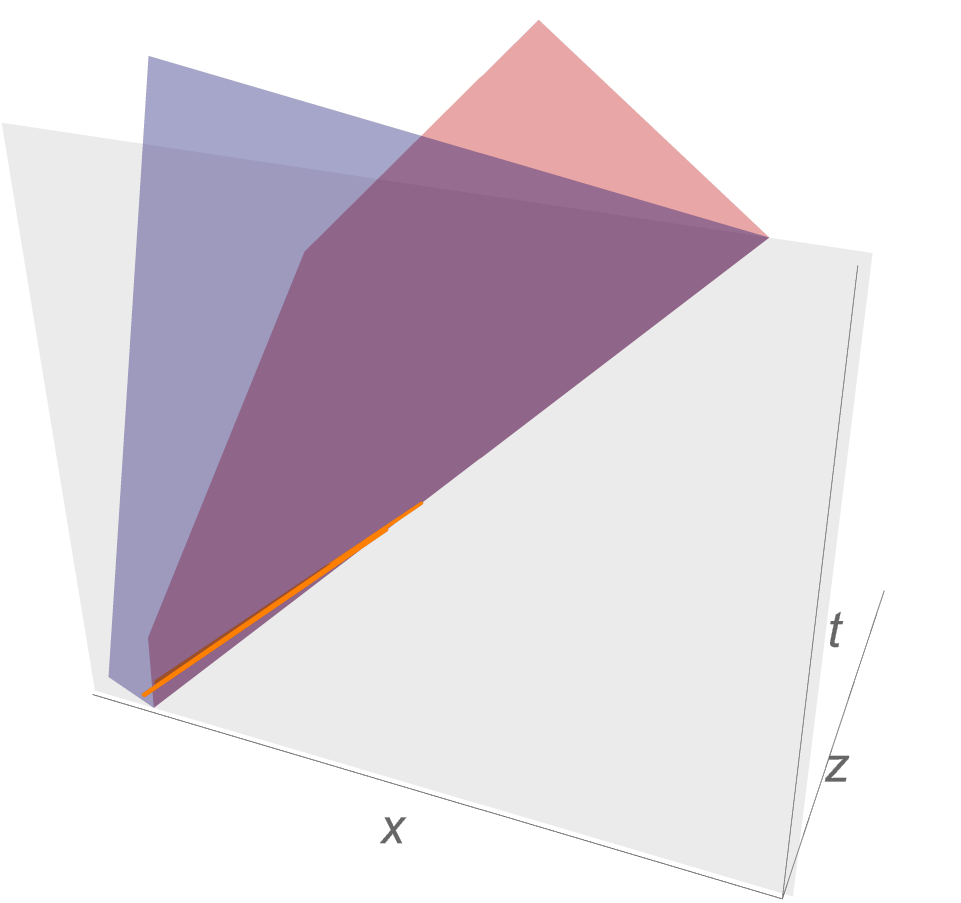}
    \end{subfigure}
    \caption{Slice $y=0$ of $\text{AdS}_4$ in Poincar\'e coordinates, showing the two branes and, in black, the section of the extremal surface \eqref{tilted_min_surf_NA} between them, for an interval with $t_0=y_0=0$, $\Delta y=2$ and $\Delta t|_{\eta=0}=0.3$. The initial tensions are $T_{\text{blue}}=0.95\,T_c$ and $T_{\text{red}}=-0.6\,T_c$. The orange curves are the projections onto this slice of the intersections of the extremal surface with the branes (for all $y$). From left to right, the joint transformation \eqref{jointlimit} is applied with $\eta=0,1,2$. In the limit $\eta\to\infty$, the intersection between the extremal surface and the branes collapses onto the generators of $\mathscr I^+$ originating from the interval endpoints.}
    \label{fig:joint limit for the min surf}
\end{figure}

We now follow the extremal surface through the limit. At each stage, before the boost, the extremal surface for the interval \eqref{initial endpoints null anchored} is eq.~\eqref{min_surf_tilted}, \ie in Poincar\'e coordinates
\begin{align}\label{tilted_min_surf_NA}
    x^2+(y-y_0)^2\left(1-\frac{\Delta t^2}{\Delta y^2}\right)+z^2=\frac{\sigma^2}{4}\,,\hspace{30pt}t-t_0=\frac{\Delta t}{\Delta y}\,(y-y_0)\,,
\end{align}
restricted to the region between the branes. For simplicity, we take the symmetric interval \eqref{symmetric_NA} with $v_-=-v_+$, so that $t_0=y_0=0$. Using $z=w/\cosh\zeta_2$ and $x=-w\tanh\zeta_2$ on $Q_2$ and applying the boost \eqref{boost_NA}, the intersection of the extremal surface with $Q_2$, which is a geodesic of the brane, becomes
\begin{align}\label{geodesic_NA}
   \left\{ \begin{aligned}
        x'(\lambda)=&-\frac{\sigma}{2}\cosh\eta\tanh\zeta_2\,\sin\lambda-\frac{\Delta t}{2}\sinh\eta\,\cos\lambda\,,\hspace{30pt}
        &y'(\lambda)=&-\frac{\Delta y}{2}\cos\lambda\,,\\
        t'(\lambda)=&-\frac{\sigma}{2}\sinh\eta\tanh\zeta_2\,\sin\lambda-\frac{\Delta t}{2}\cosh\eta\,\cos\lambda\,,
        &z'(\lambda)=&\ \frac{\sigma}{2}\,\frac{\sin\lambda}{\cosh\zeta_2}\,,
    \end{aligned}\right.
\end{align}
with $\lambda\in[0,\pi]$ and the endpoints are at $\lambda=0,\pi$. Recall that $\Delta t=\mathcal{O}(1/\cosh\eta)$ and $\sigma\to\Delta y$ in the limit. For any fixed $\lambda\in(0,\pi)$, we then have $t'+x'\to-\infty$ as $e^{\eta}$ while $t'-x'\to0^+$ as $e^{-\eta}$. By eq.~\eqref{minkowski_norm_NA}, these points recede to past timelike infinity $i^-$ of $Q_2$. The only points that remain at finite $v$ are those with $\lambda\lesssim1/\cosh\eta$ or $\pi-\lambda\lesssim1/\cosh\eta$. These have $y'\to\mp\Delta y/2$, $z'\to0$ and $x'\to t'$, and so by eq.~\eqref{change_at_asymptotic}, they lie on the null generators of $\mathscr{I}^-$ at $\phi=\phi_\mp$. In the flat limit, the intersection of the extremal surface with the brane therefore degenerates into the two null generators of $\mathscr{I}^-$ which run from the endpoints of the interval to $i^-$. This is the time reverse of the behaviour found in section \ref{geometrical_picture_section} for an interval on $\mathscr{I}^+$. In the null-anchored limit, however, the corner stays at a finite position, so that the collapse can be seen directly, as shown in Figure \ref{fig:joint limit for the min surf}.

\section{Discussion}\label{sec:disc}

Our aim in this work was to gather data about holographic Carrollian $\text{CFT}_2$ from a controlled flat limit of wedge holography, using holographic entanglement entropy (HEE) and extremal surfaces as probes.  Starting from a wedge of $\mathrm{AdS}_4$ bounded by two $\mathrm{AdS}_3$ end-of-the-world (EOW) branes, the dual description is a $\mathrm{CFT}_2$ at the asymptotic corner where the two branes meet. With a Kaluza-Klein reduction on the interval between the branes, the effective three-dimensional bulk theory becomes Einstein gravity. This can be extended to topologically massive gravity \cite{Deser:1981wh,Deser:1982vy} when a gravitational Chern-Simons interaction is included on the branes.
We presented two controlled limiting procedures where the effective three-dimensional cosmological constant vanished: the Poincar\'e-slab limit and the null-anchored limit is sections \ref{controlled_limit_section} and \ref{NA_limit_subsection}, respectively. These two approaches describe the same limit from different perspectives, both yielding a Randall-Sundrum I configuration with two flat branes sharing a common boundary at null infinity. Consequently, this flat limit corresponds to a Carrollian limit (\ie $v_{\gamma}\to0$) of the boundary theory. Hence our constructions provide an opportunity to investigate which aspects of HEE in the usual AdS/CFT  survive this Carrollian limit, and how they compare with the expectations of flat-space holography.


Our starting point was the HEE for a boundary interval on $\mathscr I^+$ with endpoints separated by angular width $\Delta\phi$ and retarded Bondi-time shift $\Delta u$, which we also interpret as the Carrollian time shift. With the Poincar\'e-slab limit, our result \eqref{combined_result} for large $\ell_{\mathrm{eff}}$ was given by
\begin{align}\label{result_discussion}
\begin{aligned}
 S={}&\frac{\ell_{\mathrm{eff}}}{2G_N^{\mathrm{eff}}}
 \log\!\left(\frac{2}{\hat\epsilon}\sin\frac{\Delta\phi}{2}\right)
 +\frac{1}{4G_N^{\mathrm{eff}}\ell_{\mathrm{eff}}}
 \left[\frac{\Delta u}{\mu_{\mathrm{eff}}}\cot\frac{\Delta\phi}{2}
 -\frac{(\Delta u)^2}{4}\,\frac{\mathcal F(\Delta\phi)}{\sin^2(\Delta\phi/2)}\right]\\
&\qquad+\text{constants}+\mathcal{O}(\epsilon, (L/\ell_{\text{eff}})^{2})\,,
\end{aligned}
\end{align}
where $\mathcal F(\Delta\phi)=\cos^2(\Delta\phi/2)$. Recall that the `constants' in the above expression vanish in the flat limit as $1/\ell_{\text{eff}}$.
Here $\ell_{\mathrm{eff}}$, $G_N^{\mathrm{eff}}$ and $\mu_{\mathrm{eff}}$ are the effective three-dimensional gravitational couplings defined in eqs.~\eqref{3d params} and \eqref{lam_eff_def}. 
With the null-anchored limit, we worked with an interval on $\mathscr I^-$ and $\Delta u$ was replaced by the advanced-time difference $\Delta v$. The analogous result in eq.~\eqref{combined_result_NA} is then the same as above, but with $\mathcal F(\Delta\phi)=1$. The difference between the two choices of $\mathcal F$ arises from a difference in the geometry of the cutoff surface at large but finite $\ell_{\mathrm{eff}}$, as explained in appendix \ref{NA_appendix}. In contrast, the displayed logarithmic and Chern-Simons terms are identical in both limits.

The natural first interpretation of eq.~\eqref{result_discussion} is to compare it directly with the entanglement entropy \eqref{CCFT_EE_intro} derived from twist operator calculations of R\'enyi entropies in a $\mathrm{CCFT}_2$ in the highest-weight representation (HWR) \cite{Bagchi:2014iea}. The functional dependence of the logarithmic term and of the term linear in $\Delta u$ agrees with the latter expression. This comparison identifies the coefficients
\begin{align}\label{entropy_inferred_charges_discussion}
 c_L^{\mathrm{EE}}=\frac{3\ell_{\mathrm{eff}}}{G_N^{\mathrm{eff}}}
 \longrightarrow\infty\,,\qquad
 c_M^{\mathrm{EE}}=\frac{3}{G_N^{\mathrm{eff}}\mu_{\mathrm{eff}}\ell_{\mathrm{eff}}}
 \longrightarrow0\,,
\end{align}
with $c_M^{\mathrm{EE}}=0$ identically when the Chern-Simons term is absent.\footnote{The limits assume that $G_N^{\mathrm{eff}}$ and $\mu_{\mathrm{eff}}$ approach finite nonzero values. See below for an alternate limit.} In this sense, the strict flat limit has the same functional form as the HWR entropy, with a divergent coefficient multiplying the logarithm and no surviving term linear in Carrollian time. The additional quadratic term in eq.~\eqref{result_discussion} is not part of the strict Carrollian formula but records the first correction away from that limit. 

We stress that the superscript ``EE'' above denotes coefficients inferred by comparison entanglement entropy \eqref{CCFT_EE_intro} for the HWR. This comparison does not assert that these are the canonical central charges of the complete limiting gravitational theory, nor that we have established the representation of the corresponding boundary states. This distinction becomes essential below. 
Our result is consistent with the direct flat-limit analysis of ordinary RT surfaces and twist operators in \cite{Caminiti:2026pma}, which likewise finds that the RT entropy retains a divergent logarithmic contribution instead of approaching the finite swing-surface result. This result is also reminiscent of the divergent central charge found for theories on null surfaces in \cite{Ciambelli:2024swv}. 

\subsection*{Symmetry algebras and the central-charge tension}

A different pair of central charges emerges in the conventional asymptotic-symmetry analysis of three-dimensional flat gravity. For an effective Einstein theory, the geometric BMS$_3$ central charges are $c_L^{\mathrm{geom}}=0$ and $c_M^{\mathrm{geom}}=3/G_N^{\mathrm{eff}}$ \cite{Barnich:2006av}. Including our gravitational Chern-Simons coupling, the standard topologically massive gravity (TMG) extension gives \cite{Bagchi:2012yk},
\begin{align}\label{geometric_charges_discussion}
 c_L^{\mathrm{geom}}=\frac{3}{G_N^{\mathrm{eff}}\mu_{\mathrm{eff}}}\,,
 \qquad c_M^{\mathrm{geom}}=\frac{3}{G_N^{\mathrm{eff}}}\,.
\end{align}
These differ sharply from eq.~\eqref{entropy_inferred_charges_discussion}. In particular, the geometric supertranslation central charge $c^{\mathrm{geom}}_M$ is independent of the Chern-Simons coupling and remains finite in the flat limit. In contrast, the coefficient $c^{\mathrm{EE}}_M$ of the linear-in-time term in the limiting HEE is proportional to $1/\mu_{\mathrm{eff}}$ and vanishes as $\ell_{\mathrm{eff}}\to\infty$.

The distinction goes beyond the algebraic form of the BMS$_3$ algebra. It concerns how the generators are realized as symmetries of the gravitational theory, which boundary conditions they preserve, and whether their central extensions arise from well-defined canonical charges. As reviewed in appendix \ref{wedge_symmetry_appendix}, the Brown-Henneaux diffeomorphisms of the $\mathrm{AdS}_3$ slices lift naturally into the four-dimensional wedge, and the conventional BMS$_3$ vector fields can be realized throughout the limiting flat wedge. Restricting to the gravitational zero-mode sector, the KK reduction over the transverse interval gives the effective Einstein action with coupling $G_N^{\mathrm{eff}}$, and its canonical charges yield eq.~\eqref{geometric_charges_discussion} (including the intrinsic Chern-Simons term on the UV brane). As explained in the appendix, this result must be taken with a grain of salt. We have not evaluated the complete four-dimensional covariant-phase-space charges with arbitrary radion and Kaluza-Klein fluctuations or all brane and corner contributions.

Let us use the conventional notation for the centrally extended BMS$_3$ algebra,
\begin{align}\label{BMS3}
 \begin{split}
 [L_m,L_n]&=(m-n)L_{m+n}+\frac{c_L}{12}m(m^2-1)\delta_{m+n,0},\\
 [L_m,M_n]&=(m-n)M_{m+n}+\frac{c_M}{12}m(m^2-1)\delta_{m+n,0},\\
 [M_m,M_n]&=0.
 \end{split}
\end{align}
To understand the different contractions, let $\mathcal L_n$ and $\bar{\mathcal L}_n$ denote the generators of the two parent Virasoro algebras, with central charges $c_+$ and $c_-$ as given in eq.~\eqref{grav_anomaly}. The standard geometric flat-gravity contraction combines oppositely oriented modes \cite{Bagchi:2012cy,Campoleoni:2016vsh},
\begin{align}\label{geometric_contraction_discussion}
 L_n^{\mathrm{geom}}=\mathcal L_n-\bar{\mathcal L}_{-n}\,,\qquad
 M_n^{\mathrm{geom}}=v_\gamma(\mathcal L_n+\bar{\mathcal L}_{-n})\,,
\end{align}
where $v_\gamma=1/\ell_{\mathrm{eff}}$ in the units used in appendix~\ref{Poin-slab-appendix}. It gives $c_L^{\mathrm{geom}}=c_+-c_-$ and $c_M^{\mathrm{geom}}=v_\gamma(c_++c_-)$, reproducing eq.~\eqref{geometric_charges_discussion} with $v_\gamma\to0$. In contrast, the contraction adapted to the ordinary relativistic entropy and its local twist operators combines the chiral sectors in the same mode orientation, up to a harmless mode-dependent automorphism detailed in \cite{Caminiti:2026pma},
\begin{align}\label{operator_contraction_discussion}
 L_n^{\mathrm{op}}=\mathcal L_n+\bar{\mathcal L}_n\,,\qquad
 M_n^{\mathrm{op}}=v_\gamma(\mathcal L_n-\bar{\mathcal L}_n)\,.
\end{align}
Its central coefficients are $c_L^{\mathrm{op}}=c_++c_-$ and $c_M^{\mathrm{op}}=v_\gamma(c_+-c_-)$, reproducing precisely those in eq.~\eqref{entropy_inferred_charges_discussion} as $v_\gamma\to0$. 

The two contractions also have different implications for the representations of BMS$_3$. Starting from ordinary highest weight representations of the parent CFT$_2$, the geometric contraction naturally leads to induced representations, while the operator frame contraction preserves the highest weight conditions \cite{Bagchi:2012cy,Campoleoni:2016vsh,Barnich:2014cwa}. This distinction also depends on the choice of representation in the parent CFT$_2$ \cite{Campoleoni:2016vsh,Caminiti:2026pma} -- see further discussion below.

The physical meaning of these two contractions was clarified by Caminiti and Pasterski~\cite{Caminiti:2026pma}. The standard contraction is natural for states on the boundary cylinder, whereas the alternative is adapted to local operator insertions. Their alternative conformal frame is well defined locally in the parent relativistic CFT$_2$, but the corresponding map between the two bases becomes singular in the Carrollian limit. The two Virasoro bases also differ in their global behaviour as gravitational diffeomorphisms. While the conventional Brown-Henneaux generators are smooth on the boundary cylinder, the alternative generators of \cite{Caminiti:2026pma} develop singularities for sufficiently large positive or negative mode numbers, as discussed in appendix~\ref{wedge_symmetry_appendix}. Therefore, they do not preserve the usual globally smooth Brown-Henneaux phase space. This observation does not invalidate the new local operator frame algebra or its Virasoro central terms. However, more work is required before identifying the latter with the charges of globally admissible gravitational asymptotic diffeomorphisms.

\subsection*{Conformal frames and representation sectors}

The distinction between the two symmetry contractions also raises the question of which representations survive the flat limit. The operator-frame contraction in eq.~\eqref{operator_contraction_discussion} maps the annihilation conditions of ordinary Virasoro primaries into highest weight conditions for the contracted generators, $L_n^{\mathrm{op}}$ and $M_n^{\mathrm{op}}$. By contrast, the geometric contraction in eq.~\eqref{geometric_contraction_discussion} mixes positive and negative Virasoro mode numbers. A generic highest weight state of the parent CFT$_2$ is therefore not a BMS$_3$ highest weight state under this contraction. The resulting states are naturally described in terms of induced BMS$_3$ representations~\cite{Campoleoni:2016vsh,Caminiti:2026pma}. Since the replica calculation leading to eq.~\eqref{CCFT_EE_intro} implicilty assumes the highest weight representation, its validity for a state reached through the geometric contraction is not automatic.

An interesting possibility is to obtain a highest weight BMS$_3$ representation from the geometric contraction by starting with the so-called flipped representation of the parent CFT$_2$ \cite{Campoleoni:2016vsh,Aggarwal:2025hji,Hao:2021urq}. 
In this representation,  one chiral sector is highest weight and the other lowest weight. In particular, the primary states are annihilated by $\mathcal L_n$ and $\bar{\mathcal L}_{-n}$ with $n>0$. 
Defining $\widetilde{\mathcal L}_n=-\bar{\mathcal L}_{-n}$, with central charge $\widetilde c_-=-c_-$, the geometric contraction in eq.~\eqref{geometric_contraction_discussion} becomes
\begin{align}\label{nameless}
L_n^{\mathrm{geom}}=\mathcal L_n+\widetilde{\mathcal L}_n\,,\qquad
M_n^{\mathrm{geom}}=v_\gamma(\mathcal L_n-\widetilde{\mathcal L}_n)\,.
\end{align}
In the flipped representation, the parent state now satisfies the highest weight conditions for $\mathcal L_n$ and $\widetilde{\mathcal L}_n$ and these conditions are inherited by the contracted BMS$_3$ generators. This mechanism was demonstrated explicitly by \cite{Hao:2021urq}, who showed that the ultrarelativistic limit of a free scalar CFT$_2$ in the flipped representation yields a highest-weight BMS$_3$ theory. Importantly, the central charges remain $c_L^{\mathrm{geom}}=c_++\widetilde c_-=c_+-c_-$ and $c_M^{\mathrm{geom}}=v_\gamma(c_+-\widetilde c_-)=v_\gamma(c_++c_-)$. Thus, the conventional geometric central charges are also compatible with a highest-weight BMS$_3$ representation. This demonstrates that the distinction between the geometric and entropy-inferred central charges cannot be attributed solely to the choice between induced and highest-weight representations. However, realizing the parent CFT$_2$ in the required flipped representation within our holographic construction remains an open question.

Let us posit a possible resolution of the tension discussed above: As explained, the asymptotic symmetry analysis suggests that the boundary CCFT$_2$ is in the induced representation with the central charges given by eq.~\eqref{geometric_charges_discussion}.  However, the entanglement entropy of a CCFT$_2$ in such an induced representation is not known. It is therefore conceivable that the entropy retains the functional form of eq.~\eqref{CCFT_EE_intro}, but with coefficients that differ from the central charges appearing in the BMS$_3$ algebra. In our construction, these coefficients would instead be those identified from the holographic entanglement entropy in eq.~\eqref{entropy_inferred_charges_discussion}. One reservation about this possibility is that Carrollian theories with induced vacua typically exhibit spatially ultralocal correlation functions, involving delta functions of the spatial separation, \eg see \cite{Hao:2021urq,Chen:2024qcs}.\footnote{Note however that in \cite{Hao:2021urq} they showed that the induced vacuum obtained from the UR limit of the parent CFT$_2$ vacuum is different from the intrinsic Carrollian one. In particular, this limiting vacuum is not ultralocal, but has divergent correlation functions.} If the entanglement entropy can be derived from correlation functions of twist operators\footnote{Deriving the entropy from twist operator correlators would require a suitable relation between states and local operators, which is subtle for induced representations -- see further discussion in \cite{Caminiti:2026pma,Hao:2021urq,Chen:2024qcs}.} and the latter had a similar structure, it would be difficult to recover the usual interval dependence in eq.~\eqref{CCFT_EE_intro}. Thus, while this possibility deserves further investigation, establishing the entanglement structure of induced BMS$_3$ representations remains an open problem, as also emphasized in \cite{Hao:2025naz}.

\subsection*{Relativistic corrections away from the Carrollian limit}

An instructive feature of our calculation is that terms which are formally second order in $v_\gamma$ are effectively first-order corrections in the limit of our holographic construction. The relativistic CFT$_2$ entropy expanded in the operator frame takes the form (see eq.~\eqref{cgamma_expansion})
\begin{align}\label{relativistic_entropy_discussion}
\begin{aligned}
 S={}&\frac{c_++c_-}{6}\log\!\left(\frac{2}{\hat\epsilon}\sin\frac{\Delta\phi}{2}\right)
 +\frac{(c_+-c_-)v_\gamma}{12}\,\Delta u\cot\frac{\Delta\phi}{2}\\
 &-\frac{(c_++c_-)\,v_\gamma^2}{48}\,\Delta u^2
 \csc^2\frac{\Delta\phi}{2}+\cdots\,.
\end{aligned}
\end{align}
For fixed parent central charges, the term which is quadratic in $\Delta u$ would be of order $v_\gamma^2$. However, the sum of the central charges itself grows with the effective curvature radius,
\begin{align}\label{enhanced_scaling_discussion}
 c_++c_-=\frac{3\ell_{\mathrm{eff}}}{G_N^{\mathrm{eff}}}
 =\frac{3}{G_N^{\mathrm{eff}}v_\gamma}\,,\qquad
 (c_++c_-)v_\gamma^2=\frac{3v_\gamma}{G_N^{\mathrm{eff}}}\,.
\end{align}
Thus, this relativistic correction to the entropy is of order $1/\ell_{\mathrm{eff}}$, rather than $1/\ell_{\mathrm{eff}}^2$, and must be retained alongside the leading Chern-Simons contribution.\footnote{We reiterate that the angular dependence of this term in eq.~\eqref{result_discussion} depends on the choice of boundary frame at finite curvature. The null-anchored limit reproduces the expected $\csc^2$ factor in eq.~\eqref{relativistic_entropy_discussion}, whereas the Poincar\'e-slab limit yields a $\cot^2$ factor. This difference is discussed further in appendix \ref{Poin-slab-appendix}.}

A similar enhancement appears in the operator frame algebra at small but finite $v_\gamma$. Using the definitions in eq.~\eqref{operator_contraction_discussion} before taking the contraction gives
\begin{align}\label{MM_correction_discussion}
 [M_m^{\mathrm{op}},M_n^{\mathrm{op}}]
 =v_\gamma^2(m-n)L_{m+n}^{\mathrm{op}}
 +\frac{v_\gamma^2\,c_L^{\mathrm{op}}}{12}
 m(m^2-1)\,\delta_{m+n,0}\,.
\end{align}
Since $c_L^{\mathrm{op}}\sim\ell_{\mathrm{eff}}=1/v_\gamma$, the central term is again effectively $O(v_\gamma)$, while the first term remains $O(v_\gamma^2)$ if the $L_n^{\mathrm{op}}$ are held fixed.\footnote{We note that in principle, its matrix elements might be enhanced in certain families of states for which the corresponding charges grow as $1/v_\gamma$.} Thus, the relativistic entropy correction and the central term in eq.~\eqref{MM_correction_discussion} both exhibit the same nonuniform power counting. At finite $v_\gamma$, this algebra is simply an invertible rewriting of the parent Virasoro algebras. The additional central term still vanishes in the Carrollian limit and does not generate a new finite central charge. For comparison, the corresponding central term in the geometric BMS$_3$ algebra vanishes as $v_\gamma^2$, since the geometric central charges in eq.~\eqref{geometric_charges_discussion} remain finite.

Although both entropy corrections appear at order $1/\ell_{\mathrm{eff}}$, they have different origins. 
The Chern-Simons contribution arises from the gravitational anomaly and is proportional to $1/\mu_{\mathrm{eff}}$. Of course, the relativistic correction comes from expanding the usual covariant RT entropy. Their relative importance is therefore controlled by $\mu_{\mathrm{eff}}$. Decreasing this coupling enhances the Chern-Simons contribution relative to the relativistic correction. Nevertheless, for fixed nonzero $\mu_{\mathrm{eff}}$, we still have $c_M^{\mathrm{EE}}\to0$. Allowing $\mu_{\mathrm{eff}}$ to scale with $\ell_{\mathrm{eff}}$ instead defines a distinct correlated limit, to which we return below. 

\subsection*{Domain of dependence and entanglement entropy}

There is an important difference between our result \eqref{result_discussion} derived from HEE in the flat limit of wedge holography and the Carrollian entanglement entropy  emerging from the swing surface proposal \cite{Jiang:2017ecm,Apolo:2020bld} usually considered in flat holography. The coefficient of the term linear in $\Delta u$ vanishes in our limit, whereas this term remains finite in the latter, which reproduces the  formula \eqref{CCFT_EE_intro} with the asymptotic symmetry value $c_M=3/G_N^{\text{eff}}$. We now explain why this difference is natural in a flat limit of ordinary AdS/CFT entanglement entropy, and what it may reveal about the resulting $\text{CCFT}_2$.

\begin{figure}
    \centering 
\begin{tikzpicture}[scale=1]
%
    \def\xL{-2}
    \def\xR{2}
    \def\uTilt{1}
%
    \fill[orange, opacity=0.25]
        (\xL,-2) rectangle (\xR,2);
%
    \draw[orange, dashed, thick]
        (\xL,-2) -- (\xL,2);
    \draw[orange, dashed, thick]
        (\xR,-2) -- (\xR,2);
%
    \draw[->, thick] (-3.6,0) -- (-2.8,0)
        node[below] {$\phi,x$};
    \draw[->, thick] (-3.6,0) -- (-3.6,0.85)
        node[above] {$u$};
%
    \draw[black, very thick]
        (\xL,0) -- (\xR,0);
%
    \draw[blue, very thick]
        (\xL,0) -- (\xR,\uTilt);
%
    \draw[decorate, decoration={brace, mirror, amplitude=6pt}]
        (\xR+0.2,0) -- (\xR+0.2,\uTilt)
        node[midway, right=8pt] {$\Delta u$};
%
    \fill[black] (\xL,0) circle (2pt);
    \fill[black] (\xR,0) circle (2pt);
    \fill[blue]  (\xR,\uTilt) circle (2pt);
\end{tikzpicture}
    \caption{We show in orange the domain of dependence of an interval in a Carrollian theory. The spacelike coordinate is $\phi$ if the theory is formulated on a cylinder ($\phi\sim\phi+2\pi$), or $x$ if it is on the plane. Two spacelike intervals that have the same spatial projection (as the black and blue intervals in the figure) share the same causal development, no matter the time shift $\Delta u$. }
    \label{fig:domain-of-dep-Carrollian}
\end{figure}
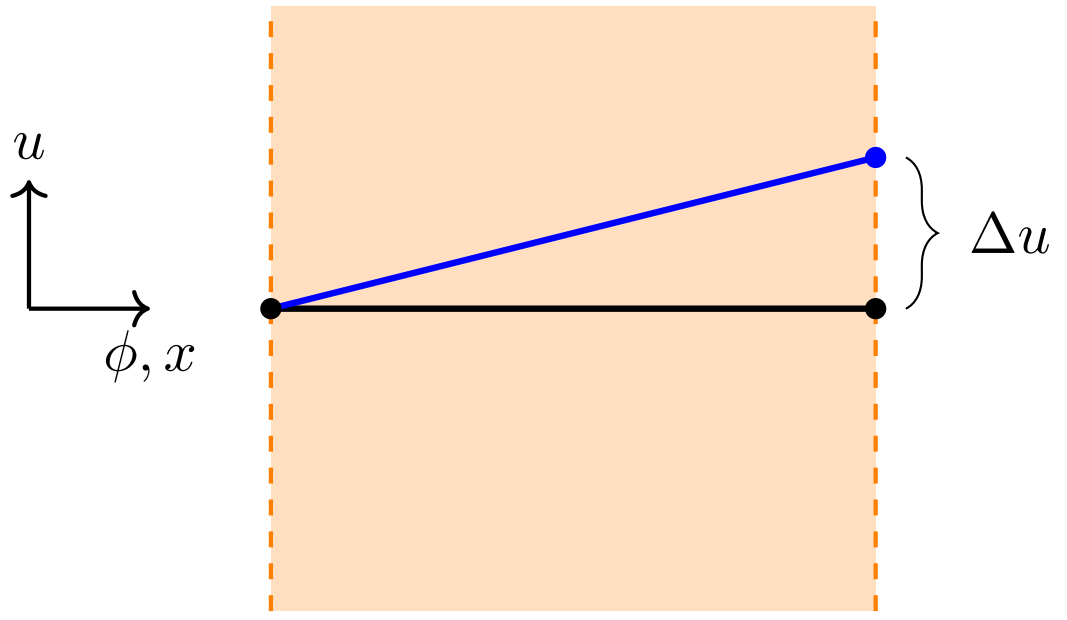

In a Carrollian theory (with $v_\gamma=0$), the light cone at any event degenerates to a vertical line. Points at distinct spatial positions are causally disconnected, regardless of their time separation. Consequently, the domain of dependence of a spacelike interval is an infinite vertical strip, as illustrated in Figure \ref{fig:domain-of-dep-Carrollian}. Therefore, two intervals with different endpoints can have the same causal development, provided they have the same spatial projection. For example, consider the horizontal interval at $u=0$ (black) and the tilted interval with the same left endpoint and right endpoint at $u=\Delta u$ (blue) in the figure. If the entropy is given by the HWR formula \eqref{CCFT_EE_intro}, or by its planar counterpart \cite{Bagchi:2014iea}\footnote{Note that consistency (\ie the combination $c_M\Delta u$ is dimensionless) requires that we use a dimensionless planar coordinate $x=y/a$. Here, the length $a$ is an arbitrary scale chosen in normalizing the spatial coordinate $x$, not a new physical coupling. It must be specified when relating the planar generators to those on the cylinder.}
\begin{align}\label{plane_carroll_EE}
    S=\frac{c_L}{6}\log\left(\frac{\Delta x}{\epsilon}\right)+\frac{c_M}{6}\frac{\Delta u}{\Delta x},
\end{align}
the EE of these two intervals differs by
\begin{align}\label{entropy_shift_carroll}
 S_{\mathrm{tilted}}-S_{\mathrm{horizontal}}
 =\begin{cases}
 \displaystyle\frac{c_M\,\Delta u}{12}\,\cot\frac{\Delta\phi}{2}\,,&\text{cylinder}\,,\\[4pt]
 \displaystyle\frac{c_M\,\Delta u}{6\,\Delta x}\,,&\text{plane}\,.
 \end{cases}
\end{align}
Thus, with $c_M\neq0$ (and fixed regulators), the entropy depends on the chosen interval representative, not only on its Carrollian domain of dependence. 

In contrast, for an ordinary relativistic CFT, the entanglement entropy of a subregion is determined by its domain of dependence. For holographic CFTs, this is reflected in the RT/HRT prescription \cite{Ryu:2006bv, Hubeny:2007xt}.\footnote{There is a subtlety regarding CFTs with gravitational anomalies. From the perspective of \cite{Castro:2014tta}, the extra entropic contribution, once the endpoint frames are specified is still dependent on the domain of dependence and not on the particular representative. However, from the perspective of \cite{Iqbal:2015vka}, one does not make a frame choice and the EE is dependent on the particular representative. These two perspectives are related by the possibility of trading a diffeomorphism anomaly for a Lorentz violation anomaly. For the purposes of this work we take the former viewpoint.} Two boundary regions with the same domain of dependence have equivalent anchoring and homology conditions for their bulk extremal surfaces, and hence the same holographic entanglement entropy. Thus, the entropy does not depend on which spacelike slice of a given causal domain is chosen to represent the boundary region. This observation offers a natural interpretation of the vanishing of the contribution which is linear in $\Delta u$ in our flat limit. At each finite stage of the limit, the relativistic entropy is associated with a domain of dependence. As the Carrollian limit is approached, distinct relativistic domains of dependence may collapse onto different representatives of the same Carrollian domain. In our construction, the entropy difference between these representatives vanishes and hence the Carrollian entanglement entropy retains this property of depending only on the domain of dependence.

Therefore, we believe that this behaviour is more robust than just a peculiarity of our particular limit. At any point in a flat limit, EE in a CFT will be a property of the domain of dependence and not  of the particular representative Cauchy slice. Moreover, the entropy of two different domains of dependence coincide as we make them more and more similar. It follows then that unless there is a discontinuous phenomenon at the limit, EE will also be a property of the domain of dependence, and therefore any $\Delta u$ dependence will vanish. As we noted above, the form of the EE for a CCFT$_2$ in the induced representation is not known. Hence, it would be interesting to determine whether the entanglement entropy of such CCFTs depends only on the Carrollian domain of dependence, even when the BMS$_3$ algebra has a nonzero $c_M$. This would contrast with the conventional HWR entropy formula, which distinguishes different interval representatives when $c_M\neq0$.\footnote{This last point might be related to CCFTs in the HWR with $c_M\neq0$ being non-unitary \cite{Grumiller:2014lna, Bagchi:2009pe}.} 

This representative dependence should not be confused with the invariance of the EE under Carroll boosts. Under a Carroll boost (generated by ${M}_0$ in the $\text{BMS}_3$ algebra eq.~\eqref{BMS3}), the coordinates change as
\begin{align}
    x'=x, \hspace{30pt}u'=u+a\,x
\end{align}
Choosing $a=\Delta u/\Delta x$ maps the black interval to the blue one in Figure \ref{fig:domain-of-dep-Carrollian}. Eq.~\eqref{entropy_shift_carroll} might then seem to imply that the entropy is not invariant under a symmetry of the theory. However, a Carroll boost also transforms the regulators. In eq.~\eqref{plane_carroll_EE}, we have omitted a cutoff-dependent term. The full $c_M$ contribution is \cite{Jiang:2017ecm, Grumiller:2019xna}
\begin{align}
    \frac{c_M}{6}\left(\frac{\Delta u}{\Delta x}-\frac{\epsilon_u}{\epsilon_x}\right)
\end{align}
Here $\epsilon_x$ is the spatial cutoff appearing in the logarithmic term, while $\epsilon_u$ is a time-direction cutoff that we have set to zero so far. Under the above Carroll boost, both $\Delta u\to\Delta u+a\Delta x$ and $\epsilon_u\to\epsilon_u+a\epsilon_x$, leaving the EE invariant. The distinction is that the representative dependence described above compares two intervals with the regulators held fixed, whereas a symmetry transformation acts on both the interval and its regulators.

\subsection*{Extremal surfaces and bulk reconstruction}

\begin{figure}
\centering

\begin{tikzpicture}[
    scale=2.5,
    line cap=round,
    line join=round,
    >=Latex,
    x={(1cm,0cm)},
    y={(0cm,0.38cm)},
    z={(0cm,0.9cm)}
]
\begin{scope}[xshift=-5cm]


\coordinate (ip)  at (0,0,1);      
\coordinate (im)  at (0,0,-1);     
\coordinate (i0R) at (1,0,0);      
\coordinate (i0L) at (-1,0,0);     
\node at (0,0,-1.4) {$\text{(a)}$};

\foreach \T/\rad in {-0.65/0.35, -0.35/0.65, 0.35/0.65, 0.65/0.35} {
    \draw[thin, dashed] (0,0,\T) ellipse [x radius=\rad, y radius=0.45*\rad];
}

\draw[line width =1pt, color=gray] (0,0,0) ellipse [x radius=0.98, y radius=0.45];

\draw[->, line width=1.2pt, color=red]
    ({-1*cos(45)}, {-0.45*1*sin(45)}, 0)
    arc[start angle=225, end angle=250, x radius=1, y radius=0.45*1];
\node[color=red] at (-0.5,-0.34,0.08) {$\phi$};

\draw[line width=1.2pt, color=blue]
    ({0.65*cos(65)}, {-0.45*0.65*sin(65)}, 0.35)
    arc[start angle=315, end angle=355, x radius=0.8, y radius=0.5*0.8];

\draw[line width=0.8pt, color=orange]
    ({0.65*cos(65)}, {-0.45*0.65*sin(65)}, 0.35)--(ip);

\draw[line width=0.8pt, color=orange]
    ({0.65*cos(39)}, {-0.6*0.65*sin(50)}, 0.48)--(ip);

\draw[line width=0.8pt, color=orange]
    ({-0.65*cos(-35)}, {-0.45*0.65*sin(-35)}, -0.35)--(im);

\draw[line width=0.8pt, color=orange]
    ({-0.65*cos(65)}, {-0.45*0.65*sin(65)}, -0.35)--(im);
    
\draw[line width=1.2pt, color=blue]
    ({-0.65*cos(-35)}, {-0.45*0.65*sin(-35)}, -0.35)
    arc[start angle=145, end angle=245, x radius=0.65, y radius=0.45*0.65];

\node[above] at (ip) {$i^+$};
\node[below] at (im) {$i^-$};
\node[right] at (i0R) {$i^0$};
\node[left] at (i0L) {$i^0$};

\node[right] at ($(im)!0.55!(i0R)$) {$\mathscr{I}^-$};

\node[left] at ($(i0L)!0.55!(ip)$) {$\mathscr{I}^+$};

\draw[line width=2pt] (im) -- (i0R) -- (ip) -- (i0L) -- cycle;
\end{scope}

\begin{scope}[xshift=-2.25cm]


\coordinate (ip)  at (0,0,1);      
\coordinate (im)  at (0,0,-1);     
\coordinate (i0R) at (1,0,0);      
\coordinate (i0L) at (-1,0,0);     
\node at (0,0,-1.4) {$\text{(b)}$};

\foreach \T/\rad in {-0.65/0.35, -0.35/0.65, 0.35/0.65, 0.65/0.35} {
    \draw[thin, dashed] (0,0,\T) ellipse [x radius=\rad, y radius=0.45*\rad];
}

\coordinate (A) at ({cos(65)}, {-0.45*sin(65)}, 0);
\coordinate (B) at ({cos(35)}, {-0.45*sin(35)}, 0);

\coordinate (C) at ({0.65*cos(35)}, {-0.45*0.65*sin(35)}, 0.35);
\coordinate (D) at ({0.65*cos(65)}, {-0.45*0.65*sin(65)}, 0.35);
\coordinate (E) at  ({-cos(-35)}, {-0.45*sin(-35)}, 0);
\coordinate (F) at ({-cos(65)}, {-0.45*sin(65)}, 0);

\fill[orange!30, opacity=0.5]
    (A)
    -- (ip)
    -- (B)
    -- cycle;

\fill[orange!30, opacity=0.5]
    (im)
    -- (E) arc[start angle=145, end angle=245,
        x radius=1, y radius=0.45]
    -- (F)
    -- cycle;
\fill[orange!30, opacity=0.5]
    (im)
    -- (-1,0,0) arc[start angle=180, end angle=190,
        x radius=1, y radius=0.45]
    -- ({-cos(51)}, {-0.45*sin(51)}, 0)
    -- cycle;
    
\draw[line width =1pt, color=gray] (0,0,0) ellipse [x radius=0.98, y radius=0.45];

\draw[line width=1.2pt, color=blue]
    ({0.65*cos(65)}, {-0.45*0.65*sin(65)}, 0.35)
    arc[start angle=315, end angle=355, x radius=0.72, y radius=0.5*0.8];

\draw[line width=0.8pt, color=orange]
    ({cos(65)}, {-0.45*sin(65)}, 0)--(ip);

\draw[line width=0.8pt, color=orange]
    ({cos(35)}, {-2*0.45*sin(50)}, 0.2)--(ip);

\draw[line width=0.8pt, color=orange]
    ({-cos(-35)}, {-0.45*sin(-35)}, 0)--(im);

\draw[line width=0.8pt, color=orange]
    ({-cos(65)}, {-0.45*sin(65)}, 0)--(im);
    
\draw[line width=1.2pt, color=blue]
    ({-0.65*cos(-35)}, {-0.45*0.65*sin(-35)}, -0.35)
    arc[start angle=145, end angle=245, x radius=0.65, y radius=0.45*0.65];

\draw[->, line width=1.2pt, color=red]
    ({-1*cos(45)}, {-0.45*1*sin(45)}, 0)
    arc[start angle=225, end angle=250, x radius=1, y radius=0.45*1];
\node[color=red] at (-0.5,-0.34,0.08) {$\phi$};
\node[above] at (ip) {$i^+$};
\node[below] at (im) {$i^-$};
\node[right] at (i0R) {$i^0$};
\node[left] at (i0L) {$i^0$};

\node[right] at ($(im)!0.55!(i0R)$) {$\mathscr{I}^-$};

\node[left] at ($(i0L)!0.55!(ip)$) {$\mathscr{I}^+$};

\draw[line width=2pt] (im) -- (i0R) -- (ip) -- (i0L) -- cycle;
\end{scope}
\end{tikzpicture}
\caption{Penrose diagram of $\mathrm{Mink}_3$ with intervals at future ($\mathscr{I}^+$) and past ($\mathscr{I}^-$) null infinity. In (a), the orange null rays show the limiting minimal surfaces found in \ref{results_setting1} and \ref{results_Setting2}, which connect the interval endpoints to $i^\pm$. In (b), the shaded regions show the intervals' boundary domains of dependence. Although $i^0$ appears as a circle in the diagram, it corresponds to a single point \cite{Wald:1984rg}.}
\label{fig:Penrose_min_surf}

\phantomsubcaption\label{fig:Penrose_min_surf:a}
\phantomsubcaption\label{fig:Penrose_min_surf:b}

\end{figure}
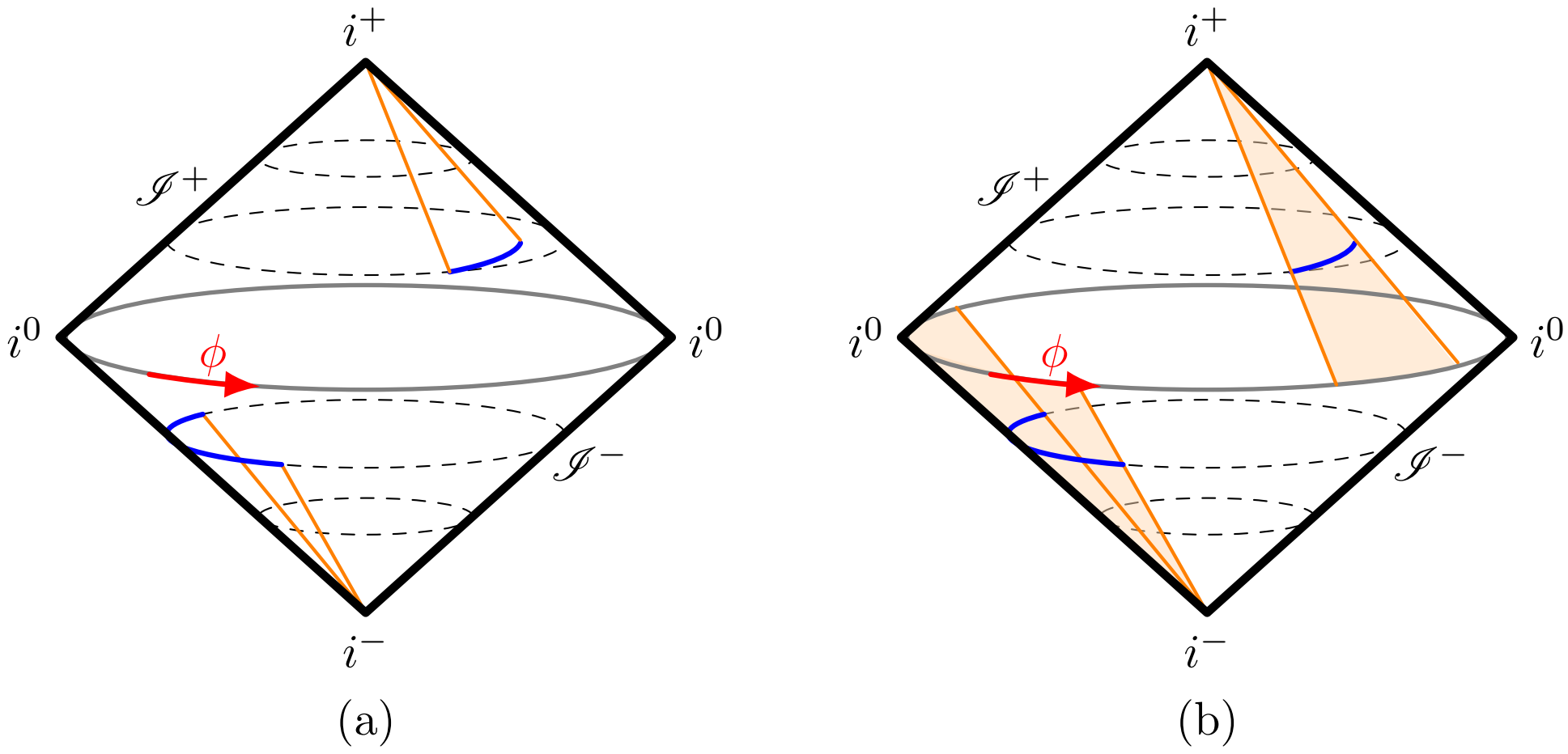

A second striking difference between our flat limit and the swing surface prescription \cite{Jiang:2017ecm,Apolo:2020bld} concerns the fate of extremal surfaces. As shown explicitly in sections~\ref{geometrical_picture_section} and \ref{min_surf_NA_sec}, the intersections of the wedge HRT surface with the branes approach null generators lying along $\mathscr I^\pm$, connecting the interval endpoints to timelike infinity $i^\pm$, as shown in Figure \ref{fig:Penrose_min_surf:a}. They do not approach swing surfaces with null ropes extending into the flat bulk and a spacelike bench joining those ropes \cite{Jiang:2017ecm,Apolo:2020bld,Hao:2025ocu}.

It is useful to interpret this behaviour from the effective three-dimensional perspective. For a spacelike boundary region $A$ in an ordinary holographic theory, the entanglement wedge $EW[A]$ is the domain of dependence of a bulk region bounded by $A$ and its HRT surface. The corresponding  causal wedge is
\begin{align}\label{causal_wedge}
 C[A]=J^+(D[A])\cap J^-(D[A])\,,
\end{align}
where $D[A]$ denotes the boundary domain of dependence and the causal sets are taken in the conformal completion of the bulk spacetime \cite{Headrick:2014cta,Hubeny:2013gba}. The two wedges coincide for a spherical boundary region in the pure AdS vacuum, \ie $C[A]=EW[A]$. A spacelike interval on the boundary of $\mathrm{AdS}_3$ is precisely such a `spherical' region.\footnote{Technically, we are applying this known three-dimensional result to the effective theory obtained by Kaluza-Klein reduction of the four-dimensional wedge, restricting to the massless graviton zero mode.}

In the flat limit, a boundary interval on $\mathscr I^+$ has Carrollian causal development
\begin{align}\label{carroll_domain_discussion}
 D[A]=\{\phi\in[\phi_-,\phi_+],\ u\in\mathbb R,\ r=\infty\}\,.
\end{align}
For an interval on $\mathscr I^-$, one replaces $u$ by $v$. These domains are shown as orange strips in Figure \ref{fig:Penrose_min_surf:b}. Within the conformal completion of the three-dimensional flat space, no interior bulk point lies to the future of the first of these regions or to the past of the second. Their corresponding causal wedges therefore do not extend into the bulk. If the equality of the vacuum causal and entanglement wedges, together with the relevant limiting causal structure, survives continuously, the limiting HRT surface must likewise be pushed toward null infinity. This is precisely what the explicit wedge calculation shows. The same conditional reasoning can be applied to spherical regions in higher dimensions, although it should not be taken as a theorem about arbitrary prescriptions for flat space EE.
\begin{figure}[t]
\centering
\begin{tikzpicture}[scale=1.05, every node/.style={font=\small}]

    \def\W{3.8}   
    \def\H{4.2}   
    \def\a{0.75}  
    \def\b{0.75}  

    \def\ycut{0.38}
    \pgfmathsetmacro{\xcut}{\a*(1-\ycut/\b)}

    \def\yB{3.9}

    \draw[thick] (-\W,-\H/3) -- (-\W,\H+0.2);
    \draw[thick] ( \W,-\H/3) -- ( \W,\H+0.2);

    \draw[dashed, ->] (0,-\H/3) -- (0,\H+0.45) node[above] {$\tau$};

    \node[rotate=90] at (-\W-0.45,0) {$\rho=\frac{\pi}{2}$};
    \node[rotate=90] at ( \W+0.45,0) {$\rho=\frac{\pi}{2}$};

    \draw[red!55, thick, dashed,->] (0,0) -- (\W,\W);
    \draw[red!55,thick, dashed,->] (0,0) -- (-\W,\W);

    \fill[orange!20] (0,\b) -- (\a,0) -- (0,-\b) -- (-\a,0) -- cycle;
    \draw[thick, orange!70!black] (0,\b) -- (\a,0) -- (0,-\b) -- (-\a,0) -- cycle;


    \node[right=4pt] at (0,\b) {$\mathscr I^+_{\rm flat}$};
    \node[left=4pt] at (0,-\b) {$\mathscr I^-_{\rm flat}$};

    \fill[red!80!black] (-\xcut,\ycut) circle (1.3pt);
    \fill[red!80!black] ( \xcut,\ycut) circle (1.3pt);

    \node[red!80!black, above left=2pt] at (-\xcut,\ycut)
        {$u=\mathrm{const}$};
    \node[red!80!black, above right=2pt] at (\xcut,\ycut) {};

    \draw[blue!70!black, thick]
        plot[smooth] coordinates {
            (-\W,\W)
            (-\W+\W* 1/3,\W-0.08)
            (-\W+\W* 2/3,\W-0.1)
            (\W-\W*2/3,\W+0.1)
            (\W-\W*1/3,\W+0.08)
            (\W,\W)
        };

    \node[blue!70!black, above] at (-0.5,\W) {$\Sigma$};


    \fill[blue!20, opacity=0.2] (-\W,\W) -- (0,0) -- (\W,\W) -- (0,2*\W) -- cycle;
    \draw[thick, blue!70!black,opacity=0.2] (-\W,\W) -- (0,0) -- (\W,\W) -- (0,2*\W) -- cycle;

    \draw[red!80!black, thick] (-\W-0.10,\W) -- (-\W+0.10,\W);
    \draw[red!80!black, thick] ( \W-0.10,\W) -- ( \W+0.10,\W);


    \node[red!80!black, left] at (-\W,\yB+0.05)
        {$\tau=\frac{\pi}{2}+\frac{u}{L}$};

    \node at (\W/2,-\W/3+0.15) {global AdS$_3$};

\end{tikzpicture}
\caption{Schematic Penrose diagram of global AdS$_3$ with a small approximately flat causal diamond at its center, we are suppressing the angular coordinate. A fixed-$u$ cut on future null infinity of the flat region, $\mathscr I^+_{\rm flat}$, is represented by the two red points on the upper null edges of the inner diamond. In the flat limit, these map to cuts on the AdS boundary at global times $\tau=\pi/2+u/L$. By AdS/CFT, knowing the state on this boundary cut is equivalent to knowing the state on the bulk slice $\Sigma$. If we allow for transparent boundary conditions, we can only time evolve the state on $\Sigma$ to all the points inside the blue causal diamond.}
\label{fig:AdS_flat_diamond}
\end{figure}
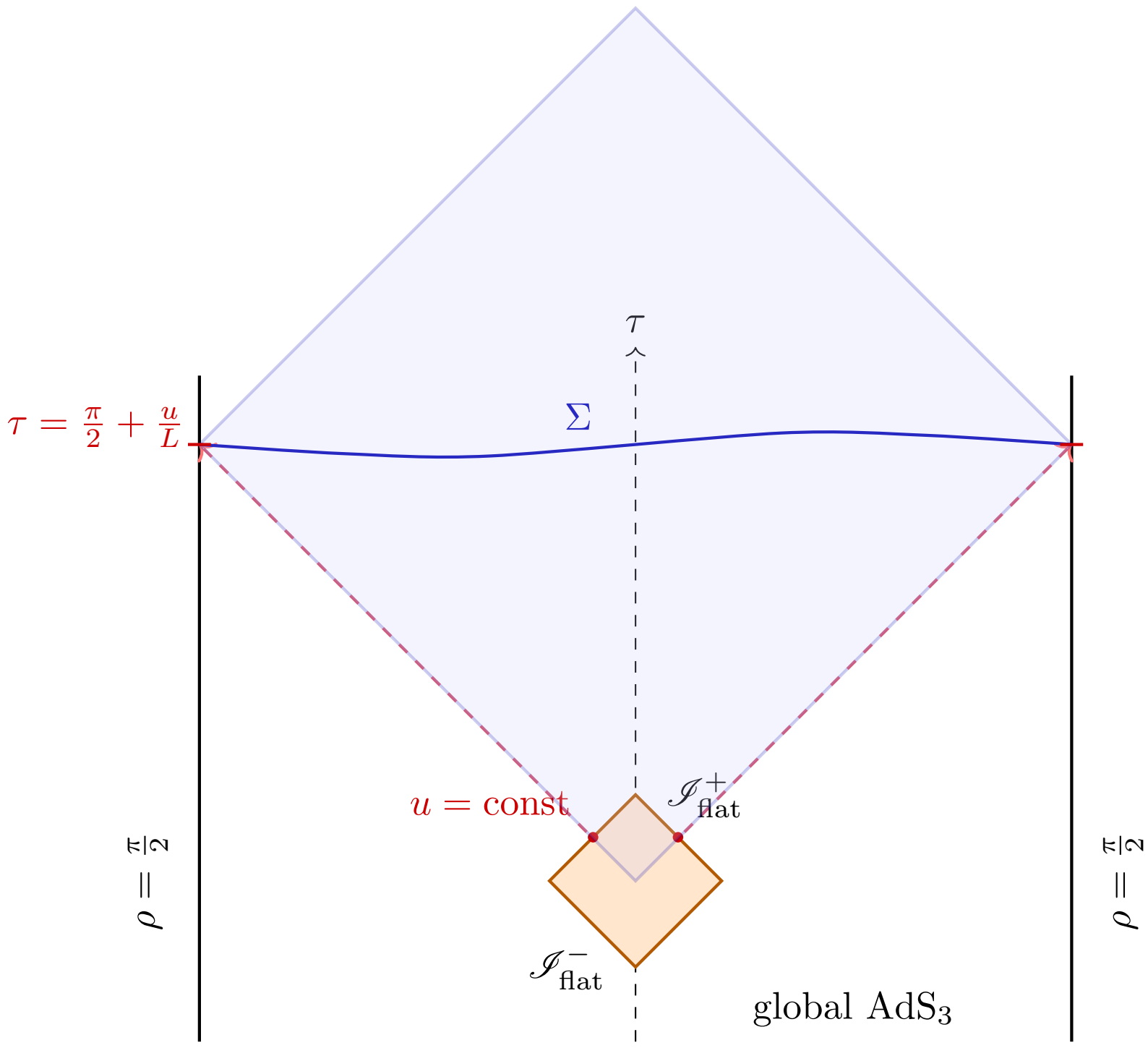

This conclusion has an interesting implication for bulk reconstruction. In ordinary AdS/CFT, the entanglement wedge describes the bulk region reconstructible from the corresponding boundary algebra~\cite{Dong:2016eik}. If this interpretation persists in our flat limit, an angular interval of one component of null infinity would not reconstruct an open bulk region through its entanglement wedge. We note that this conclusion does not extend to complete cuts spanning the entire angular circle on null infinity \cite{Chen:2023tvj,Chen:2024kuq} -- see discussion below. Our conclusion might be compared with recent approaches to bulk reconstruction in flat holography, where bulk scalar excitations were reconstructed using highest weight \cite{Chen:2023naw} or induced \cite{Hao:2025btl} representations of the boundary CCFT$_2$. These constructions do not establish reconstruction from the operator algebra of a specified boundary subregion. However, there is some tension between our conclusion and the construction of \cite{Chen:2023naw}, which expresses bulk fields as an infinite sum of local boundary operators and their derivatives. Understanding how these reconstruction prescriptions relate to entanglement wedges in the flat limit is worthy of further investigation.

Our conclusion agrees with the recent algebraic field theory analysis of \cite{Chen:2026zxc}, which finds that a partial cut of null infinity does not reconstruct bulk operators. There is a difference, however, in the boundary region associated with the subregion algebra. Our construction inherits the full Carrollian domain of dependence from the parent AdS/CFT description. By contrast, ref.~\cite{Chen:2026zxc} can only reconstruct operators to the future of the cut (on future null infinity) and cannot reconstruct observables to its past, since radiation that has already escaped through null infinity is no longer accessible. This difference also points to the distinct boundary conditions underlying the two approaches, \ie reflecting boundary conditions in AdS/CFT versus the transparent, radiative boundary conditions natural in asymptotically flat gravity.

Ref.~\cite{Laddha:2020kvp} first showed that in four-dimensional gravity on asymptotically flat spacetimes the information of the quantum state at null infinity is nested in the null direction. That is, the information present on complete cut at any $u_1$ is included in any earlier cut $u_0<u_1$. Ref.~\cite{Chen:2026zxc} reproduced this nesting result and showed that the information on such a cut of null infinity can reconstruct its associated spacelike bulk wedge, while excluding radiation that has escaped through null infinity before the cut. As the cut approaches past infinity on $\mathscr I^+$, the algebra recovers the full operator algebra in the bulk spacetime, again coinciding with \cite{Laddha:2020kvp}.  
A heuristic picture of this result can also be obtained from a flat limit of AdS -- see Figure~\ref{fig:AdS_flat_diamond}. In a flat limit centered on a small region of global AdS, a complete cut of future null infinity maps to a complete cut of the AdS boundary \cite{Caminiti:2026pma,Bagchi:2025vri,Hijano:2020szl}. In standard AdS/CFT, the state on this cut specifies data on a bulk Cauchy slice $\Sigma$, which can be used to determine the global evolution. If radiation is instead allowed to escape through the AdS boundary and no additional data are supplied, the slice determines only its causal development. The intersection of the latter with the nearly flat region is then precisely of the kind reconstructed in \cite{Chen:2026zxc}.\footnote{As previously mentioned, in \cite{Laddha:2020kvp, Chen:2026zxc} they show that the algebra associated with a cut at $u_0$ is strictly included in the algebra associated with a cut at $u_1>u_0$. A heuristic way to reproduce this behaviour from this limit would be to trace out the outgoing modes at the transparent AdS boundary. In that case, one could reconstruct the full future of $\Sigma$ in Figure \ref{fig:AdS_flat_diamond}. And the bulk regions would be properly nested. }

A related open question concerns the union of two boundary regions, one on $\mathscr I^+$ and one on $\mathscr I^-$. In the flat spacetime, the two components of null infinity are distinct, while their preimages in the AdS boundary can be timelike related, making such a configuration difficult to describe using the usual spacelike HRT prescription. Nevertheless, the joint causal past and future of the two components can encompass a much larger part of the bulk than either one alone. This suggests a possible relationship with the observation of \cite{Laddha:2020kvp,Chen:2026zxc} that the algebra at spatial infinity may reconstruct the full spacetime. We leave the precise algebraic formulation of this possibility for future investigation.

\subsection*{Two intervals and phase transitions}

The differences between the limiting RT prescription and the swing prescription also appear in the entropy of two disjoint intervals. Consider intervals $(1,2)$ and $(3,4)$ at $\mathscr I^+$, with angular coordinates ordered as $\phi_1<\phi_2<\phi_3<\phi_4$. For classical HRT surfaces the two candidate configurations pair the endpoints as
\begin{align}\label{two_interval_pairings_discussion}
 \text{Disconnected:}\quad (1,2)\cup(3,4),\qquad
 \text{Connected:}\quad (1,4)\cup(2,3).
\end{align}
The leading entropy \eqref{result_discussion} inherited from our flat limit is dominated by the logarithmic term, and therefore selects the connected phase when
\begin{align}\label{two_interval_RT_discussion}
 \sin\frac{\phi_{41}}{2}\sin\frac{\phi_{32}}{2}
 <\sin\frac{\phi_{21}}{2}\sin\frac{\phi_{43}}{2}\,,
\end{align}
where $\phi_{ij}=\phi_i-\phi_j$ and the positive interval widths are understood. This is the familiar angular dependence inherited from the parent relativistic CFT. For the swing prescription with the standard Einstein gravity values $c_L=0$, $c_M=3/G_N^{\mathrm{eff}}$, the corresponding condition instead depends on the Carrollian time separations:
\begin{align}\label{phase_transition_flat}
 \left| u_{41}\cot\frac{\phi_{41}}{2}\right|
 +\left| u_{32}\cot\frac{\phi_{32}}{2}\right|
 <\left| u_{21}\cot\frac{\phi_{21}}{2}\right|
 +\left| u_{43}\cot\frac{\phi_{43}}{2}\right|\,,
\end{align}
where $u_{ij}=u_i-u_j$, where the absolute values ensure the positive swing length~\cite{Jiang:2017ecm}. Thus a change in Carrollian time separations can alter the dominant swing configuration even when the angular intervals are held fixed. In particular, it may favor a connected configuration for angular intervals that would be far into the disconnected phase of the RT prescription. This is a qualitatively distinct entropic prediction and provides another useful comparison of the two prescriptions.

\subsection*{Alternative scaling limits and outlook}

So far, we have taken $\ell_{\mathrm{eff}}\to\infty$ while keeping the effective Newton and Chern-Simons couplings finite. The wedge construction also admits correlated limits in which the brane separation changes as the curvature radius diverges. For the Poincar\'e-slab parameters, the effective couplings are given by eqs.~\eqref{RS1_eff_G} and \eqref{mu_eff_flat},
\begin{align}\label{couplings_discussion}
 G_N^{\mathrm{eff}}=\frac{G_N}{L(1-z_2/z_1)}\,,\qquad
 \mu_{\mathrm{eff}}=\mu L(1-z_2/z_1)\,.
\end{align}
Hence, the product $G_N^{\mathrm{eff}}\mu_{\mathrm{eff}}=G_N\mu$ remains fixed. By bringing the branes closer together (\ie $z_2/z_1\to1$) as $\ell_{\mathrm{eff}}\to\infty$, one can arrange for $\ell_{\mathrm{eff}}/G_N^{\mathrm{eff}}$ to approach a finite, nonzero value. In the effective three-dimensional theory, this sends $G_N^{\mathrm{eff}}\to\infty$ and $\mu_{\mathrm{eff}}\to0$, while keeping both $G_N^{\mathrm{eff}}\mu_{\mathrm{eff}}$ and $\mu_{\mathrm{eff}}\ell_{\mathrm{eff}}$ finite.\footnote{It also makes the gap between the massless graviton and the first KK massive tensor mode grow to infinity, see Appendix \ref{brane description}.} The Einstein term is then suppressed relative to the Chern-Simons interaction, as in the double-scaling limit defining flat-space chiral gravity~\cite{Bagchi:2012yk}.

An interesting feature of this limit is that both parent AdS$_3$ central charges $c_\pm$ in eq.~\eqref{grav_anomaly}  remain finite as $\ell_{\mathrm{eff}}\to\infty$. Moreover, the ratio in eq.~\eqref{eq:cdiffer} also remains finite, so the gravitational anomaly survives the flat limit. Further we observe from eq.~\eqref{entropy_inferred_charges_discussion} that $c_L^{\mathrm{EE}}=3\ell_{\mathrm{eff}}/G_N^{\mathrm{eff}}$ approaches a finite value while $c_M^{\mathrm{EE}}\to0$. Similarly, from eq.~\eqref{geometric_charges_discussion} we find a finite $c_L^{\mathrm{geom}}=3/(G_N^{\mathrm{eff}}\mu_{\mathrm{eff}})$ while $c_M^{\mathrm{geom}}\to0$. The two surviving central charges are therefore finite but generally different. They agree only when $\mu_{\mathrm{eff}}\ell_{\mathrm{eff}}\to1$, corresponding to the chiral point of the parent AdS$_3$ TMG theory, \ie $c_+=3\ell_{\mathrm{eff}}/G_N^{\mathrm{eff}}$ and $c_-=0$ in eq.~\eqref{grav_anomaly}. Thus, while the correlated limit removes the divergence in the entropy-inferred central charge and preserves a nonvanishing gravitational anomaly, it does not generally reconcile the entanglement-based identification of the central charges with those obtained from the geometric BMS$_3$ contraction.

Our results suggest that the controlled flat limit of AdS/CFT retains structures that cannot be described by a simple contraction of the boundary symmetry algebra. As described in appendix \ref{wedge_symmetry_appendix}, the conventional BMS$_3$ diffeomorphisms arise naturally in the limiting gravitational geometry, which would naturally describe the dual Carrollian theory in the induced representation with the standard contraction \eqref{geometric_contraction_discussion}. 
In contrast, the flat limit of ordinary RT entropy yields a geometric form  associated with highest weight representation, and the entropy coefficients \eqref{entropy_inferred_charges_discussion} can be identified as the central charges of the corresponding contraction \eqref{operator_contraction_discussion}. 
A related distinction appears in the geometry of the extremal surfaces. Our RT surfaces collapse onto null rays at null infinity rather than approaching swing surfaces, which have instead been related to different information theoretic observables of the parent holographic CFT in \cite{Caminiti:2026pma}. Understanding which Carrollian theories, representations, and information theoretic observables emerge from these different limits remains an interesting open problem in flat holography.

\acknowledgments

We thank David Blanco, Jackie Caminiti, Horacio Casini, Damian Galante, Joaquim Llorens Giralt, Juan Maldacena, Pedro Jorge Martinez, Prahar Mitra, Dominik Neuenfeld, Sabrina Pasterski, Guillem P\'erez-Nadal, Wei Song and Tadashi Takayanagi for useful discussions and feedback. Research at Perimeter Institute is supported in part by the Government of Canada through the Department of Innovation, Science and Economic Development Canada and by the Province of Ontario through the Ministry of Colleges and Universities.  RCM is also supported in part by a Discovery Grant from the Natural Sciences and Engineering Research Council of Canada, and by funding from the BMO Financial Group. BS is supported by a CONICET PhD fellowship. BS also thanks the Perimeter Scholars International program, which he attended during the early stages of this research.

\appendix

\section{Three-dimensional brane description}\label{brane description}

In this appendix, we discuss the gravitational theory obtained by KK-reducing over the transverse interval in wedge holography. That is, we think of the AdS$_{d+1}$ wedge as $\text{AdS}_d\times [\zeta_1,\zeta_2]$ where $\zeta$ is the slicing coordinate defined in eq.~\eqref{change of variables} and we perform a Kaluza-Klein reduction over the $\zeta$-interval. We carry out the analysis for the case $d=3$, as that is the one that concerns us here, but it can be easily extended to $d\geq3$. We first study the massless tensor sector, and then comment on the full spectrum and its behaviour as we take the flat limit of the branes.

\subsubsection*{Massless tensor sector and gravitational parameters}

To understand the massless tensor sector of the reduced gravitational theory, we consider the following four dimensional ansatz
\begin{align}\label{metric_slices}
    ds^2=L^2 d\zeta^2+\cosh^2\!\zeta\,\,\gamma_{ab}(x)\,dx^a dx^b,
\end{align}
where $L$ is the AdS$_4$ curvature radius, $\zeta$ is the interval-valued coordinate over which we want to reduce, and $\gamma_{ab}$ is a three-dimensional metric that does not depend on $\zeta$. In the wedge geometry \eqref{Poincare slicing} considered in the main text, $\gamma_{ab}(x)$ would be the AdS$_3$ metric with radius $L$, but for the purposes of this appendix we consider a general metric.

This ansatz is obtained by promoting the metric of the AdS$_3$ slices to a general three-dimensional metric, while keeping the warp factor of the background geometry fixed. This corresponds to retaining the massless tensor mode of the KK reduction. In particular, we omit the mixed components of the metric and fluctuations of the transverse proper distance, which are not required to describe this sector. As a result, the reduction below identifies the Einstein–Hilbert action governing the massless graviton, but does not capture the complete gravitational spectrum. For a more detailed analysis see \cite{Hu:2022lxl, Miao:2020oey}.\footnote{See also \cite{Geng:2025rov} for a careful study of the graviton spectrum in the Karch-Randall braneworld.}

The four-dimensional gravitational action is given by eq.~\eqref{action} where a Gibbons-Hawking-York term supplements the tension term for each of the EOW branes.  Using the ansatz \eqref{metric_slices}, the relevant metric functions split as
\begin{align}
\begin{aligned}
    \sqrt{-g}=&L\cosh^3\!\zeta\,\sqrt{-\gamma},\hspace{50pt}&R=&\frac{R^{(3)}[\gamma_{ab}]}{\cosh^2\!\zeta}+\frac{6}{L^2\cosh^2\!\zeta}-\frac{12}{L^2},\\
    K^{(1)}_{ab}=&-\frac{\sinh\!\zeta_1\cosh\!\zeta_1}{L}\gamma_{ab}, \hspace{40pt}&K^{(2)}_{ab}=&\frac{\sinh\!\zeta_2\cosh\!\zeta_2}{L}\gamma_{ab}.
    \end{aligned}
\end{align}
where $R^{(3)}[\gamma_{ab}]$ is the Ricci scalar of the three-dimensional metric $\gamma_{ab}$, and $K_{ab}^{(i)}$ is the extrinsic curvature of the $Q_i$ brane. Putting everything together and imposing Neumann boundary conditions (see eq.~\eqref{brane eom}), the gravitational action becomes 
\begin{align}\label{bulk_reduced}
    \frac{L(\sinh\!\zeta_2-\sinh\!\zeta_1)}{16\pi G_N}\int d^3x\,\,\sqrt{-\gamma}\left(R^{(3)}[\gamma_{ab}]+\frac{2}{L^2}\right).
\end{align}
From this action, we could already read the three-dimensional Newton's constant together with the effective three-dimensional AdS radius. However, since $\gamma_{ab}(x)$ does not measure physical distances on any given $\zeta$-slice, doing so would obscure the flat limit of the branes. Therefore, before reading off the effective parameters of the three-dimensional description, we rewrite the action in terms of $h^{(2)}_{ab}(x)=\cosh^2\!\zeta_2\,\,\gamma_{ab}(x)$, that is, the induced metric on the $Q_2$ brane. Using $R^{(3)}[\gamma_{ab}]=R^{(3)}[h^{(2)}_{ab}]\cosh^2\!\zeta_2$, we finally get
\begin{align}\label{induced_description}
    \frac{L}{16\pi G_N}\,\frac{\sinh\!\zeta_2-\sinh\!\zeta_1}{\cosh\!\zeta_2}\int d^3x \,\sqrt{-h^{(2)}}\left(R^{(3)}[h^{(2)}_{ab}]+\frac{2}{L^2\cosh^2\!\zeta_2}\right)\,.
\end{align}
This is the three-dimensional Einstein-Hilbert action with negative cosmological constant, where the effective Newton's constant and curvature scale are
\begin{align}\label{brane_params_appendix}
\begin{aligned}
    G_N^{\text{eff}}=&\frac{G_N}{L}\frac{\cosh\!\zeta_2}{\sinh\!\zeta_2-\sinh\!\zeta_1}=\frac{G_N}{L}\frac{T_c}{\sqrt{T_c^2-T_2^2}}\frac{1}{\frac{T_2}{\sqrt{T_c^2-T_2^2}}+\frac{T_1}{\sqrt{T_c^2-T_1^2}}}\,,\\
    \ell_{\text{eff}}=&L\cosh\!\zeta_2=L\frac{T_c}{\sqrt{T_c^2-T_2^2}}\,,
    \end{aligned}
\end{align}
as presented in eq.~\eqref{3d params}.
We write the expressions in terms of the tension of the branes here using eq.~\eqref{tensions_mu}. Choosing the induced metric on $Q_2$ as representative is arbitrary. We could have instead chosen the metric on $Q_1$ thereby modifying the individual values of $G_N^{\text{eff}}$ and $\ell_{\text{eff}}$ but not their ratio. In fact, we see that the central charge appearing in the entropy of an interval in the corner theory in wedge holography (see eq.~\eqref{holo_central_charge}), is the Brown-Henneaux central charge of the reduced theory  \cite{Brown:1986nw}
\begin{align}
    c=\frac{3\ell_{\text{eff}}}{2G_N^{\text{eff}}}.
\end{align}
Both the Poincar\'e-slab limit and the null-anchored limit keep $G_N^{\text{eff}}$ finite while $\ell_{\text{eff}}\to\infty$.

The reduction above identifies the Einstein–Hilbert action governing the massless tensor sector, but it does not capture the complete three-dimensional gravitational theory. In particular, a more general description of the induced theory on the branes includes higher curvature corrections and a matter contribution \cite{Hu:2022lxl,Miao:2020oey}. With Neumann boundary conditions on both branes, the gravitational spectrum contains an exactly massless graviton, together with a tower of massive Kaluza–Klein modes. In addition, the scalar sector contains a radion associated with fluctuations of the separation between the branes, which becomes massless in the flat limit. We now examine this spectrum and its behaviour in the flat limit.

\subsubsection*{Kaluza--Klein spectrum and the flat limit}

To study the full tensor spectrum we can modify the ansatz \eqref{metric_slices} as 
\begin{align}\label{1st_order_ansatz}
    ds^2=L^2 d\zeta^2+\cosh^2\!\zeta\,\,[\gamma^{(0)}_{ab}(x)+\epsilon \,q_{ab}(\zeta,x)]\,dx^a dx^b,    
\end{align}
where for simplicity we now consider $\gamma^{(0)}_{ab}$ to be the AdS$_3$ metric with radius $L$, and $q_{ab}(\zeta,x)$ is the first order perturbation. 

Due to the background being a warped product, the linearized tensor perturbations admit a separation into radial eigenfunctions and three dimensional tensor modes
\begin{align}
     q_{ab}(\zeta,x)&=\sum_{n\geq0} H_n(\zeta)q_{ab}^{(n)}(x).
\end{align}
 
For the bulk Einstein
theory without additional intrinsic brane terms and with Neumann boundary conditions, transverse
traceless tensor perturbations satisfy \cite{Hu:2022lxl}
\begin{align}
\begin{aligned}
    \left(\Box +\frac{2-\nu_n^2}{L^2}\right)q^{(n)}_{ab}(x)&=0\\
    H_n''+3\tanh\!\zeta\,H_n'
    +\nu_n^2\operatorname{sech}^2\!\zeta\,H_n&=0,
    \qquad H_n'(\zeta_1)=H_n'(\zeta_2)=0,
    \end{aligned}
\end{align}
where primes denote $\zeta$ derivatives and $\Box$ is constructed with $\gamma^{0}_{ab}(x)$. The physical masses in
the $Q_2$ description are
$m_n^2=\nu_n^2/\ell_{\text{eff}}^2$ with
$\ell_{\text{eff}}=L\cosh\!\zeta_2$. The constant profile $H_0$ (with $\nu_0=0$) corresponds to an exactly massless, normalizable graviton, which is the zero mode retained in the reduction leading to eq.~\eqref{induced_description}.

For $d=3$, the remaining tensor spectrum can be obtained
explicitly. Introducing
$\theta=\arctan(\sinh\zeta)$ and
$\Phi_n=\cosh^2\!\zeta\,H_n'$, the nonzero modes obey
\begin{align}
    -\partial_\theta^2\Phi_n=(\nu_n^2+1)\Phi_n,
    \qquad \Phi_n(\theta_1)=\Phi_n(\theta_2)=0.
\end{align}
Furthermore, the masses are given by
\begin{align}\label{KK_spectrum_appendix}
    m_0^2=0,\qquad
    m_n^2=\frac{1}{\ell_{\text{eff}}^2}
    \left[\frac{n^2\pi^2}{(\Delta\theta)^2}-1\right],
    \qquad n=1,2,\ldots,
\end{align}
where
$\Delta\theta=\arctan(\sinh\zeta_2)
-\arctan(\sinh\zeta_1)$.

Considering the Poincar\'e-slab limit, writing $X=x_0e^\eta$,
we have $\sinh\!\zeta_i=X/z_{i}$, with
$z_{1}>z_{2}>0$. Although both $\zeta_i$ diverge in the limit, the proper
interval length approaches the finite value
\begin{align}
    D=L(\zeta_2-\zeta_1)
    \longrightarrow D_\infty
    =L\log\frac{z_{1}}{z_{2}}.
\end{align}
Moreover,
$\Delta\theta=(z_{1}-z_{2})/X+O(X^{-3})$
and $\ell_{\text{eff}}\sim LX/z_{2}$.
Consequently, the massive tensor modes retain a finite gap:
\begin{align}\label{KK_flat_spectrum_appendix}
    m_n\longrightarrow
    \frac{n\pi z_{2}}{L(z_{1}-z_{2})}
    =\frac{n\pi}{L(e^{D_\infty/L}-1)},
    \qquad n\geq1.
\end{align}
This is the tensor spectrum of the limiting four-dimensional
Randall-Sundrum~I slab~\cite{Randall:1999ee}.
The dimensionless eigenvalues $\nu_n$ diverge in the limit,
compensating the divergence of $\ell_{\text{eff}}$.

Let us briefly comment on how different the resulting theory would be if instead of landing on the RSI model in the flat limit, we forced both flat branes to have positive tension.\footnote{This is not possible in the Poincar\'e slab limit, but it would be possible to achieve in the null-anchored limit where the corner theory would end up at future null infinity for $Q_2$ but at past null infinity for $Q_1$. In this case, the $k$ parameter in eq.~\eqref{anomalous_spectrum} would be $k^2=\frac{L}{\pi}\left(1+\frac{\beta_2}{|\beta_1|}\right)$.} In this case, we would also have $\ell_{\text{eff}}\to\infty$ in the limit, but now $\Delta\theta\to\pi$ at the same rate. Thus, in this limiting theory, setting $\Delta\theta=\pi-\pi k^2/\ell_{\text{eff}}$ for some constant $k$, we would have
\begin{align}\label{anomalous_spectrum}
    m_0=0,\hspace{30pt} m^2_1\sim\frac{2k^2}{\ell_{\text{eff}}^3}, \hspace{30pt}m_n^2\sim\frac{n^2-1}{\ell^2_{\text{eff}}}, \quad n>1.
\end{align}
That is, we would obtain a continuous spectrum, and the first mode becomes an anomalously light graviton ($m_1/m_2\to0$).\footnote{See \cite{Kogan:2000vb} for a more comprehensive analysis of the AdS$_5$ case with two positive tension AdS$_4$ branes. }

The scalar sector also contains a radion describing fluctuations
of the proper separation between the two branes. For curved
branes, this separation is fixed by the junction conditions and
the radion acquires a curvature-induced mass -- see
\cite{Chacko:2001radion} for the corresponding analysis in five
bulk dimensions. In the present three-dimensional brane
description, its linearized equation is
\begin{align}\label{radion_mass_appendix}
    \left(\Box_{(2)}-\frac{3}{\ell_{\text{eff}}^2}\right)
    \varphi=0,
    \qquad
    m_{\mathrm{rad}}^2=\frac{3}{\ell_{\text{eff}}^2}
    \longrightarrow0,
\end{align}
where $\Box_{(2)}$ is constructed from $h^{(2)}_{ab}$.
At the critical flat-brane tensions, the separation becomes a
modulus. Holding its background value fixed while taking the
limit does not remove the corresponding fluctuations.
Thus, below the massive tensor gap, the limiting theory contains
three-dimensional Einstein gravity coupled to a massless
radion, unless the latter is separately stabilized or excluded.
The Einstein sector itself has no local propagating tensor
degrees of freedom in three dimensions.


\section{Poincar\'e-slab limit: detailed computations}\label{Poin-slab-appendix}

In this appendix, we present some of the detailed computations regarding the Poincar\'e-slab limit and its HEE, together with the frame dependence of the anomalous entropy functional used in sections \ref{chern-simons} and \ref{null-anchored-appendix}.

\subsection*{Poincar\'e-slab limit in global coordinates} 

\begin{figure}
    \centering

    \begin{subfigure}[t]{0.3\textwidth}
        \centering
        \includegraphics[width=\linewidth]{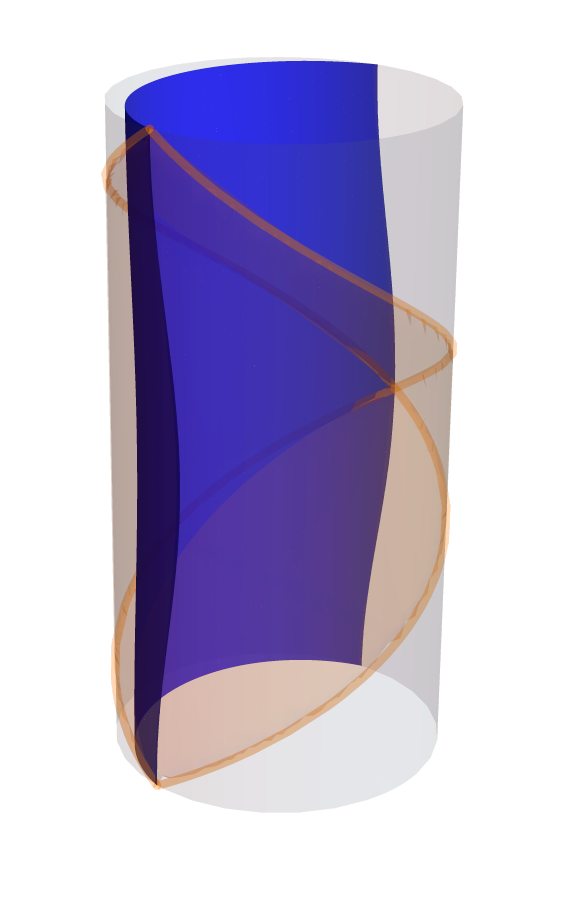}
        \caption{}
        \label{fig:Poinc-slab-lim-global-a}
    \end{subfigure}
    \hfill
    \begin{subfigure}[t]{0.3\textwidth}
        \centering
        \includegraphics[width=\linewidth]{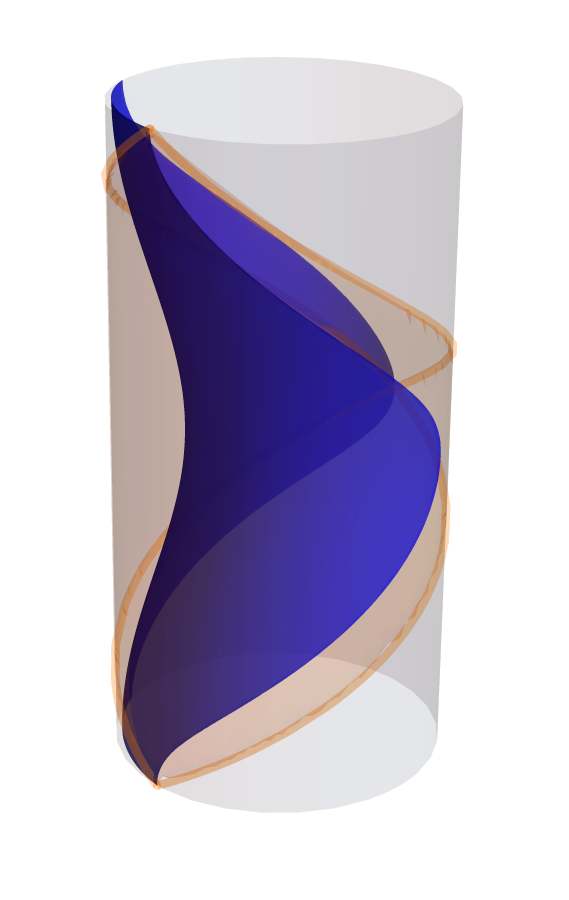}
        \caption{}
        \label{fig:Poinc-slab-lim-global-b}
    \end{subfigure}
    \hfill
    \begin{subfigure}[t]{0.3\textwidth}
        \centering
        \includegraphics[width=\linewidth]{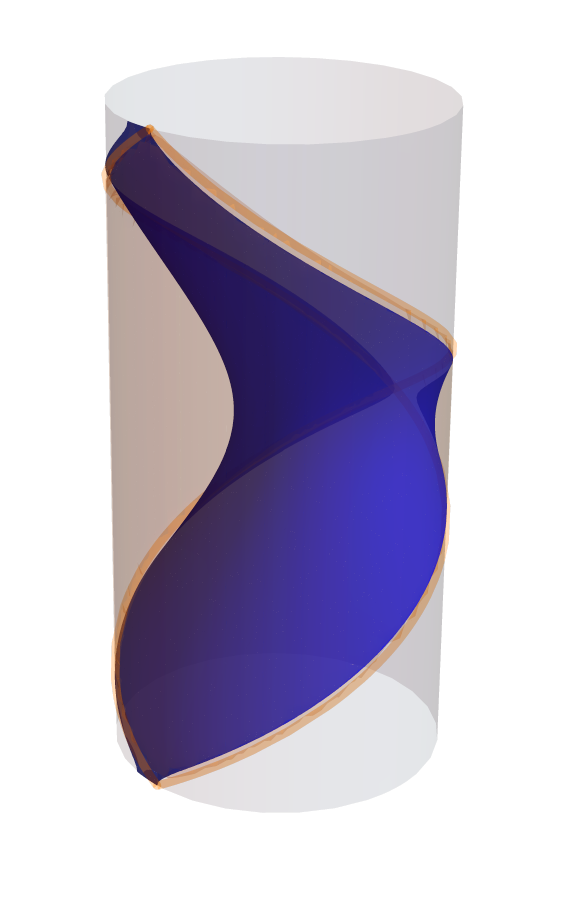}
        \caption{}
        \label{fig:Poinc-slab-lim-global-c}
    \end{subfigure}

    \caption{Fixed $\varphi=\pi/2$ slice of $\text{AdS}_4$ in global coordinates \eqref{change_to_global}. In orange we show the Poincar\'e patch covered by the coordinates \eqref{change_to_global}, and in blue we show branes of the form \eqref{brane_eq_joint_global} with $x_*=-3$, $z_*=1$, and different values of $x_0e^{\eta}$. (a) $x_0e^{\eta}=0.1$, (b) $x_0e^{\eta}=1.5$, and (c) $x_0e^{\eta}=10$.}
    \label{fig:Poinc-slab-lim-global}
\end{figure}

\begin{figure}
    \centering  \hspace{80pt}\includegraphics[width=0.65\linewidth]{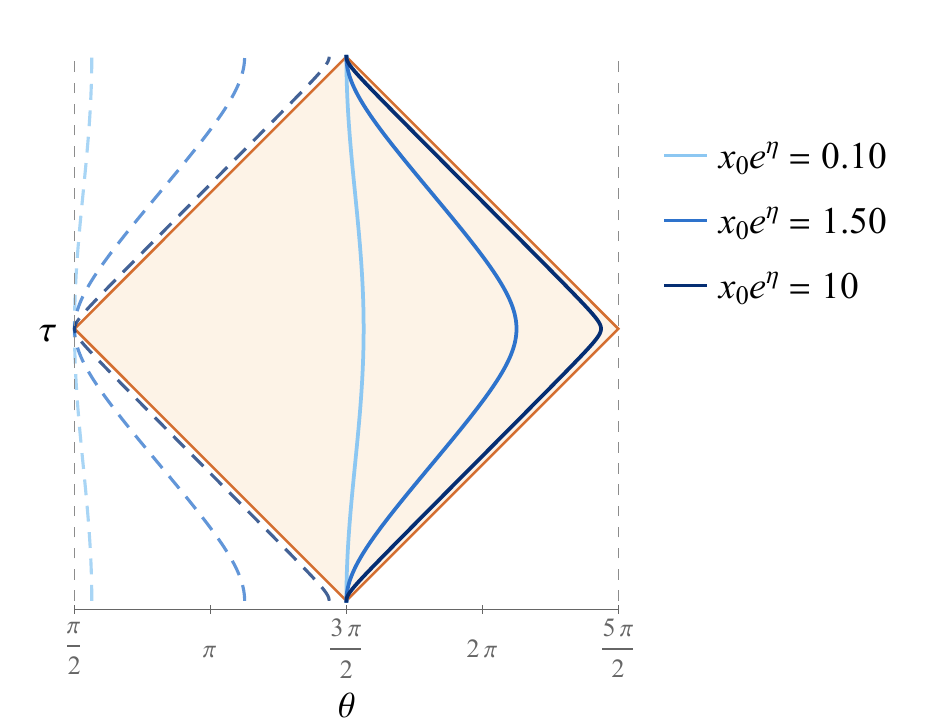}
    \caption{Unwrapped boundary of global $\text{AdS}_4$ for $\varphi=\pi/2$, we show the identified segments as vertical dashed lines. In orange we show the boundary of the Poincar\'e patch in eq.~\eqref{change_to_global} and in different shades of blue, we show the intersection of the branes \eqref{brane_eq_joint_global} with the AdS boundary for different values of $x_0e^{\eta}.$ The dashed lines correspond to the brane boundary lying outside of the Poincar\'e patch.}
    \label{fig:bound_inter_poin_slab}
\end{figure}

In section \ref{controlled_limit_section}, we discussed the controlled flat limit in Poincar\'e coordinates. It is also instructive, however, to understand this transformation in global coordinates. Making the coordinate change
\begin{align}\label{change_to_global}
    \begin{aligned}
        \frac{z}{2}+\frac{1}{2z}(L^2+x^2+y^2-t^2) =& L \cosh\rho\,\cos\tau\,,\\
        \frac{z}{2}-\frac{1}{2z}(L^2-x^2-y^2+t^2)=& L \sinh\rho\, \sin\theta\,\sin\varphi\,,\\
        L\frac{y}{z}=& L \sinh\rho\, \sin\theta\, \cos\varphi\,,\\
        L\frac{x}{z}=& L \sinh\rho\,\cos\theta\,,\\
        L\frac{t}{z}=&L \cosh\rho\,\sin\tau\,,
    \end{aligned}
\end{align}
and fixing $y=0$ ($\varphi=\pi/2$) since the configuration is invariant under translations in $y$, the brane equation \eqref{brane_eq_joint} in global coordinates becomes
\begin{align}\label{brane_eq_joint_global}
    \sinh\rho\,\cos\theta=\frac{x_*-x_0e^{\eta}}{z_*}+\frac{x_0e^{\eta}}{L}\left(\cosh\rho\,\cos\tau-\sinh\rho\,\sin\theta\right)\,.
\end{align}
We show these branes for $x_*=-3$, $z_*=1$ and different values of $x_0e^{\eta}$ in Figure \ref{fig:Poinc-slab-lim-global}.\footnote{The Penrose diagrams in Figure \ref{fig:Poinc-slab-lim-global} are made with the usual conformal compactification $\sinh\rho=\tan\psi$.} We also show the Poincar\'e patch to which we have access while working in the Poincar\'e coordinates. The figure makes clear that the limiting configuration is an RS1 configuration. Moreover, from Figure \ref{fig:Poinc-slab-lim-global} it is clear that the boundary of the branes becomes null infinity of the Poincar\'e patch as we take $\eta\to\infty$. This behaviour is also evident in Figure \ref{fig:bound_inter_poin_slab} which shows the intersection of the brane with the asymptotic AdS$_4$ boundary for various values of $x_0e^\eta$.

%
%
%

\subsection*{General intervals in the Poincar\'e-slab limit} 

In section \ref{results_Setting2}, we examined the HEE for intervals at null infinity which were symmetric about $\phi=0$, as in eq.~\eqref{symmetric_interval}. Here we describe the calculations for a general interval with endpoints $(\tilde u_\pm,\phi_\pm)$. Our analysis makes clear that while extra terms appear at intermediate steps, they are artifacts of the regulator. We also evaluate the non-universal constants and show how they are absorbed into the regulator $\hat\epsilon$ of eq.~\eqref{regulator_def}.

\paragraph{The entropy of a general interval:}
Using eq.~\eqref{endpoints_poinc_slab}, the invariant separation of the endpoints for a general interval on null infinity can be written  as
\begin{align}\label{sigma_general}
\begin{aligned}
    \sigma^2=&\,X^2(\tan\phi_+-\tan\phi_-)^2-\left[\Delta\tilde u+X\left(\frac{1}{\cos\phi_+}-\frac{1}{\cos\phi_-}\right)\right]^2\\
    =&\,X^2\left[\frac{4\sin^2\frac{\Delta\phi}{2}}{\cos\phi_+\cos\phi_-}-\frac{2\Delta\tilde u}{X}\left(\frac{1}{\cos\phi_+}-\frac{1}{\cos\phi_-}\right)-\frac{\Delta\tilde u^2}{X^2}\right]\,,
    \end{aligned}
\end{align}
where the second line follows from expanding the squares and using standard trigonometric identities. Recall that $X=x_0\,e^\eta$ from eq.~\eqref{Xdef}. Expanding in the flat limit $X\to\infty$,
\begin{align}\label{intermediate_log_PS_gen}
\begin{aligned}
    \log\left(\frac{\sigma}{\epsilon}\right)=\log\left[\frac{2X}{\epsilon}\frac{\sin\frac{\Delta\phi}{2}}{\sqrt{\cos\phi_+\cos\phi_-}}\right]&-\frac{\Delta \tilde{u}}{2X}\frac{\sin \bar{\phi}}{\sin\frac{\Delta\phi}{2}}\\
    &-\frac{\Delta \tilde{u}^2}{8X^2}\frac{\cos^2\frac{\Delta\phi}{2}+\sin^2\bar{\phi}}{\sin^2\frac{\Delta\phi}{2}}+\mathcal{O}(X^{-3})\,,
    \end{aligned}
\end{align}
with $\bar\phi=(\phi_++\phi_-)/2$ the angular midpoint of the interval. By setting $\phi_\pm=\pm\Delta\phi/2$ as in eq.~\eqref{symmetric_interval} (which produces $\bar\phi=0$), one recovers eq.~\eqref{intermediate_log_PS}.

The Weyl factor relating the boundary at fixed $x$ to null infinity at fixed $\tilde r$ was quoted in eq.~\eqref{standard_null_metric}. This factor follows by examining the induced metrics on the two families of cutoff surfaces, which in the Bondi coordinates \eqref{null_coordinates} read
\begin{align}\label{fixed_x_metric}
\begin{aligned}
    ds^2\big|_{\tilde r=\tilde r_0}=&-d\tilde u^2+\tilde r_0^2\,d\phi^2\,,\\
    ds^2\big|_{x=x_0}=&-d\tilde u^2-\frac{2x_0\sin\phi}{\cos^2\phi}\,d\tilde u\,d\phi+\frac{x_0^2}{\cos^2\phi}\,d\phi^2\,.
    \end{aligned}
\end{align}
In both cases, we have stripped off the overall factor $L^2/z^2$ common to both. Dividing by $\tilde r_0^2$ and $x_0^2$ respectively and taking the limits $\tilde r_0,x_0\to\infty$ yields eq.~\eqref{standard_null_metric}, so that the two metrics differ by $e^{2w}=\sec^2\!\phi$.

Applying eq.~\eqref{weyl_entropy_contribution} with $w=\log\sec\phi$ at the two endpoints, we get
\begin{align}\label{shift_weyl_factor_gen}
    S_{\text{Fixed }x}-S_{\text{Fixed }\tilde r}=-\frac{c}{3}\log\sqrt{\cos\phi_+\cos\phi_-}\,,
\end{align}
where $S_{\text{Fixed }x}$ is the entropy obtained from eq.~\eqref{intermediate_log_PS_gen} and $S_{\text{Fixed }\tilde r}$, that which we compare with the Carrollian result in eq.~\eqref{CCFT_EE_intro}. Note that the overall factor $L^2/z_2^2$ relating the induced metric on $Q_2$ to $\text{Mink}_3$ multiplies both metrics in eq.~\eqref{fixed_x_metric}, and so does not enter their relative Weyl factor \cite{Jiang:2017ecm, Apolo:2020bld, Hao:2025ocu}. Combining eqs.~\eqref{int_entropy}, \eqref{c_and_params_eta}, \eqref{intermediate_log_PS_gen} and \eqref{shift_weyl_factor_gen}, the entropy of a general interval is
\begin{align}\label{S_PS_provisional}
\begin{aligned}
    S=&\,\frac{L^2}{2G_N}X\left(\frac{1}{z_2}-\frac{1}{z_1}\right)\log\left(\frac{2X}{\epsilon}\sin\frac{\Delta\phi}{2}\right)
    -\frac{L^2}{4G_N}\left(\frac{1}{z_2}-\frac{1}{z_1}\right)\Delta \tilde{u}\,\frac{\sin \bar{\phi}}{\sin\frac{\Delta\phi}{2}}\\
    &-\frac{L^2}{16G_N X}\left(\frac{1}{z_2}-\frac{1}{z_1}\right)\Delta \tilde{u}^2\,\frac{\cos^2\frac{\Delta\phi}{2}+\sin^2\bar{\phi}}{\sin^2\frac{\Delta\phi}{2}}+\mathcal{O}(X^{-2})+\text{constants}\,.
    \end{aligned}
\end{align}

The first term is the one kept in the main text. The remaining two depend on $\bar\phi$, \ie on the position of the interval and not only on its angular width, and so break the rotational invariance expected of the flat-space answer \eqref{CCFT_EE_intro}. Worse, the second term survives the flat limit. Using eqs.~\eqref{bondi_physical} and \eqref{RS1_eff_G}, it becomes
\begin{align}\label{spurious_term}
    -\frac{1}{4G_N^{\text{eff}}}\,\Delta u\,\frac{\sin\bar\phi}{\sin\frac{\Delta\phi}{2}}\,,
\end{align}
which is finite and linear in $\Delta u$, and could easily be mistaken for a $c_M$ contribution with $c_M\neq0$.

\paragraph{The angle-dependent cutoff:}
The origin of these terms is in the shape of the cutoff surface. From eq.~\eqref{wedge_branes_eta}, the intersection of the cutoff surface $z=\epsilon$ with the brane sits at some $x=X_\epsilon<X$, which in Bondi coordinates is the cut
\begin{align}\label{angle_dependent_cutoff}
    R_\epsilon(\phi)=\frac{X_\epsilon}{\cos\phi}\,,
\end{align}
rather than the cut at fixed $\tilde r$ that one would naturally impose in flat space. Each endpoint of the interval is therefore regulated at its own radius $R_\pm=X_\epsilon/\cos\phi_\pm$, and the factor which breaks rotational invariance in eq.~\eqref{S_PS_provisional} is precisely a measure of the difference,
\begin{align}\label{sin_phibar_cutoff}
    \sin\bar{\phi}=\frac{X_{\epsilon}}{2\sin\frac{\Delta\phi}{2}}\left(\frac{1}{R_-}-\frac{1}{R_+}\right).
\end{align}
To compare with the result \eqref{CCFT_EE_intro} obtained in flat space with a fixed $\tilde r=R_\epsilon$ cutoff, one must therefore choose the endpoints so that $R_-=R_+$, which is the symmetric choice $\phi_+=-\phi_-=\Delta\phi/2$ of eq.~\eqref{symmetric_interval}. With that choice eq.~\eqref{S_PS_provisional} reduces to eq.~\eqref{S_PS_sym} and both offending terms disappear. That is, the term \eqref{spurious_term} vanishes, while $\cos^2\frac{\Delta\phi}{2}+\sin^2\bar\phi\to\cos^2\frac{\Delta\phi}{2}$ converts the last term into the $\cot^2(\Delta\phi/2)$ term of eq.~\eqref{S_PS_sym}.

We emphasize that this is a statement about the regulator and not about the state. The entropy of a general interval is of course well defined, and is given by eq.~\eqref{S_PS_provisional} with the cutoff \eqref{angle_dependent_cutoff}. What the symmetric choice buys us is that the regulator is the rotationally invariant one implicitly used in the Carrollian result \eqref{CCFT_EE_intro}. The $\bar\phi$-dependence of the anomalous Chern-Simons contribution of section \ref{chern-simons} has a different origin. It is insensitive to this Weyl rescaling, and it is removed by the choice of endpoint frames rather than by the choice of cutoff -- see appendix \ref{Poin-slab-appendix} and eq.~\eqref{arctanh_identity}.

\paragraph{The non-universal constants:}
Finally, we evaluate the terms denoted ``constants'' above and show that they are absorbed into $\hat\epsilon$. They come from the second term in eq.~\eqref{int_entropy}, which can be integrated exactly,
\begin{align}\label{constants_exact}
    -\frac{L^2}{2G_N}\int_{\zeta_1}^{\zeta_2}\!\!d\zeta\,\cosh\zeta\log(\cosh\zeta)=-\frac{L^2}{2G_N}\Big[\sinh\zeta\,\log\cosh\zeta-\sinh\zeta+\arctan(\sinh\zeta)\Big]_{\zeta_1}^{\zeta_2}.
\end{align}
Throughout the Poincar\'e-slab limit, the branes sit at $\sinh\zeta_i=X/z_i$ and $\cosh\zeta_i=\ell_{\text{brane},i}/L$, so that at large $X$
\begin{align}\label{blow-up-constants}
    -\frac{L^2}{2G_N}\int_{\zeta_1}^{\zeta_2}\!\!d\zeta\,\cosh\zeta\log(\cosh\zeta)=-\frac{c}{3}\left(\Lambda-1\right)+\mathcal{O}(1/X)\,,\hspace{20pt}\Lambda=\frac{z_1\log\frac{X}{z_2}-z_2\log\frac{X}{z_1}}{z_1-z_2}\,,
\end{align}
with $c$ given in eq.~\eqref{c_and_params_eta}. Every term here is proportional to $c\propto X= x_0e^\eta$, and hence diverges in the flat limit; but for the same reason all of them can be absorbed into a redefinition of the cutoff inside the logarithm of eq.~\eqref{S_PS_provisional}, since a rescaling $\epsilon\to\lambda\epsilon$ shifts the entropy by $-\frac{c}{3}\log\lambda$. Explicitly, adding eq.~\eqref{blow-up-constants} to the first term of eq.~\eqref{S_PS_provisional} gives
\begin{align}\label{absorbing_constants}
    \frac{c}{3}\left[\log\left(\frac{2X}{\epsilon}\sin\frac{\Delta\phi}{2}\right)-\Lambda+1\right]=\frac{c}{3}\log\left(\frac{2}{\hat\epsilon}\sin\frac{\Delta\phi}{2}\right),\hspace{20pt}\hat\epsilon=\frac{\epsilon}{X}e^{\Lambda-1}=\epsilon\,e^{-1-\kappa}\,,
\end{align}
with $\kappa=(z_1\log z_2-z_2\log z_1)/(z_1-z_2)$, which is the definition \eqref{regulator_def} quoted in the main text. Note that $\hat\epsilon$ is independent of $X$, so that holding the bulk cutoff $\epsilon$ fixed along the limit is equivalent to holding $\hat\epsilon$ fixed. Since $\epsilon$ here is the cutoff in the dilated coordinates \eqref{dilatation}, this corresponds to a cutoff $\epsilon\,e^{-\eta}$ in the original coordinates, \ie to finite brane cutoffs $\epsilon_i$, as in section \ref{wedge_holo_section} and in eq.~\eqref{regulator_NA} for the null-anchored limit. From the perspective of the dilated coordinates eq.~\eqref{effective_cutoff_0} is still satisfied even though $\epsilon_i\to\infty$ there, since the proper size of the interval $\sigma$ also diverges. 

\paragraph{Correction around the Carrollian limit:}
As noted below eq.~\eqref{final_result_set2}, the second term there is not a failure to reproduce the Carrollian answer, but rather a correction away from the Carrollian limit.
To see this, we can restore the speed of light $v_\gamma$ in the relativistic $\text{CFT}_2$ answer for a tilted interval on the cylinder and expand around $v_{\gamma}\to0$. For an interval with displacements $(\Delta u, \Delta \phi)$ for the CFT vacuum on a cylinder identified with $(u,\phi) = (u+2\pi s/v_{\gamma}, \phi+2\pi)$ with $|s|<1$:
\begin{equation}\label{eq:complicated}
\begin{aligned}
S
={}&
\frac{c_-+c_+}{12}\,
\log\!\left[
\frac{4(1-s^2)}{\hat{\epsilon}^2}
\left|
\sin\!\left(
\frac{\Delta \phi-v_{\gamma}\Delta u}{2(1-s)}
\right)
\sin\!\left(
\frac{\Delta \phi+v_{\gamma}\Delta u}{2(1+s)}
\right)
\right|
\right]
\\[2mm]
&\qquad+
\frac{c_--c_+}{12}\,
\log\!\left[
\frac{1-s}{1+s}
\left|
\frac{
\sin\!\left(
\dfrac{\Delta \phi-v_{\gamma}\Delta u}{2(1-s)}
\right)}
{
\sin\!\left(
\dfrac{\Delta \phi+v_{\gamma}\Delta u}{2(1+s)}
\right)}
\right|
\right].
\end{aligned}
\end{equation}

Expanding with $v_{\gamma}\ll1$ and fixed $(\Delta u,\Delta\phi)$ in the case $s=0$, we find
\begin{align}\label{cgamma_expansion}
    S=\frac{c_-+c_+}{6}\log\left(\frac{2}{\hat\epsilon}\sin\frac{\Delta\phi}{2}\right)+\frac{c_+-c_-}{12}v_{\gamma}\Delta u \cot\frac{\Delta\phi}{2}-\frac{c_-+c_+}{48}\,v_\gamma^2\,\Delta u^2\csc^2\frac{\Delta\phi}{2}+\mathcal{O}(v_\gamma^3)\,.
\end{align}
With $(c_-+c_+)/2=3\ell_{\text{eff}}/(2G_N^{\text{eff}})$ and\footnote{Note that since the angular coordinate $\phi$ is dimensionless
and $u$ has dimensions of length, $v_\gamma$ has dimensions
of inverse length. With a cylinder of circumference $a$, its relation with the usual dimensionless speed of light is $v_{\gamma}=2\pi\,c_{\gamma}/a$. \label{footy33}}
\begin{align}\label{cgamma_identification}
    v_\gamma=\frac{1}{\ell_{\text{eff}}}\,,
\end{align}
the third term in the previous expression becomes $-\Delta u^2\csc^2(\Delta\phi/2)/(16G_N^{\text{eff}}\ell_{\text{eff}})$, which matches the term appearing in eq.~\eqref{final_result_set2} up to the replacement $\cot^2\to\csc^2$. The identification \eqref{cgamma_identification} is the statement that the flat limit of the branes is an ultrarelativistic limit of the boundary theory, which we return to in section \ref{sec:disc}. On the other hand, the linear term in $v_{\gamma}$ does not appear in eq.~\eqref{S_PS_provisional} because so far we have been working with $c_-=c_+$. This term will be relevant once we include the gravitational Chern-Simons term and its anomalous entropy contribution, as seen in appendix \ref{Poin-slab-appendix}.

The remaining discrepancy, $\csc^2=\cot^2+1$, is instructive: the difference between the two is $\Delta u^2/(16G_N^{\text{eff}}\ell_{\text{eff}})$, which depends on the interval and so is not a constant. Indeed, the null-anchored limit of section \ref{null-anchored-appendix} returns the $\csc^2$ form in eq.~\eqref{rot inv NA cl}. The two limits reach the same flat configuration and necessarily agree on the leading term, but the way they approach the flat limit differs at finite $\ell_{\text{eff}}$. To see this in a concrete way, we first recall that the asymptotic boundary of the branes in the Poincar\'e-slab limit is at a fixed value of $x=X$ throughout the limit, and $\ell_{\text{eff}}\to\infty$ is achieved with $X\to\infty$. With this in mind, let us compare the metric on the boundary of the branes in the Poincar\'e-slab limit after removing the Weyl factor discussed around eq.~\eqref{fixed_x_metric} with the metric on a Minkowski cylinder of unit radius, which is the one implicitly used for eq.~\eqref{cgamma_expansion}:
\begin{align}\label{bound_metric_finite_l_PS}
\begin{aligned}
    ds^2\big|_{\text{cyl}}=&-\frac{1}{\ell_{\text{eff}}^2}\,du^2+\,d\phi^2\,,\\
    \frac{\cos^2\phi}{X^2} \,ds^2\big|_{x=X}=
    &-\frac{1}{\ell_{\text{eff}}^2}\,du^2+\left(d\phi-\frac{\sin\phi}{\ell_{\text{eff}}}\, du\right)^2+\mathcal{O}(\ell^{-3}_{\text{eff}})
    \,,
    \end{aligned}
\end{align}
where we used eq.~\eqref{leff_X} to relate $X$ to $\ell_{\text{eff}}$, and eq.~\eqref{bondi_physical} to go from the tilde coordinates in eq.~\eqref{fixed_x_metric} to the physical Bondi coordinates.

Making the identification \eqref{cgamma_identification} we see that the two metrics coincide in the limit $v_{\gamma}\to0$ but differ at non-zero $v_{\gamma}$ (finite $\ell_{\text{eff}}$). This is a clear indication that the subleading terms in the entropy need not coincide. In particular, there are two differences between the two subleading terms of the metrics: a cross term $dud\phi$ and an extra $\cos^2\phi$ factor multiplying $du^2$. For the symmetric interval the cross term ends up not contributing since it is antisymmetric in $\phi$, on the other hand, the extra $\cos^2\phi$ in the $du^2$ piece is exactly what is making the $\csc^2\to\cot^2$ in our final entropy eq.~\eqref{final_result_set2}. The $\mathcal{O}(\Delta u^2/\ell_{\text{eff}})$ term should therefore be read as depending on how the boundary frame is fixed by the particular limit, rather than as a robust feature of the construction. The leading term, and the $c_M$ term found in section \ref{chern-simons}, agree in the two limits.


\subsection*{Frame dependence of the anomalous contribution} 

Because the endpoint frames are part of the definition of the anomalous entropy functional \eqref{CS entropy}, we need to know how the anomalous contribution changes when they are changed. We record the general result here. It is used for the Poincar\'e-slab limit in section \ref{chern-simons} and appendix \ref{Poin-slab-appendix}, and for the null-anchored limit in section \ref{null-anchored-appendix}.

Suppose two choices of endpoint frames ($n_{i,f},\tilde n_{i,f}$) and ($m_{i,f},\tilde m_{i,f}$) are related by local boosts,
\begin{align}\label{related_normal_frames}
\left\{
\begin{aligned}
    n_i=&\cosh\chi_i\,\,m_i+\sinh\chi_i \,\,\tilde{m}_i\\
    \tilde{n}_i=&\cosh\chi_i\,\,\tilde{m}_i+\sinh\chi_i\,\,m_i
    \end{aligned}\right.\hspace{30pt}
    \left\{
\begin{aligned}
    n_f=&\cosh\chi_f\,\,m_f+\sinh\chi_f \,\,\tilde{m}_f\\
    \tilde{n}_f=&\cosh\chi_f\,\,\tilde{m}_f+\sinh\chi_f\,\,m_f
    \end{aligned}\right.
\end{align}
where $i,f$ label the two endpoints and tildes denote spacelike normals. Expanding each of the four frames in the parallel transported frame $(q,\tilde q)$ as in eq.~\eqref{q_expansion}, with rapidities $\vartheta^{(n)}_{i,f}$ and $\vartheta^{(m)}_{i,f}$, eq.~\eqref{related_normal_frames} immediately gives $\vartheta^{(n)}_i=\vartheta^{(m)}_i+\chi_i$ at the initial endpoint. At the final endpoint the bulk parallel transport reverses the orientation of the spacelike normal \cite{Castro:2014tta}, \ie $\tilde n_f=-\cosh\vartheta^{(n)}_f\tilde q_f-\sinh\vartheta^{(n)}_f q_f$ and likewise for $\tilde m_f$, so that instead $\vartheta^{(n)}_f=\vartheta^{(m)}_f-\chi_f$. From eq.~\eqref{CS_in_rapidity}, the two anomalous contributions are therefore related by
\begin{align}\label{entropy_rel_dif_frames}
    S_{\text{CS}}^{(n)}=S_{\text{CS}}^{(m)}-\frac{1}{4G_N^{\text{eff}}\mu_{\text{eff}}}(\chi_f+\chi_i)\,.
\end{align}
Note that it is the sum of the two rapidities that appears, again because of the orientation reversal. In particular, if the endpoint frames are boosted by equal and opposite rapidities, $\chi_f=-\chi_i$, the anomalous contribution is unchanged. The latter is the situation for the symmetric interval in section \ref{chern-simons}.

\subsection*{Chern-Simons contribution for a general interval}\label{CS_general_appendix}

In section \ref{chern-simons}, we evaluated the anomalous contribution for the symmetric interval \eqref{symmetric_interval}, for which the endpoint frames $n=\partial_t$ and the Carrollian frame give the same answer. Here we consider a general interval with endpoints \eqref{endpoints_poinc_slab} at angles $\phi_\pm$. In this case, the two frames differ and, as we now show, only the Carrollian frame yields a rotationally invariant result in the flat limit.

\paragraph{The anomalous contribution in the frame $\partial_t$:} The displacements between the endpoints \eqref{endpoints_poinc_slab} are
\begin{align}\label{displacements_general_slab}
    \Delta t=\Delta\tilde u+\frac{2X\sin\bar\phi\,\sin\frac{\Delta\phi}{2}}{\cos\phi_+\cos\phi_-}\,,\hspace{30pt}\Delta y=\frac{X\sin\Delta\phi}{\cos\phi_+\cos\phi_-}\,,
\end{align}
with $\bar\phi=(\phi_++\phi_-)/2$. Hence we have 
\begin{align}\label{displacement_ratio_slab}
    \frac{\Delta t}{\Delta y}=A+B\,,\hspace{30pt}A\equiv\frac{\sin\bar\phi}{\cos\frac{\Delta\phi}{2}}\,,\hspace{20pt}B\equiv\frac{\Delta\tilde u}{2X}\,\frac{\cos\phi_+\cos\phi_-}{\sin\frac{\Delta\phi}{2}\cos\frac{\Delta\phi}{2}}\,.
\end{align}
Using $\cos\phi_+\cos\phi_-=\cos^2\frac{\Delta\phi}{2}-\sin^2\bar\phi$, the second term can be written as $B=b\,(1-A^2)$, with
\begin{align}\label{b_def_slab}
    b\equiv\frac{\Delta\tilde u}{2X}\cot\frac{\Delta\phi}{2}\,.
\end{align}
For the symmetric interval, $A=0$ and $B=b$, from which we recover eq.~\eqref{displacement_ratio_sym}. Eq.~\eqref{CS tilted} now gives
\begin{align}\label{intermediate_CS_PS_S}
\begin{aligned}
    S^{(t)}_{\text{CS}}=&\frac{1}{2G_N^{\text{eff}}\mu_{\text{eff}}}\,\text{arctanh}\left(A+B\right)\\
    =&\frac{1}{2G_N^{\text{eff}}\mu_{\text{eff}}}\,\text{arctanh}\left(\frac{\sin\bar{\phi}}{\cos\frac{\Delta\phi}{2}}\right)+\frac{1}{4G_N^{\text{eff}}\mu_{\text{eff}}X}\,\Delta\tilde{u}\cot\frac{\Delta\phi}{2}+\mathcal{O}(X^{-2})\,.
    \end{aligned}
\end{align}
The second term is of the expected $c_M$ form and coincides with eq.~\eqref{CS_t_sym}. The first is an anomalous piece which is not rotationally invariant, since it depends on the position $\bar\phi$ of the interval, and which survives the flat limit. It cannot be attributed to the angle-dependent cutoff discussed in appendix \ref{Poin-slab-appendix}. The anomalous contribution to eq.~\eqref{TMG HEE} is insensitive to Weyl rescalings. Nor is it an artifact of the slab limit. In the null-anchored limit of section \ref{null-anchored-appendix}, placing both endpoints on the same cut of null infinity still leaves an analogous mismatch, as shown in eq.~\eqref{fraction_NA_appendix}.

\paragraph{The Carrollian frame:} The resolution lies in the choice of endpoint frames. Eq.~\eqref{CS tilted} uses $n=\partial_t$, which is natural for the $\text{CFT}_2$ but not for the limiting $\text{CCFT}_2$, where one wants the unit normal to the constant-$u$ cuts. Since the boundary sits at fixed $x=X$ throughout the limit, and the constant-$\tilde u$ cuts there are $t=\tilde u+\sqrt{X^2+y^2}$, that normal frame is
\begin{align}\label{local_Lorentz_poinc_slab}
\begin{aligned}
    n_{u}=&\sec\phi\,\partial_{t}+\tan\phi\,\partial_{y}, \hspace{20pt}&n_{u}\cdot n_{u}=-1\\
    \tilde{n}_{\phi}=&\tan\phi\,\partial_{t}+\sec\phi\,\partial_{y},\hspace{20pt}&\tilde{n}_{\phi}\cdot\tilde{n}_{\phi}=1
    \end{aligned}
\end{align}
with the inner products taken in the boundary metric. We stress that eq.~\eqref{local_Lorentz_poinc_slab} is exact at finite $X$, rather than only in the limit. Comparing with eq.~\eqref{related_normal_frames} with $m=\partial_t$ and $\tilde m=\partial_y$, the Carrollian frame is related to the one used in eq.~\eqref{CS tilted} by a local Lorentz transformation with rapidity
\begin{align}\label{chi_slab}
    \tanh\chi(\phi)=\sin\phi\,.
\end{align}

Using eqs.~\eqref{entropy_rel_dif_frames} and \eqref{chi_slab}, the anomalous contribution in the Carrollian frame is
\begin{align}\label{CS_frame_change_general}
    S^{(u)}_{\text{CS}}=S^{(t)}_{\text{CS}}-\frac{1}{4G_N^{\text{eff}}\mu_{\text{eff}}}\left[\text{arctanh}(\sin\phi_+)+\text{arctanh}(\sin\phi_-)\right].
\end{align}
The identity
\begin{align}\label{arctanh_identity}
    \text{arctanh}(\sin\phi_+)+\text{arctanh}(\sin\phi_-)=2\,\text{arctanh}\left(\frac{\sin\bar\phi}{\cos\frac{\Delta\phi}{2}}\right),
\end{align}
which follows from the addition formula for $\text{arctanh}$ together with $1+\sin\phi_+\sin\phi_-=\cos^2\frac{\Delta\phi}{2}+\sin^2\bar\phi$, shows that the frame change removes the first term in eq.~\eqref{intermediate_CS_PS_S} exactly. Combining eqs.~\eqref{intermediate_CS_PS_S}--\eqref{arctanh_identity} and using the subtraction formula for $\text{arctanh}$ together with $B=b(1-A^2)$, we find the closed form
\begin{align}\label{CS_u_general_exact}
    S^{(u)}_{\text{CS}}=\frac{1}{2G_N^{\text{eff}}\mu_{\text{eff}}}\left[\text{arctanh}(A+B)-\text{arctanh}(A)\right]=\frac{1}{2G_N^{\text{eff}}\mu_{\text{eff}}}\,\text{arctanh}\left(\frac{b}{1-A\,b}\right).
\end{align}
Expanding in the flat limit, with $\Delta u=(L/z_2)\Delta\tilde u$ and eq.~\eqref{leff_X},
\begin{align}\label{CS_u_general_expanded}
    S^{(u)}_{\text{CS}}=\frac{1}{4G_N^{\text{eff}}\mu_{\text{eff}}\ell_{\text{eff}}}\,\Delta u\,\cot\frac{\Delta\phi}{2}+\frac{\Delta u^2}{8G_N^{\text{eff}}\mu_{\text{eff}}\ell_{\text{eff}}^2}\,\frac{\sin\bar\phi\,\cos\frac{\Delta\phi}{2}}{\sin^2\frac{\Delta\phi}{2}}+\mathcal{O}\left((L/\ell_{\text{eff}})^3\right).
\end{align}

\paragraph{Discussion:} A few comments are in order. First, the leading term in eq.~\eqref{CS_u_general_expanded} is independent of $\bar\phi$ and coincides with eq.~\eqref{CS_slab_result}. Hence, once the Carrollian frame is used, the $c_M$ term \eqref{CS_slab_result} is obtained for an arbitrary interval, not merely a symmetric one.

Second, the identity \eqref{arctanh_identity} shows why the first term in eq.~\eqref{intermediate_CS_PS_S} should be regarded as an artifact of the frame choice. It is a sum of two terms, each of which depends on the position of a single endpoint. This is precisely the structure produced by a change of local data at the endpoints, as in eq.~\eqref{entropy_rel_dif_frames}. In contrast, the $c_M$ term depends on both endpoints through $\Delta u\cot\frac{\Delta\phi}{2}$, and cannot be removed in this way.

Third, the Carrollian frame does not remove all of the $\bar\phi$ dependence since the second term in eq.~\eqref{CS_u_general_expanded} still remains. However, this term is of $\mathcal{O}((L/\ell_{\text{eff}})^2)$, which is beyond the accuracy of eq.~\eqref{combined_result}, and it vanishes in the strict flat limit. Its presence is not surprising, since at finite $X$, the configuration is not invariant under rotations in $\phi$, \ie the corner of the wedge sits at fixed $x=X$ rather than at fixed $\tilde r$, and so rotational invariance is only expected to emerge in the limit. For the symmetric interval, $A=0$ and eq.~\eqref{CS_u_general_exact} reduces to $\text{arctanh}(b)$, so that the first correction to eq.~\eqref{CS_slab_result} is $\mathcal{O}((L/\ell_{\text{eff}})^3)$.

Finally, we note that the $\bar\phi$ dependence of the Chern-Simons contribution and that of the area contribution in eq.~\eqref{S_PS_provisional} have different origins. The latter comes from the angle-dependent cutoff \eqref{angle_dependent_cutoff} and is removed by choosing the endpoints on a cut of fixed $\tilde r$, \ie by the symmetric interval. The former is insensitive to the cutoff and is removed by the choice of endpoint frames. The symmetric interval used in the main text is the one for which neither issue arises at the order we work.

%
%

\section{Null-anchored limit: detailed computations}\label{NA_appendix}

In this appendix, we collect the details of the null-anchored limit of section \ref{null-anchored-appendix}. We describe the branes in global coordinates, construct the Minkowski and Bondi coordinates on the limiting branes, compute the entropy of a general interval together with the regulator \eqref{regulator_NA}, and derive the change of endpoint frame used for the anomalous contribution.

\subsection*{Branes in global coordinates} 

\begin{figure}
    \centering
    \begin{subfigure}[t]{0.26\textwidth}
        \centering
        \includegraphics[width=1.2\linewidth]{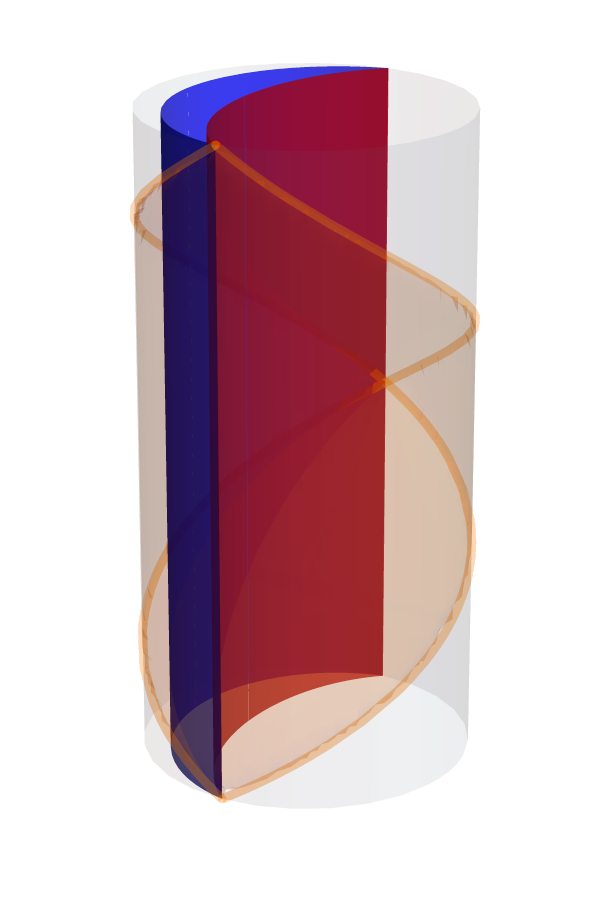}
        \caption{}
        \label{fig:branes-global-a}
    \end{subfigure}
    \hfill
    \begin{subfigure}[t]{0.26\textwidth}
        \centering
        \includegraphics[width=1.2\linewidth]{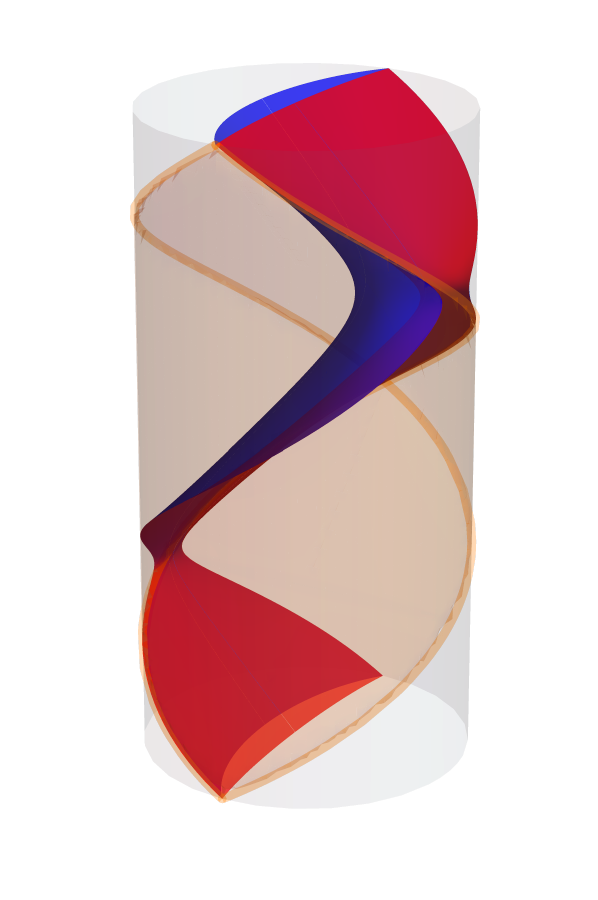}
        \caption{}
        \label{fig:branes-global-b}
    \end{subfigure}
    \hfill
    \begin{subfigure}[t]{0.26\textwidth}
        \centering
        \includegraphics[width=1.2\linewidth]{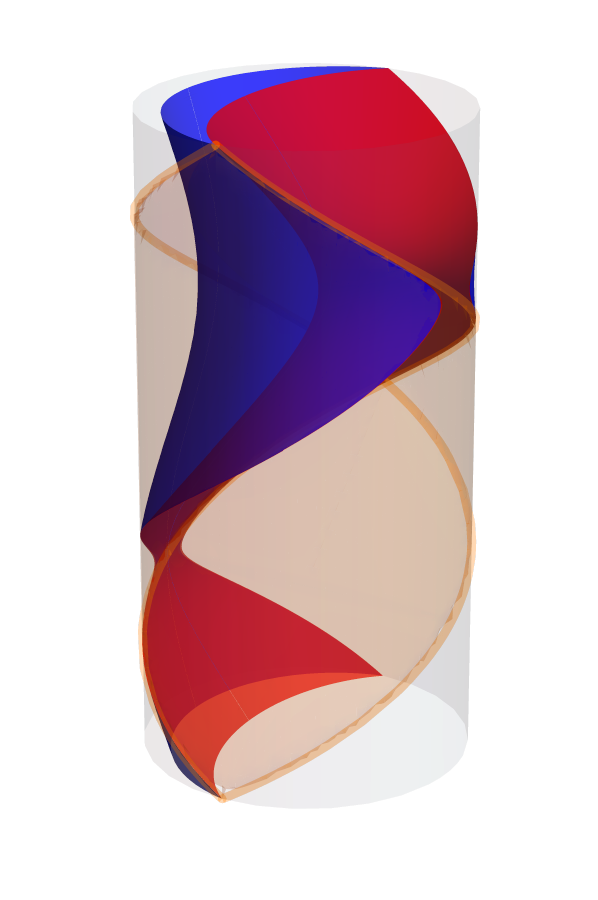}
        \caption{}
        \label{fig:branes-global-c}
    \end{subfigure}
    \caption{Slice $\varphi=\pi/2$ (\ie $y=0$) of $\text{AdS}_4$ in the global coordinates \eqref{change_to_global}, with the branes of Figure \ref{fig:poincare-branes-comparison}, of tensions $T_{\text{blue}}=0.95\,T_c$ and $T_{\text{red}}=0.6\,T_{\text{c}}$. The Poincar\'e patch is shown in orange. (a) $\eta=0$. (b) The boost \eqref{boosted branes in global} alone with $\eta=2.5$. (c) The joint limit \eqref{brane_eq_global_NA} with $\eta=2.5$. The Penrose diagrams are drawn as in appendix \ref{Poin-slab-appendix}.}
    \label{fig:all-global-branes}
\end{figure}

Using the change of coordinates \eqref{change_to_global}, a brane at $x=-\sinh\zeta\,z$ is described in global coordinates by
\begin{align}\label{brane position global}
    \sinh\rho\,\cos\theta=-\sinh\zeta\,.
\end{align}
These branes are shown for two values of the tension\footnote{Recall eqs.~\eqref{alpha(T)} and \eqref{tensions_mu}.} in Figure \ref{fig:branes-global-a}. As $|T|\to T_c$, they again approach the asymptotic boundary. After the boost \eqref{boost_NA}, the brane \eqref{brane equation poincare v2} becomes
\begin{align}\label{boosted branes in global}
    \cosh\eta\,\sinh\rho\,\cos\theta-\sinh\eta\,\cosh\rho\,\sin\tau=-\sinh\zeta\,.
\end{align}
At the asymptotic boundary $\rho\to\infty$, this reduces to $\cos\theta=\tanh\eta\,\sin\tau$. Hence, the boundary of the brane becomes increasingly null and is null in the limit $\eta\to\infty$, as shown in Figure \ref{fig:branes-global-b}. In the joint limit \eqref{jointlimit}, the branes are
\begin{align}\label{brane_eq_global_NA}
    \sinh\rho\,\cos\theta-\tanh\eta\,\cosh\rho\,\sin\tau=-\beta_i\,,
\end{align}
which become $\sinh\rho\,\cos\theta-\cosh\rho\,\sin\tau=-\beta_i$ as $\eta\to\infty$.
The slice $\tau=0$ of each brane is unchanged throughout, as shown in Figure \ref{fig:branes-global-c}.

\begin{figure}
    \centering
    \hspace{80pt}\includegraphics[width=0.65\linewidth]{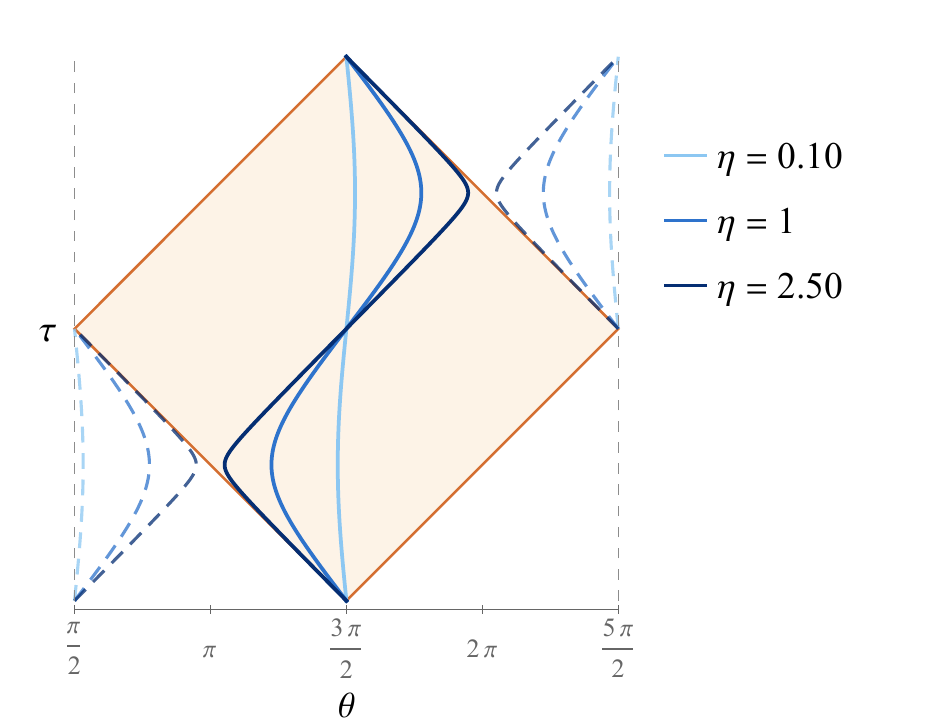}
    \caption{Unwrapped boundary of global $\text{AdS}_4$ for $\varphi=\pi/2$, we show the identified segments as vertical dashed lines. In orange we show the boundary of the Poincar\'e patch covered by the Poincar\'e coordinates eq.~\eqref{change_to_global} and in different shades of blue we show the intersection of the branes eq.~\eqref{brane_eq_global_NA} with the AdS boundary for different values of $\eta$.}
    \label{fig:bound_inter_NA}
\end{figure}

Moreover, in Figure \ref{fig:bound_inter_NA} we show the intersection of these branes with the asymptotic boundary of the $\text{AdS}_4$ spacetime for various values of $\eta$. We can see that, in contrast with the Poincar\'e-slab case, one piece of the boundary remains anchored to the middle of the Poincar\'e patch throughout the limit, while the rest go to null infinity on the boundary patch.

\paragraph{Relation with the Poincar\'e-slab branes:}
We can understand the relationship between the null-anchored limit and the Poincar\'e-slab limit by relating the Poincar\'e patches covered in both descriptions. To do this, it is simplest to work at the boundary of AdS, in which the asymptotic boundary of the branes satisfy the following equations in each limit:
\begin{align}
\begin{aligned}
    &\text{Poincar\'e-slab: } \cos\theta+\frac{x_0e^{\eta_P}}{L}\sin\theta-\frac{x_0e^{\eta_P}}{L}\cos\tau=0,\\
    &\text{Null-anchored: }\cos\theta-\tanh\eta_N\sin\tau=0,
    \end{aligned}
\end{align}
where we distinguished the $\eta$ parameter as $\eta_P$ for the Poincar\'e-slab and $\eta_N$ for the null-anchored limit since they need not be the same. 

As written, these two equations seem to describe different curves at the boundary. However, we could rename the coordinates involved in the first equation as $(\tilde{\tau},\tilde{\theta})$, and ask if there is a simple coordinate change $(\tilde{\tau},\tilde{\theta})\to(\tau,\theta)$ such that both equations agree. Renaming $X\equiv x_0e^{\eta_P}$ and $\eta\equiv\eta_N$, the relation:
\begin{align}\label{boundary_map}
    \tilde\tau=\tau-\frac{\pi}{2}, \hspace{40pt} \tilde{\theta}=\theta+\arctan(\sinh\eta),\hspace{20pt}\sinh\eta=\frac{X}{L},
\end{align}
maps the two equations to each other. This transformation is a rigid time shift, and an $\eta$-dependent rotation. 

\begin{figure}
    \centering
    \includegraphics[width=\linewidth]{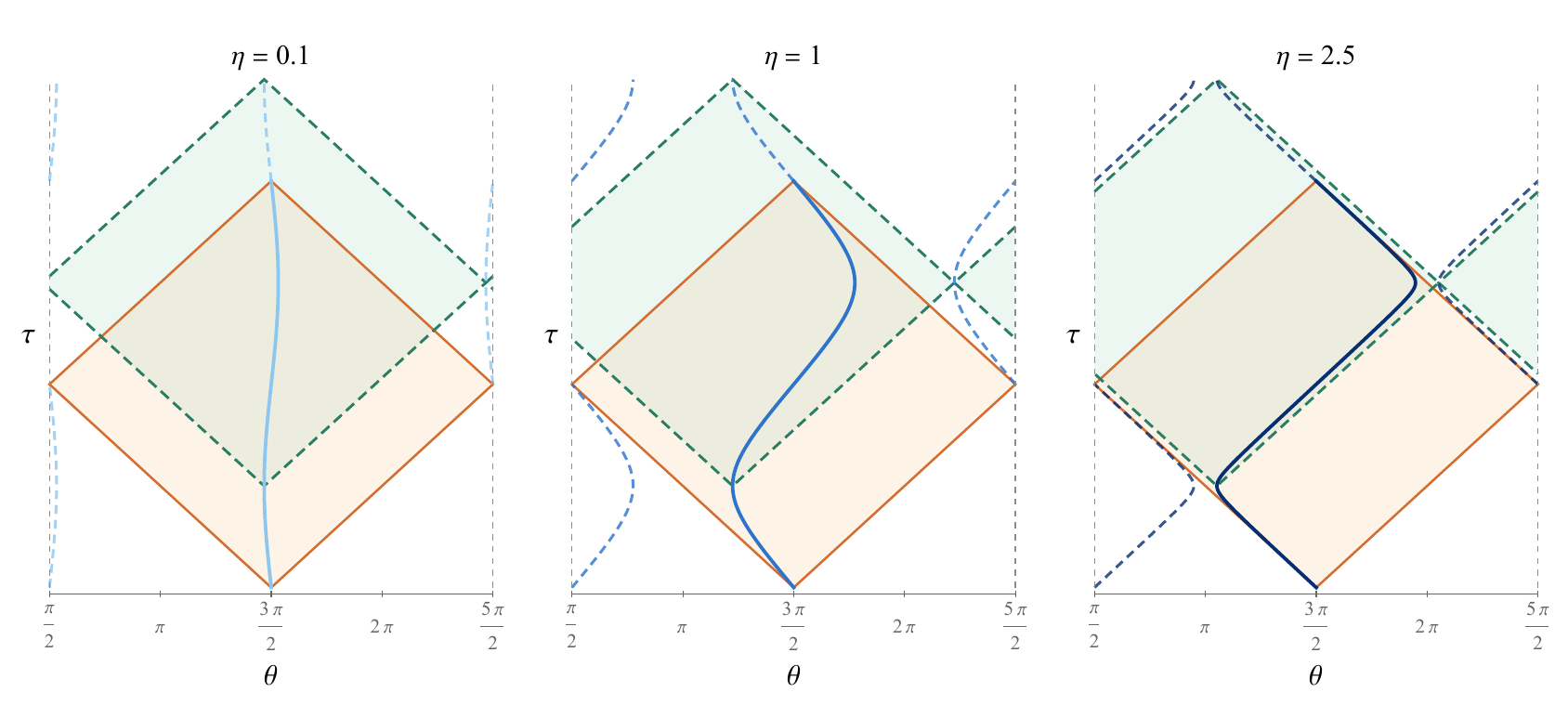}
    \caption{Unwrapped boundary of global $\text{AdS}_4$ for $\varphi=\pi/2$, we show the identified segments as vertical dashed lines. In orange we show the boundary of the Poincar\'e patch covered by the Poincar\'e coordinates eq.~\eqref{change_to_global} for the null-anchored limit, and in green the analogous for the Poincar\'e-slab limit. In different shades of blue we show the intersection of the branes eq.~\eqref{brane_eq_global_NA} with the AdS boundary for different values of $\eta$.}
    \label{fig:poinc_patch_relation}
\end{figure}

Now that we can map the two brane boundaries to each other, we can follow what happens to the two Poincar\'e patches as we take the limit. Recalling that at the boundary these patches are the regions $\cos\tau-\sin\theta>0$ and $\cos\tilde\tau-\sin\tilde\theta>0$, the transformation \eqref{boundary_map} makes the Poincar\'e patch of the Poincar\'e-slab limit look in the coordinates of the null-anchored description as shown in Figure \ref{fig:poinc_patch_relation}. 
Therefore, from the boundary perspective, the two descriptions differ in which Poincar\'e patch they cover throughout the limit. In fact, the transformation \eqref{boundary_map} can also be extended to the bulk by keeping $\tilde\rho=\rho$, in which case eq.~\eqref{brane_eq_joint_global} takes the form of eq.~\eqref{brane_eq_joint_null_anchored}  with $\beta_i=(L/z_i)\tanh\eta$. Since $\beta_i$ is held fixed in the original null-anchored construction, this does not establish an exact equivalence between the two families at finite $\eta$, although they become equivalent in the flat limit.\footnote{There exists, however, a more involved $\eta$-dependent AdS isometry that maps the two families exactly at corresponding stages of the limit, while mapping the Poincar\'e-slab patch to the same boundary region shown in Figure \ref{fig:poinc_patch_relation}.}

\subsection*{Minkowski and Bondi coordinates on the limiting branes} 

We consider the limiting brane $z=(t-x)/\beta$ of eq.~\eqref{brane_eq_Setting1_flat}, where $\beta=\beta_2$ for $Q_2$ (and the same construction applies to $Q_1$ with $\beta=\beta_1$). Its induced metric is eq.~\eqref{induced_NA}. Introducing
\begin{align}
    q\equiv t-x\,,\hspace{40pt}p\equiv t+x-\frac{q}{\beta^2}\,,
\end{align}
it becomes
\begin{align}\label{induced_NA_qp}
    ds^2=L^2\beta^2\,\frac{-dq\,dp+dy^2}{q^2}\,.
\end{align}
The conformal factor is removed by
\begin{align}\label{minkowski_coords_NA}
    X^0=\frac{1}{2}\left[p-\frac{y^2+L^2\beta^2}{q}\right],\hspace{20pt}X^1=\frac{1}{2}\left[p-\frac{y^2-L^2\beta^2}{q}\right],\hspace{20pt}X^2=L\beta\,\frac{y}{q}\,,
\end{align}
in terms of which $ds^2=-(dX^0)^2+(dX^1)^2+(dX^2)^2$. We have chosen the sign of $X^2$ so that the map $(t,x,y)\to(X^0,X^1,X^2)$ preserves the orientation. This matters for the parity-odd Chern-Simons contribution, and it is the reason why the angle $\phi$ below increases with $y$. 

Since $q>0$ on the brane, the Poincar\'e patch covers the half $X^0-X^1=-L^2\beta^2/q<0$ of the brane. In advanced Bondi coordinates, $X^0=v-r$, $X^1=r\cos\phi$ and $X^2=r\sin\phi$, the metric is $ds^2=-dv^2+2\,dv\,dr+r^2d\phi^2$, and the inverse transformation is
\begin{align}\label{change_to_mink}
    \begin{aligned}
        t=&\frac{v^2-2vr-L^2 (1+\beta^2)}{2\left[v-r(1+\cos\phi)\right]}\,,\hspace{40pt}
        &y=&-\frac{L\beta\, r\sin\phi}{v-r(1+\cos\phi)}\,,\\
        x=&\frac{v^2-2vr-L^2 (1-\beta^2)}{2\left[v-r(1+\cos\phi)\right]}\,,
        &z=&\ \frac{L^2\beta}{r(1+\cos\phi)-v}\,.
    \end{aligned}
\end{align}
The region $q>0$ is $v<r(1+\cos\phi)$, and so it contains all of $\mathscr{I}^-$ (\ie $r\to\infty$ at fixed $v$) except the generator $\phi=\pi$, but none of $\mathscr{I}^+$. Taking $r\to\infty$ in eq.~\eqref{change_to_mink} gives eq.~\eqref{change_at_asymptotic}, using $\tan(\phi/2)=\sin\phi/(1+\cos\phi)$, and solving $z=\epsilon$ for $r$ gives the cutoff surface \eqref{angle_dep_cutoff_NA}. Finally, the expressions in eq.~\eqref{minkowski_coords_NA} yield the relation
\begin{align}\label{minkowski_norm_NA}
    (X^0)^2-(X^1)^2-(X^2)^2=L^2\left[1+\beta^2\,\frac{t+x}{x-t}\right],
\end{align}
which is used in section \ref{min_surf_NA_sec}. Points with $t+x\to-\infty$ and $t-x\to0^+$ have $(X^0)^2-r^2\to+\infty$, while both $X^0\pm X^1\to-\infty$. They therefore approach past timelike infinity $i^-$.

\subsection*{General intervals and the regulator} 

For a general interval with endpoints $(v_\pm,\phi_\pm)$, eq.~\eqref{initial endpoints null anchored} gives the displacements
\begin{align}\label{displacements_NA}
    \Delta t=\frac{1}{2\cosh\eta}\left(\frac{v_+}{\cos^2\frac{\phi_+}{2}}-\frac{v_-}{\cos^2\frac{\phi_-}{2}}\right),\hspace{30pt}\Delta y=L\beta_2\,\frac{\sin\frac{\Delta\phi}{2}}{\cos\frac{\phi_+}{2}\cos\frac{\phi_-}{2}}\,.
\end{align}
Their ratio is eq.~\eqref{fraction_NA_appendix}, which follows from writing $\cos\frac{\phi_\mp}{2}/\cos\frac{\phi_\pm}{2}=\cos\frac{\Delta\phi}{2}\pm\tan\frac{\phi_\pm}{2}\sin\frac{\Delta\phi}{2}$. Since $\Delta t/\Delta y=\mathcal{O}(1/\cosh\eta)$,
\begin{align}\label{log_general_NA}
\begin{aligned}
    \log\left(\frac{\sigma}{\epsilon}\right)=&\log\left(\frac{L\beta_2}{\epsilon}\,\frac{\sin\frac{\Delta\phi}{2}}{\cos\frac{\phi_+}{2}\cos\frac{\phi_-}{2}}\right)\\
    &-\frac{1}{8L^2\beta_2^2\cosh^2\eta}\left[\Delta v\cot\frac{\Delta\phi}{2}+v_+\tan\frac{\phi_+}{2}+v_-\tan\frac{\phi_-}{2}\right]^2+\mathcal{O}(\cosh^{-4}\eta)\,.
\end{aligned}
\end{align}
In contrast to eq.~\eqref{intermediate_log_PS_gen}, there is no term linear in the time separation, and so no analogue of the spurious term \eqref{spurious_term} survives the flat limit here.

As explained in section \ref{results_setting1}, the corner metric restricted to $\mathscr{I}^-$ is $dy^2=(L\beta_2/2)^2\sec^4(\phi/2)\,d\phi^2$. Applying eq.~\eqref{weyl_entropy_contribution} with $w=\log\sec^2(\phi/2)$ at the two endpoints removes the factor $\cos\frac{\phi_+}{2}\cos\frac{\phi_-}{2}$ from the leading logarithm. The leading term then becomes $\frac{c}{3}\log\big(\frac{L\beta_2}{\epsilon}\sin\frac{\Delta\phi}{2}\big)$, which is rotationally invariant for any interval.

\paragraph{The non-universal constants:} The second term in eq.~\eqref{int_entropy} is given exactly by eq.~\eqref{constants_exact}. Throughout the null-anchored limit, $\sinh\zeta_i=\beta_i\cosh\eta$, and so at large $\eta$
\begin{align}\label{constants_NA}
    -\frac{L^2}{2G_N}\int_{\zeta_1}^{\zeta_2}\!\!d\zeta\,\cosh\zeta\log(\cosh\zeta)=-\frac{c}{3}\left(\Lambda-1\right)+\mathcal{O}(1/\cosh\eta)\,,
\end{align}
with $c$ and $\Lambda$ given in eqs.~\eqref{c_and_params_NA} and \eqref{regulator_NA}. Here we used $\log\cosh\zeta_i=\log\sinh\zeta_i+\mathcal{O}(\cosh^{-2}\eta)$, and that the difference of the $\arctan(\sinh\zeta)$ terms is $\mathcal{O}(1/\cosh\eta)$ for $\zeta_{1,2}>0$. The $\mathcal{O}(1/\cosh\eta)$ remainder is independent of the interval but of the same order as the $\Delta v^2$ term below, and we include it in the constants. As in the Poincar\'e-slab limit, the leading part of eq.~\eqref{constants_NA} is proportional to $c$, and is absorbed together with the factor $L\beta_2/2$ into the regulator \eqref{regulator_NA},
\begin{align}
    \frac{c}{3}\left[\log\left(\frac{L\beta_2}{\epsilon}\sin\frac{\Delta\phi}{2}\right)-\Lambda+1\right]=\frac{c}{3}\log\left(\frac{2}{\hat\epsilon}\sin\frac{\Delta\phi}{2}\right).
\end{align}
Explicitly, $\hat\epsilon=(2\epsilon\cosh\eta/L\beta_2)\,e^{\kappa-1}$ with $\kappa=(\beta_2\log\beta_2-\beta_1\log\beta_1)/(\beta_2-\beta_1)$, so that $\hat\epsilon$ is held fixed if $\epsilon\cosh\eta$ is, and thus satisfying eq.~\eqref{effective_cutoff_0}. 

\paragraph{The entropy of a general interval:} Combining the above with eq.~\eqref{c_and_params_NA} yields
\begin{align}\label{S_NA_general}
\begin{aligned}
    S=&\,\frac{\ell_{\text{eff}}}{2G_N^{\text{eff}}}\log\left(\frac{2}{\hat\epsilon}\sin\frac{\Delta\phi}{2}\right)-\frac{1}{16G_N^{\text{eff}}\ell_{\text{eff}}}\left[\Delta v\cot\frac{\Delta\phi}{2}+v_+\tan\frac{\phi_+}{2}+v_-\tan\frac{\phi_-}{2}\right]^2\\
    &\qquad+\text{constants}+\mathcal{O}\big(\epsilon,(L/\ell_{\text{eff}})^{2}\big)\,.
\end{aligned}
\end{align}
The subleading term depends on $v_+$ and $v_-$ separately, and so it is not invariant under translations $v_\pm\to v_\pm+a$ along $\mathscr{I}^-$, unless $\tan\frac{\phi_+}{2}+\tan\frac{\phi_-}{2}=0$, \ie unless $\phi_+=-\phi_-$. This is precisely the condition that both endpoints lie on the same cut \eqref{angle_dep_cutoff_NA} as $\epsilon\to0$, and so this dependence is again an artifact of the angle-dependent cutoff. For the symmetric interval \eqref{symmetric_NA}, the identity $\cot2\theta+\tan\theta=\csc2\theta$ with $\theta=\Delta\phi/4$ reduces the expression in the square brackets to $\Delta v\csc(\Delta\phi/2)$, and eq.~\eqref{S_NA_general} reduces to eq.~\eqref{result_setting_1}.

\paragraph{Correction around the Carrollian limit:}
As we did for the Poincar\'e-slab limit in eq.~\eqref{bound_metric_finite_l_PS}, it is instructive to analyze how the boundary metric looks at finite $\ell_{\text{eff}}$ in this limit. Using the coordinate relation \eqref{initial endpoints null anchored}, removing the corresponding Weyl factor, and using eq.~\eqref{c_and_params_NA} to write the expression in terms of $\ell_{\text{eff}}$ we are left with
\begin{align}
    \left(\frac{4}{L^2\beta^2}\cos^4\frac{\phi}{2}\right)\,\,ds^2=d\phi^2-\frac{1}{\ell^2_{\text{eff}}}\left(dv+v\tan\frac{\phi}{2} d\phi\right)^2+\mathcal{O}(\ell_{\text{eff}}^{-4}).
\end{align}
Again, identifying $1/\ell_{\text{eff}}$ with $v_{\gamma}$ we see that in the limit $v_{\gamma}\to0$ we recover the Carrollian limit of the cylinder metric in eq.~\eqref{bound_metric_finite_l_PS}. However, the subleading terms differ with respect to those of the cylinder and to those in the boundary metric of the Poincar\'e-slab limit. In this case, we also have an antisymmetric cross term, but the leading $g_{vv}$ is the same as for the cylinder, which eventually leads to the same $\csc^2$ subleading term in the entanglement entropy.

\subsection*{Change of frame for the anomalous contribution} 

With large but finite $\eta$ as we approach the flat limit, eq.~\eqref{initial endpoints null anchored} identifies points on the corner with points on $\mathscr{I}^-$ through $v=(1+\cos\phi)\cosh\eta\;t$ and $y=L\beta_2\tan(\phi/2)$. Using $1+\cos\phi=2/(1+y^2/L^2\beta_2^2)$, the constant-$v$ cuts are the curves
\begin{align}
    t=\frac{v}{2\cosh\eta}\left(1+\frac{y^2}{L^2\beta_2^2}\right)
\end{align}
on the corner. In the boundary metric $-dt^2+dy^2$, their future-pointing unit normal is
\begin{align}
    n_v=\cosh\chi\,\partial_t+\sinh\chi\,\partial_y\,,\hspace{40pt}\tanh\chi=\frac{v\,y}{L^2\beta_2^2\cosh\eta}=\frac{v}{L\beta_2\cosh\eta}\tan\frac{\phi}{2}\,,
\end{align}
which is eq.~\eqref{chi_NA}. This is of the form \eqref{related_normal_frames} with $m=\partial_t$ and $\tilde m=\partial_y$, as in the Poincar\'e-slab case \eqref{local_Lorentz_poinc_slab}. Since $y$ increases with $\phi$ (as shown above),
the spatial orientation is the same as there. Eq.~\eqref{entropy_rel_dif_frames} then gives
\begin{align}
    S^{(v)}_{\text{CS}}=\frac{1}{2G_N^{\text{eff}}\mu_{\text{eff}}}\,\text{arctanh}\left(\frac{\Delta t}{\Delta y}\right)-\frac{1}{4G_N^{\text{eff}}\mu_{\text{eff}}}\left(\chi_++\chi_-\right).
\end{align}
Expanding both terms with eqs.~\eqref{fraction_NA_appendix} and \eqref{chi_NA}, the contributions $v_\pm\tan(\phi_\pm/2)$ cancel at $\mathcal{O}(1/\cosh\eta)$, leaving
\begin{align}
    S^{(v)}_{\text{CS}}=\frac{1}{4G_N^{\text{eff}}\mu_{\text{eff}}}\,\frac{\Delta v}{L\beta_2\cosh\eta}\cot\frac{\Delta\phi}{2}+\mathcal{O}(\cosh^{-3}\eta)\,.
\end{align}
With $\ell_{\text{eff}}=L\beta_2\cosh\eta\,\big(1+\mathcal{O}(\cosh^{-2}\eta)\big)$ from eq.~\eqref{c_and_params_NA}, this is eq.~\eqref{CS_NA_result}. For the symmetric interval with $v_-=-v_+$, the endpoint rapidities are equal, $\chi_+=\chi_-$, rather than opposite as in the Poincar\'e-slab case. This is why the change of frame is needed here, even for the symmetric interval.

\section{Asymptotic symmetries and the flat limit of the wedge}
\label{wedge_symmetry_appendix}

In this appendix, we examine the realization of asymptotic symmetries as diffeomorphisms in our wedge holography framework and their behaviour under the controlled flat limit. In particular, we distinguish the conventional BMS$_3$ generators associated with asymptotic gravitational states from those associated with the Virasoro generators adapted to local operator insertions, as derived by Caminiti and Pasterski~\cite{Caminiti:2026pma}. Our objective is to exhibit their three- and four-dimensional geometrical realizations and the standard charge algebra in the gravitational zero-mode sector. We do not attempt a complete four-dimensional covariant phase space analysis that allows fluctuations of the radion and massive Kaluza--Klein modes. 

\subsection*{Brown-Henneaux diffeomorphisms and their four-dimensional lift}

The original wedge metric \eqref{Poincare slicing} is
\begin{align}\label{sym_wedge_metric}
 ds_4^2=L^2d\zeta^2+\cosh^2\!\zeta\,\gamma_{ab}(x)dx^a dx^b\,,
 \qquad \zeta_1\leq\zeta\leq\zeta_2\,,
\end{align}
where $\gamma_{ab}$ is the $\mathrm{AdS}_3$ metric with curvature radius $L$. The two EOW branes are at fixed $\zeta=\zeta_i$ (with $i=1,2$), and the induced geometry on the brane $Q_2$ has radius $\ell_{\mathrm{eff}}=L\cosh\zeta_2$. With Poincar\'e coordinates, we write
\begin{align}
 \gamma_{ab}\,dx^a\, dx^b=\frac{L^2}{w^2}
 \big(dw^2+dx^+\,dx^-\big)\,,\qquad x^\pm=y\pm t\,.
\end{align}
The leading Brown-Henneaux asymptotic diffeomorphisms in AdS$_3$ are parameterized by two independent chiral functions $f(x^+)$ and $g(x^-)$, with bulk representatives that preserve Fefferman-Graham gauge given by \cite{Brown:1986nw,Compere:2015knw}
\begin{align}\label{sym_BH_vector}
\begin{aligned}
 \xi^+&=f(x^+)-\frac{w^2}{2}g''(x^-)\,,\\
 \xi^-&=g(x^-)-\frac{w^2}{2}f''(x^+)\,,\\
 \xi^w&=\frac w2\big(f'(x^+)+g'(x^-)\big)\,.
\end{aligned}
\end{align}
These representatives give the two chiral Witt algebras at the conformal boundary, with their ordinary Lie brackets understood modulo asymptotically trivial subleading transformations. The centrally extended Virasoro algebras arise from the canonical charges, rather than from the vector-field brackets themselves~\cite{Brown:1986nw}.

For each such vector field, we define its lift into the wedge by
\begin{align}\label{sym_four_lift}
 \hat\xi^\zeta=0\,,\qquad
 \hat\xi^a=\xi^a(x)\,,\qquad
 \partial_\zeta\hat\xi^a=0\,.
\end{align}
Hence the action of these diffeomorphisms on the four-dimensional metric satisfy
\begin{align}\label{sym_lie_lift}
 \mathcal L_{\hat\xi}g_{\zeta\zeta}=0\,,\qquad
 \mathcal L_{\hat\xi}g_{\zeta a}=0\,,\qquad
 \mathcal L_{\hat\xi}g_{ab}
 =\cosh^2\!\zeta\,\mathcal L_\xi\gamma_{ab}\,.
\end{align}
Thus the lift preserves the transverse foliation. Further, these vectors are tangent to both EOW branes, and they respect the covariant Neumann boundary conditions there. This establishes the kinematic realization of both chiral asymptotic symmetry algebras throughout the wedge. It does not, by itself, determine the central extension of the  four-dimensional charge algebra. We return to this question in a moment.

\subsection*{Two conformal frames and the local-operator generators}

Before proceeding, we would like to compare with the Virasoro generators adapted to local operator insertions, as proposed in \cite{Caminiti:2026pma}. For this purpose, we imagine compactifying the corner of the wedge to a Lorentzian cylinder with dimensionless coordinates $(\tau,\phi)$ and define the null coordinates $y^\pm=\phi\pm\tau$. In the conventional state frame, the Virasoro generators can be chosen as smooth Fourier modes on the cylinder. Caminiti and Pasterski propose an alternative conformal frame adapted to local operator insertions at $y^+=y^-=0$, with chiral vector fields (see their eq.~(3.28))
\begin{align}\label{sym_CP_vectors}
 \ell_n=-\sin y^+\left(i\tan\frac{y^+}{2}\right)^n\partial_+\,,
 \qquad
 \bar\ell_n=-\sin y^-\left(-i\tan\frac{y^-}{2}\right)^n\partial_-\,.
\end{align}
These vector fields separately realize the two Witt algebras (wherever the fields are regular). The chiral functions appearing as the prefactors in these vector fields can be inserted into eq.~\eqref{sym_BH_vector}, after the usual change to a Fefferman-Graham chart on the cylinder, and then lifted to the four-dimensional wedge by eq.~\eqref{sym_four_lift}. Thus, the alternative Virasoro generators \eqref{sym_CP_vectors} admit Brown-Henneaux extensions into the AdS$_3$ slices, which can in turn be lifted into the four-dimensional wedge. However, they are generally well-defined only locally because of singularities in the boundary generators, as we emphasize below.

The relevant Carrollian contraction of these local-operator generators is~\cite{Caminiti:2026pma}
\begin{align}\label{sym_CP_contraction}
 L_n^{\mathrm{op}}=\ell_n+(-1)^n\bar\ell_n\,,\qquad
 M_n^{\mathrm{op}}=v_\gamma\big(\ell_n-(-1)^n\bar\ell_n\big)\,.
\end{align}
Writing $y^\pm=\phi\pm v_\gamma v$, this gives in the limit $v_\gamma\to0$\footnote{Compare with eqs.~\eqref{sym_flat_boundary_field} and \eqref{sym_flat_boundary_field22} below.}
\begin{align}\label{sym_CP_BMS}
 L_n^{\mathrm{op}}\longrightarrow
 F_n(\phi)\partial_\phi+vF_n'(\phi)\partial_v\,,\qquad
 M_n^{\mathrm{op}}\longrightarrow F_n(\phi)\partial_v\,,
\end{align}
where
\begin{align}\label{sym_CP_F}
 F_n(\phi)=-\sin\phi\left(i\tan\frac\phi2\right)^n\,.
\end{align}
The central charges of this contraction are $c_L^{\mathrm{op}}=c_++c_-$ and $c_M^{\mathrm{op}}=v_\gamma(c_+-c_-)$, in agreement with the coefficients inferred from the limiting RT entropy. By contrast, the conventional geometric contraction in eq.~\eqref{geometric_contraction_discussion}, which pairs $\mathcal L_n$ with $\bar{\mathcal L}_{-n}$, yields $c_L^{\mathrm{geom}}=c_+-c_-$ and $c_M^{\mathrm{geom}}=v_\gamma(c_++c_-)$, reproducing the central charges in eq.~\eqref{geometric_charges_discussion}. The two choices correspond to distinct prescriptions of the conformal frame and need not define globally equivalent limits~\cite{Caminiti:2026pma}.

However, one might be concerned that the vector fields in eq.~\eqref{sym_CP_vectors} are not globally regular. For $n\geq2$, the positive-index chiral modes have a pole at $y^+=\pi$ (and similarly for the barred sector), and for $n\leq-2$, the pole is at $y^+=0$.\footnote{Of course, the singularities discussed here are reminiscent of those of the conventional Virasoro generators $\ell_n=-z^{n+1}\partial_z$ on the complex plane, which have poles at $z=0$ for $n\leq-2$ and at $z=\infty$ for $n\geq2$. Through a series of conformal transformations, ref.~\cite{Caminiti:2026pma} moves these singularities to $\tau_\mt{E}=0$ and $\phi=0,\pi$ on the Euclidean cylinder for their operator frame generators.} Thus the full basis is adapted to local operator insertions rather than to smooth global vector fields on the Lorentzian boundary cylinder. We can make the consequences for the holographic framework explicit by considering $n=2$ as an example.

Consider global AdS$_3$ in Fefferman-Graham (FG) coordinates\cite{Compere:2015knw},\footnote{In the convention of eq.~\eqref{sym_wedge_metric}, this metric is $\gamma_{ab}dx^adx^b$. That is, we do not include the $\cosh^2\!\zeta$ factor in the three-dimensional metric, and thus the relevant AdS radius is $L$ and not $\ell_{\text{eff}}$.}
\begin{align}\label{sym_global_FG}
\begin{aligned}
 ds_3^2
 &=\frac{L^2}{z^2}\left[\,
 dz^2-\frac{z^2}{4}\,{dy^+}^2-\frac{z^2}{4}\,{dy^-}^2+\left(1+\frac{z^4}{16}\right)dy^+ dy^-\, \right]
 \,.
 \end{aligned}
\end{align}
For a chiral boundary field $f(y^+)\,\partial_+$, one exact representative preserving the FG gauge is\footnote{Here and below the primes on $f$ denote derivatives with respect to $y^+$.}
\begin{align}\label{sym_global_FG_lift}
 \xi^+=f-\frac{z^4}{8D(z)}f''\,,\quad
 \xi^-=-\frac{z^2}{2D(z)}f''\,,\quad
 \xi^z=\frac z2 f'\,,\quad{\rm with}\ \  D(z)=1-\frac{z^4}{16}\,.
\end{align}
These expressions are valid only within the region where the Fefferman--Graham coordinate chart is regular. In particular, we restrict attention to regions where the coordinates are nonsingular and the corresponding vector fields are smooth, without assuming that they extend globally across AdS$_3$. 

Now, direct substitution gives
\begin{align}\label{sym_global_variation}
 (\mathcal L_\xi g)_{++}
 =-\frac{L^2}{2}\big(f'''+f'\big)\,,\qquad
 (\mathcal L_\xi g)_{+-}
 =\frac{L^2z^2}{16}\big(f'''+f'\big)\,,\qquad
 (\mathcal L_\xi g)_{--}=0\,.
\end{align}
For the $n=2$  generator in eq.~\eqref{sym_CP_vectors}, we have
\begin{align}
 f(y^+)&=\sin y^+\,\tan^2\frac{y^+}2
 =-\frac4{y^+-\pi}+O(y^+-\pi)\,,\\
 f'''+f'&=\frac32\,\sec^4\frac{y^+}2=\frac{24}{(y^+-\pi)^4}+\frac4{(y^+-\pi)^2}+O(1)\,,
\end{align}
where we have shown the singular behaviour of these expressions as $y^+\to\pi$.
Now consider the $2\pi$-periodic chiral functions $\mathcal L_\pm(y^{\pm})=g_{\pm\pm}/L^2$, which characterize the boundary graviton degrees of freedom in the Ba\~nados family of asymptotically AdS$_3$ geometries \cite{Banados:1998gg,Compere:2015knw}
\begin{align}\label{Banados_family}
  {ds_3^2=\frac{L^2}{z^2}\left[
 dz^2+z^2\mathcal{L}_+{dy^+}^2+z^2\mathcal{L}_-{dy^-}^2+\left(1+z^4\mathcal{L}_+\mathcal{L}_-\right)dy^+ dy^- \right]
 \,
 }.
\end{align}
For global AdS$_3$ as in eq.~\eqref{sym_global_FG}, they take the constant vacuum values $\mathcal L_+=\mathcal L_-=-1/4$, corresponding to the Casimir energy of the dual CFT$_2$ on the cylinder. Hence, under the chiral diffeomorphism generated by the $n=2$ generator above, we obtain
\begin{align}\label{sym_CP_pole}
 \delta_{\ell_2}\mathcal L_+
 =-\frac34\sec^4\frac{y^+}{2}
 \underset{y^+\to\pi}{\sim}
 -\frac{12}{(y^+-\pi)^4}\,.
\end{align}
The transformation obeys the local radial Brown--Henneaux falloffs away from the pole, but it fails to preserve the usual globally smooth phase space of periodic $\mathcal L_\pm$. The natural four-dimensional lift inherits the same singularity,
\begin{align}
 (\mathcal L_{\hat\ell_2}g^{(4)})_{++}
 =-\frac{3L^2\cosh^2\!\zeta}{4}
 \sec^4\frac{y^+}{2}\,.
\end{align}
Thus the singularity is present on both branes as well as in the interior of the transverse interval.

The obstruction also appears in the usual canonical charges. The variation of the canonical charge associated with a smooth chiral generator $g(y^+)$ includes
\begin{align}
 \delta_{\ell_2} Q_g=\frac{\ell_{\text{eff}}}{8\pi G_N^{\rm eff}}
 \int_0^{2\pi}d\phi\,g(y^+)\delta_{\ell_2}\mathcal L_+(y^+)\,,
\end{align}
where $G_N^{\rm eff}$ is the effective three-dimensional Newton constant obtained by dimensional reduction in eq.~\eqref{sym_effective_G}.
On a constant-$\tau$ cut, let $g=1$ be the smooth zero-mode generator. Regulating the singular point symmetrically yields
\begin{align}\label{sym_pole_charge}
\begin{aligned}
\delta_{\ell_2} Q_{g}\sim& \int_{-\pi+\delta}^{\pi-\delta}d\phi\,
 \delta_{\ell_2}\mathcal L_+
 =-\frac34\int_{-\pi+\delta}^{\pi-\delta}d\phi\,
 \sec^4\frac\phi2\\
 &=-\cot^3\frac\delta2-3\cot\frac\delta2
 =-\frac8{\delta^3}-\frac4\delta+O(\delta)\,.
\end{aligned}
\end{align}
Even the variation of a smooth charge diverges. Consequently, $\ell_2$ is not tangent to the conventional smooth Brown-Henneaux gravitational phase space. 

Similar singularities arise for all higher modes. As noted above, for $n\geq2$, the functions $f_n(y^+)$ have poles of order $n-1$ at $y^+=\pi$, and hence the corresponding variations of the boundary graviton data develop poles of order $n+2$. Analogous singularities arise for $n\leq-2$ at $y^+=0$, with poles of order $|n|+2$ in the variations of the boundary graviton data. However, these singularities do not necessarily imply that every associated charge integral diverges. For example, when $n\geq3$ is odd, the singular variation is odd about $y^+=\pi$, and its integral against an even smearing function vanishes under a symmetric principal value prescription. Such cancellations do not resolve the failure of these generators to preserve the standard smooth Brown-Henneaux phase space. Establishing well-defined canonical charges would require a consistent prescription for the singular boundary data, together with an analysis of charge integrability and the resulting symmetry algebra.

This does not mean that the local operator frame Virasoro algebra of \cite{Caminiti:2026pma} is inconsistent. Rather, realizing these generators as gravitational symmetries would require extending the usual Brown--Henneaux phase space to accommodate their singular behaviour, for example, by allowing punctures on the boundary or defining charges locally around the singularities. Such a construction would require specifying appropriate boundary conditions and demonstrating that the associated charges are finite, integrable, and satisfy the expected symmetry algebra. In particular, although the alternative Virasoro generators admit a well-defined algebraic contraction, the resulting central terms cannot automatically be identified with the canonical charges of asymptotically flat three-dimensional gravity. Establishing that identification would require an independent analysis of the corresponding gravitational phase space and its surface charges.

\subsection*{The null-anchored limit and BMS$_3$ on the flat wedge}

The null-anchored limit described in section~\ref{NA_limit_subsection} gives a direct geometrical realization of the conventional BMS transformations. Let us define
\begin{align}
 \varepsilon_\eta=\frac1{\cosh\eta}\qquad{\rm and}\qquad
 A(\phi)=1+\cos\phi\,.
\end{align}
At the corner, eq.~\eqref{initial endpoints null anchored} becomes
\begin{align}\label{sym_corner_map}
 t=\frac{\varepsilon_\eta v}{A(\phi)}\,,\qquad
 y=L\beta_2\,\tan\frac\phi2\,.
\end{align}
Consider a boundary conformal vector field $\xi=f(x^+)\,\partial_++g(x^-)\,\partial_-$, where as above, $x^\pm=y\pm t$. Since $\delta y=(f+g)/2$ and $\delta t=(f-g)/2$, finiteness at fixed $(v,\phi)$ requires $f-g=O(\varepsilon_\eta)$ while $f+g=O(1)$. A convenient family which realizes this scaling is
\begin{align}
 f(s)=h(s)+\varepsilon_\eta k(s)\,,\qquad
 g(s)=h(s)-\varepsilon_\eta k(s)\,.
\end{align}
Expanding at $x^\pm=y\pm t$ and using eq.~\eqref{sym_corner_map}, one obtains
\begin{align}\label{sym_flat_boundary_field}
 \xi\ \ \longrightarrow \ \ Y(\phi)\partial_\phi
 +\big[T(\phi)+vY'(\phi)\big]\partial_v\,,
\end{align}
with
\begin{align}\label{sym_flat_boundary_field22}
 Y(\phi)=\frac{A(\phi)}{L\beta_2}\, h(y(\phi))\qquad{\rm and}\qquad
 T(\phi)=A(\phi)k(y(\phi))\,.
\end{align}
This is the standard BMS$_3$ boundary vector field. We note that in the conventional geometric contraction, the reflection of one chiral mode index arises because the two Virasoro algebras have opposite angular Fourier conventions. Pairing $\mathcal L_n$ with $\bar{\mathcal L}_{-n}$ ensures that both contributions have the same angular dependence $e^{in\phi}$ in the Carrollian limit, yielding the standard BMS$_3$ generators.

Further, we observe that the stereographic relation $y=L\beta_2\,\tan(\phi/2)$ in eq.~\eqref{sym_corner_map} provides a direct connection with the generators \eqref{sym_CP_vectors} of \cite{Caminiti:2026pma}. In particular, choosing $h(y)=-y(iy/L\beta_2)^n$ in eq.~\eqref{sym_flat_boundary_field} reproduces the angular functions $F_n(\phi)$ in eq.~\eqref{sym_CP_F}. The singularities of the positive-index generators at $\phi=\pi$ correspond to $y\to\infty$. The functions $Y,T$ need not be globally smooth when $h,k$ are chosen to reproduce the singular local-operator basis.

For the limiting four-dimensional geometry, it is convenient to use coordinates shared by all flat slices. Starting with the null-anchored branes $t-x=\beta_i z$ and the AdS$_4$ Poincar\'e coordinates of eq.~\eqref{Poincare metric}, define
\begin{align}\label{sym_flat_coordinates}
 \beta=\frac{t-x}{z}\,,\qquad q=t-x\,,\qquad
 p=t+x-\frac{q}{\beta^2}\,,\qquad a=L\beta_2\,.
\end{align}
The branes are at constant $\beta=\beta_{1,2}$. Defining three-dimensional Cartesian coordinates as in eq.~\eqref{minkowski_coords_NA} with the substitution $\beta\to\beta_2$,
we find the exact warped-product metric
\begin{align}\label{sym_flat_warp}
{\displaystyle
 ds_4^2=\frac{L^2}{\beta^2}d\beta^2+
 \frac{\beta^2}{\beta_2^2}
 \big[-(dX^0)^2+(dX^1)^2+(dX^2)^2\big]\,.}
\end{align}
The induced metric on the UV brane is precisely Minkowski space. Introducing advanced Bondi coordinates $X^0=v-r$, $X^1=r\cos\phi$, $X^2=r\sin\phi$ yields
\begin{align}\label{sym_flat_bondi}
 ds_4^2=\frac{L^2}{\beta^2}d\beta^2+
 \frac{\beta^2}{\beta_2^2}
 \big(-dv^2+2\,dv\,dr+r^2d\phi^2\big)\,.
\end{align}
The asymptotic three-dimensional Bondi residual diffeomorphisms can be written as
\begin{align}\label{sym_BMS_bondi}
 \begin{aligned}
 \xi^v&=T+vY'\,,\\
 \xi^\phi&=Y+\frac{T'+vY''}{r}+O(r^{-2})\,,\\
 \xi^r&=-rY'-T''-vY'''+O(r^{-1})\,.
 \end{aligned}
\end{align}
Their leading action at $\mathscr I^-$ is eq.~\eqref{sym_flat_boundary_field}. By setting $\hat\xi^\beta=0$ and $\partial_\beta\hat\xi^a=0$, the entire BMS$_3$ family admits an asymptotic lift throughout the limiting four-dimensional wedge. These transformations preserve the two brane positions and the warped-product structure \eqref{sym_flat_bondi}, just as in eq.~\eqref{sym_lie_lift}. Their asymptotic brackets close into the desired BMS$_3$ symmetry, up to trivial subleading diffeomorphisms. Smooth representatives may be modified in the interior, but a pole already present in $Y(\phi)$ or $T(\phi)$ at null infinity cannot be removed by such a modification. Recall that the Poincar\'e chart of the null-anchored limit omits the generator $\phi=\pi$ (see appendix~\ref{NA_appendix}). Therefore, statements about global Fourier charges require the smooth completion of the angular circle.

\subsection*{Canonical charges in the gravitational zero-mode sector}

Eq.~\eqref{sym_flat_warp} also supplies a direct reduction of the gravitational zero-mode sector. Let us replace the flat three-dimensional metric in eq.~\eqref{sym_flat_warp} by a $\beta$-independent metric $h_{ab}(X)$, 
\begin{align}\label{sym_flat_warp_hh}
{\displaystyle
 ds_4^2=\frac{L^2}{\beta^2}d\beta^2+
 \frac{\beta^2}{\beta_2^2}
 \,h_{ab}\,dX^a\,dX^b\,.}
\end{align}
When $h_{ab}$ is locally Ricci flat, the metric ansatz \eqref{sym_flat_warp_hh} satisfies the four-dimensional Einstein equations. Furthermore, the extrinsic curvature of each constant-$\beta$ brane satisfies $K_{ab}=(T_i/2)h_{ab}$, with the critical tensions $T_1=-2/L$ and $T_2=2/L$. These geometries therefore form a family of exact solutions in the gravitational zero-mode sector. Now the coefficient of $R^{(3)}[h]$ from the four-dimensional Einstein action is 
\begin{align}\label{sym_effective_G}
 \frac1{G_N^{\mathrm{eff}}}
 =\frac1{G_N}\int_{\beta_1}^{\beta_2}
 \frac{L\,d\beta}{\beta_2}
 =\frac{L}{G_N}\left(1-\frac{\beta_1}{\beta_2}\right)\,.
\end{align}
This reproduces eq.~\eqref{RS1_eff_G}, using the identification $\beta_1/\beta_2=z_2/z_1$. Thus, the dimensional reduction of the limiting wedge directly recovers the effective Newton constant previously obtained by taking the flat limit of the AdS$_3$ brane theory. 

Hence the KK reduction to three dimensions yields the standard Einstein-Hilbert action with Newton constant $G_N^{\rm eff}$. Within this reduced theory, the canonical charges consequently have the usual normalization and reproduce the standard BMS$_3$ central extension \cite{Barnich:2006av,Barnich:2012aw}. 
Note that as in appendix~\ref{brane description}, our considerations here fix the radion and do not excite the massive KK tower. Hence establishing that the full four-dimensional symplectic structure reduces to this form, including all brane and corner contributions, would require a separate covariant phase-space analysis.

For definiteness, an advanced-Bondi family of three-dimensional locally flat solutions may be written \cite{Barnich:2006av,Barnich:2012aw}
\begin{align}\label{sym_general_flat_metric}
 ds_3^2=\Theta(\phi)dv^2+2\,dv\,dr+r^2d\phi^2
 +2\left[\Xi(\phi)+\frac v2\Theta'(\phi)\right]dv\,d\phi\,.
\end{align}
Here, $\Theta(\phi)$ and $\Xi(\phi)$ play an analogous role as $\mathcal{L}_{\pm}$ in eq.~\eqref{Banados_family}. Three-dimensional Minkowski space is recovered by $\Theta=-1$ and $\Xi=0$.

The corresponding Einstein charges are then \cite{Barnich:2006av,Barnich:2012aw}
\begin{align}\label{sym_BMS_charge}
 Q[T,Y]=\frac1{16\pi G_N^{\mathrm{eff}}}
 \int_0^{2\pi}d\phi\,
 \big[T(\phi)\Theta(\phi)+2Y(\phi)\Xi(\phi)\big]\,.
\end{align}
In these conventions, their coadjoint action includes
\begin{align}\label{sym_coadjoint}
 \begin{aligned}
 \delta_{T,Y}\Theta&=Y\Theta'+2Y'\Theta-2Y'''\,,\\
 \delta_{T,Y}\Xi&=Y\Xi'+2Y'\Xi
 +\tfrac12T\Theta'+T'\Theta-T'''\,.
 \end{aligned}
\end{align}
The third-derivative terms give the familiar central extension. 
The resulting BMS$_3$ algebra \eqref{BMS3} has central charges\footnote{As usual, we have shifted the zero modes by $L_0\to L_0+c_L/24$ and $M_0\to M_0+c_M/24$, so that the Minkowski vacuum (with $\Theta=-1$ and $\Xi=0$) has vanishing charges.}
\begin{align}\label{sym_central_geom}
 c_L^{\mathrm{geom}}=0\,,\qquad
 c_M^{\mathrm{geom}}=\frac{3}{G_N^{\mathrm{eff}}}\,.
\end{align}
When an intrinsic gravitational Chern-Simons term is added on the UV brane, the effective coupling of the massless graviton is $\mu_{\mathrm{eff}}$ given in eq.~\eqref{lam_eff_def}. The standard TMG charges then give~\cite{Castro:2014tta,Bagchi:2012yk}
\begin{align}\label{sym_central_TMG}
 c_L^{\mathrm{geom}}=\frac{3}{G_N^{\mathrm{eff}}\mu_{\mathrm{eff}}}\,,\qquad
 c_M^{\mathrm{geom}}=\frac{3}{G_N^{\mathrm{eff}}}\,,
\end{align}
with the sign of the first term determined by the parity and null-time orientation conventions. These central charges agree with the geometric contraction of the finite-AdS$_3$ Virasoro algebras described in eq.~\eqref{geometric_contraction_discussion}. In this contraction, the two chiral sectors are combined with opposite mode orientations, yielding $c_L^{\mathrm{geom}}=c_+-c_-$ and $c_M^{\mathrm{geom}}=v_\gamma(c_++c_-)$, where $v_\gamma=1/\ell_{\mathrm{eff}}$. Substituting the finite-AdS$_3$ central charges from eq.~\eqref{grav_anomaly} reproduces eq.~\eqref{geometric_charges_discussion}, in agreement with the canonical charges of the reduced gravitational theory.

To close, we emphasize again that our calculations are deliberately restricted in scope, and our conclusions should be understood accordingly. Eqs.~\eqref{sym_central_geom} and \eqref{sym_central_TMG} follow from the induced three-dimensional gravitational sector and its familiar canonical charges, rather than from an independent calculation of the complete four-dimensional surface-charge two-form using the covariant phase-space methods of \cite{Iyer:1994ys,Barnich:2001jy}.
Such a calculation would need to include the Neumann brane terms, the localized Chern-Simons action, possible corner terms and the allowed radion and KK fluctuations, and to establish finiteness and integrability on the resulting phase space. Finally, the coefficients inferred from RT entropy by comparison with the HWR CCFT$_2$ formula are those of the local-operator contraction, not automatically these geometric charges. Determining whether the corresponding limiting observables can be accommodated in a common global phase space remains an open question.

\bibliographystyle{style_files/utphys}
\bibliography{references}

\end{document}